\documentclass{bmcart}

\usepackage{amsthm,amsmath}
\usepackage{url}
\RequirePackage{hyperref}
\usepackage[utf8,latin1]{inputenc} 
\usepackage{graphicx} 
\usepackage{dcolumn} 
\usepackage{bm} 
\usepackage{hyperref} 
\usepackage{float} 
\usepackage{mathtools} 
\usepackage{caption} 
\usepackage{subcaption} 
\usepackage{diagbox} 
\usepackage{braket}
\usepackage{calc}
\usepackage{color,soul}
\usepackage{tabularx}

\begin{document}

\begin{frontmatter}

\begin{fmbox}
\dochead{Research}


\title{Benchmarking Machine Learning Algorithms for Adaptive Quantum Phase Estimation with Noisy Intermediate-Scale Quantum Sensors}

\author[
 addressref={aff1},                   
 email={nelsoncosta.jr17@gmail.com}   
]{\inits{N.F.C.}\fnm{Nelson Filipe} \snm{Costa}}
\author[
 addressref={aff1,aff2},
 email={yasser.omar@tecnico.ulisboa.pt}
]{\inits{Y.O.}\fnm{Yasser} \snm{Omar}}
\author[
  addressref={aff3},
	corref={aff3},
	email={aidar.sultanov@aalto.fi}
]{\inits{A.S.}\fnm{Aidar} \snm{Sultanov}}

\author[
addressref={aff3},
email={sorin.paraoanu@aalto.fi}
]{\inits{G.S.P.}\fnm{Gheorghe Sorin} \snm{Paraoanu}}

\address[id=aff1]{
  \orgdiv{Instituto Superior T\'{e}cnico},             
  \orgname{Universidade de Lisboa},          
  \city{Lisbon},                              
  \cny{Portugal}                                    
}

\address[id=aff2]{%
  \orgname {Instituto de Telecomunica\c{c}\~oes},
  \orgdiv{Physics of Information and Quantum Technologies Group},
  \city{Lisbon},
  \cny{Portugal}
}

\address[id=aff3]{
  \orgdiv{QTF  Centre  of  Excellence,  Department  of  Applied  Physics},             
  \orgname{Aalto  University  School  of  Science},          
  \city{Helsinki},                              
  \cny{Finland}                                    
}


\end{fmbox}


\begin{abstractbox}
	
	\begin{abstract} 
		
		Quantum phase estimation is a paradigmatic problem in quantum sensing and metrology.  Here we show that adaptive methods based on classical machine learning algorithms can be used to enhance the precision of quantum phase estimation when noisy non-entangled qubits are used as sensors. We employ the Differential Evolution (DE) and Particle Swarm Optimization (PSO) algorithms to this task and we identify the optimal feedback policies which minimize the Holevo variance. We benchmark these schemes with respect to scenarios that include Gaussian and Random Telegraph fluctuations as well as reduced Ramsey-fringe visibility due to decoherence. We discuss their robustness against noise in connection with real experimental setups such as Mach-Zehnder interferometry with optical photons and Ramsey interferometry in trapped ions, superconducting qubits and nitrogen-vacancy (NV) centers in diamond.
	\end{abstract}
	
	
	
	
\end{abstractbox}
%

\end{frontmatter}



\section{Introduction}

Precision measurements play a fundamental role in physics, as they constitute a key ingredient of many state-of-the-art applications and experiments testing the limits of scientific theories.
But the accuracy to which such measurements can be performed is itself governed by the laws of physics - and, ultimately, by those of quantum mechanics \cite{giovannetti2004quantum, giovannetti2006quantum, giovannetti2011advances,toth2014quantum}.


A generic measurement protocol for estimating the value of an unknown parameter consists of preparing a probe in a desired initial state, allowing it to interact with the physical system whose state depends on the parameter and, finally, obtaining a measurement result that encapsulates the information about it. This process, however, is often affected by systematic and statistical errors. While the source of the former may stem from  imperfect calibration of the measurement instruments, the origin of the latter can either be accidental, due to insufficient control of the measurement chain, or fundamental, deriving from the nature of the physical measurement \cite{helstrom1969quantum, holevo2011probabilistic}. Fortunately, statistical errors, regardless of their origin, can be minimized by repeating the process and averaging the resulting outcomes, as a consequence of the central limit theorem \cite{cramer2016mathematical}. This theorem states that, given a large number $N$ of independent measurement results, their average will converge to a Gaussian distribution with standard deviation scaling as $1/\sqrt{N}$. In metrology, this  is referred to as the Standard Quantum Limit (SQL). 
For many practical applications that involve noisy and error-prone systems, it is essential to devise protocols that yield precision levels close to the SQL. However, it is known that this limit is not fundamental -- the  ultimate limit set by the laws of quantum physics is the Heisenberg Limit (HL) \cite{braginski1975, braginsky1980quantum, braginsky1995quantum, ozawa1989realization}.

Recently, several systems have achieved sufficient levels of technological maturity to allow the experimental exploration of these limits. Due to the non-idealities present in these systems, they are typically referred to as NISQ (noisy intermediate-scale quantum). Largely, the motivation for this development has been rooted in quantum information processing, where efficient and precise measurements and state manipulation is required \cite{Buluta_2011}, and indeed significant progress towards the implementation of quantum gates and algorithms has been made in optical setups \cite{Pryde_2019, Flamini_2018}, NV centers in diamond \cite{Wrachtrup_2006, Prawer_2018}, trapped ions \cite{Lange2012, Sage_2019} and superconducting circuits \cite{Paraoanu2014, Nori_review_2011}.  
Since it employs similar control and measurement techniques as quantum computing, the exploration of quantum-enhancing techniques has grown as a separate field, usually referred to as quantum metrology \cite{degen_2017,paris2009quantum}, with applications such as ultrasensitive force detection \cite {biercuk_2010}, adaptive environment sensing \cite{Scerri_2020}, near-surface electric field measurements \cite{brownnutt_2015}, sensing of weak signals \cite{review_Naderi} and even detection of gravitational waves
\cite{abadie2011gravitational}.

The paradigmatic protocol in quantum metrology is quantum phase estimation. The quantum phase is a parameter that cannot be directly measured and yet, it contains information about other quantities of interest, such as electric or magnetic fields. 
The traditional quantum metrology approach to phase estimation has been through the use of highly entangled states such as NOON states \cite{giovannetti2004quantum, giovannetti2011advances, Smerzi_RMP}, as well as other specially optimized states \cite{berry2000optimal,berry2001optimal,berry2009perform}. However, highly entangled states tend to be very sensitive to noise: for NOON states even the loss of a single qubit (photon) results in a separable state. Thus, the implementation of this method is challenging with respect to real-life applications. Fortunately, it has been later realized  that precise quantum phase estimation does not necessarily require entanglement. In optics, it was demonstrated that adaptive homodyne phase measurements yield uncertainties close to the quantum uncertainty of coherent states \cite{Wiseman1995,Mabuchi2002}. It was also shown how to adaptively estimate the angle of a half waveplate that characterizes the linear polarization of photons \cite{Fujiwara_2006,Takeuchi2012} as well as how to perform phase-shift estimation with polarization-encoded qubits \cite{Paris2010}. Furthermore, the inverse quantum Fourier transform,  which is the last step in Shor's factorization algorithm and which typically requires gate entangling, can be carried out using local measurements and feedback \cite{PhysRevLett.76.3228}. This approach has been used to break the SQL in experiments with photons \cite{higgins_2007}, superconducting qubits \cite{danilin2018quantum} and NV centers in diamond \cite{Hanson2016}.

The incorporation of machine learning techniques in estimation protocols is a natural step forward. Seminal theoretical work on employing reinforcement learning algorithms has demonstrated the potential of these methods for reaching sensitivities below the SQL when used in conjunction with entanglement \cite{Sanders2010, Sanders2011, Lovett2013, palittpongarnpim2016single, palittapongarnpim2016controlling, palittapongarnpim2017learning, palittapongarnpim2018robustness}. Recently, some of these methods have been tested experimentally in an optics setup with their estimation precision being limited by the SQL \cite{Lumino2018}. This only upholds the potential of applying machine learning for the optimization of parameter estimation under resources and noise level constraints, but the foundations of these methods are still poorly understood. In mathematical statistics, the limits of machine learning algorithms are an active area of investigation -- for example deriving bounds on the number of samples needed to reach a certain accuracy \cite{Wossnig2020}.  However, the issue at stake is that typically all the mathematical results, including the SQL, are derived from the so-called central limit $N \gg 1$ and for ideal noiseless systems. Yet the relevant regime for the present status of quantum technologies in the NISQ era is that of moderate $N$ and significant noise. In this context no general unbiased estimator has been found yet \cite{Smerzi_RMP}.

The objective of this work is to employ two machine learning algorithms, the DE and the PSO algorithm, in the design of adaptive estimation schemes capable of driving the measurement precision close to the the SQL. We numerically show that the SQL is still a valid bound even in the regime of not too large $N$. We also demonstrate that machine learning algorithms can be used even in the presence of various experimental noises, consequently providing robust policies with better performance than non-adaptive protocols. For practical devices such as magnetometers based on single qubits, these methods can be directly applied. More precisely, we observe that in order to increase the precision of these devices, one straightforward option is to increase the sensitivity to the parameter to be estimated by, for example, using higher-energy states or by increasing the response  using a different biasing point. However, both of these strategies result in an increased coupling to noise sources and, therefore, a compromise needs to be reached in order to achieve the maximum performance. This will be further detailed in the paper when analyzing the experimental realizations. 

Both the DE and the PSO algorithm can also be employed as subroutines in order to enhance other quantum algorithms. For example, algorithms that use variable sensing times, or multipass techniques, are in principle able to breaks the SQL and reach the HL. Instead of averaging at every sensing time, which is the typical approach used one can further increase the sensitivity by using our technique. Beyond quantum metrology, machine learning protocols could become useful in other quantum-information paradigmatic problems that involve phase estimation, such as factorization, sampling, and computation of molecular spectra \cite{NielsenChuang}. In particular, our calculations are relevant for NISQ quantum technologies, where the number of qubits is limited and subject to errors and noise. Overall, by benchmarking these two important protocols for the task of quantum phase estimation, we hope that machine learning algorithms will be more prominently employed in applications such as optical interferometry and magnetometry, where the increase of precision is essential and where the aim is set on reaching the HL.

The paper is organized in the following sections. Section II describes the general concept of adaptive quantum phase estimation in Ramsey interferometry, discussing the updating procedure as well as the relevant sources of noise. Section III presents the two machine learning algorithms, the DE and PSO algorithm. Section IV presents our main results, where we show how the two algorithms allow us to approach the SQL. In this section we also provide an analysis on the effects of Gaussian noise, Random Telegraph noise (RTN) and quantum decoherence on the performance of the algorithms. Section V discusses the implementation of our protocol on several experimental platforms, namely Mach-Zehnder optical interferometers, superconducting qubits, trapped ions and defects in diamond. Finally, we conclude our results in Section VI.

\section{Adaptive Quantum Phase Estimation Scheme}

To perform the estimation of an unknown phase $\phi$, we use a generic adaptive scheme that is able to adjust the value of a control phase $\theta$ to match the value of $\phi$, based on the results of previous measurements in a Ramsey interferometric sequence. 

\begin{figure}[h!]
	\centering
	\includegraphics[width=300pt]{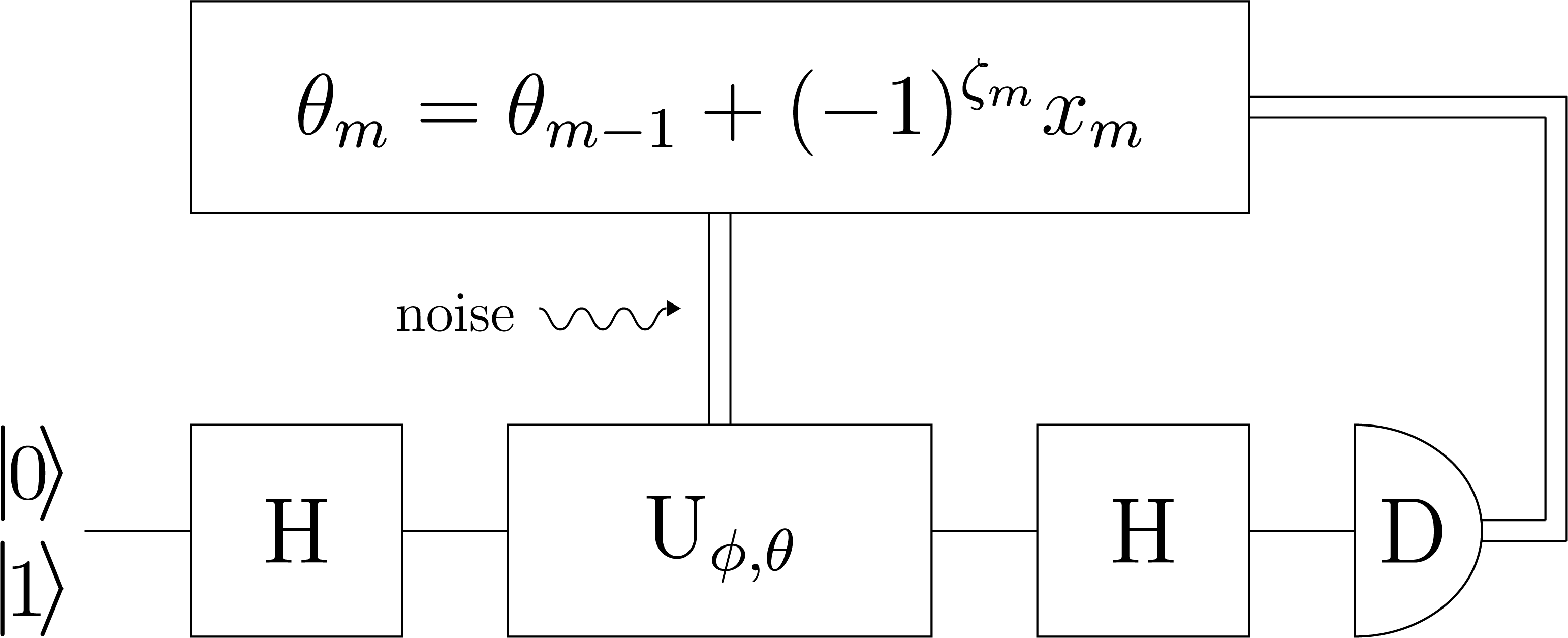}
	\caption{Adaptive quantum phase estimation scheme. A qubit $m$ is injected into the quantum circuit either in the quantum state $\ket{0}$ or $\ket{1}$ and its measured outcome $\zeta_m \in \{0,1\}$ is used together with some policy phase instruction $x_m \in \left[0,2\pi\right[$ to prepare the circuit for the next round of measurements. The quantum circuit consists of two Hadamard gates $\mathrm{H}$, a phase shift gate $\mathrm{U}_{\phi ,\theta}$, a detector D, and a processing unit with phase update instructions for the controllable phase shifter.}
	\label{fig:aqem}
\end{figure}

The schematic representation of the adaptive quantum phase estimation scheme displayed in Fig. \ref{fig:aqem} consists of a quantum circuit made of a Hadamard gate $\mathrm{H}$, a phase shift gate $\mathrm{U}_{\phi ,\theta}$, another Hadamard gate $\mathrm{H}$ and a detector augmented with a processing unit to calculate the value of $\theta$ for the next round of measurements. Using standard notations in quantum information, the Hadamard and phase shift gates can be respectively defined as
\begin{equation}
\mathrm{H} = \frac{1}{\sqrt{2}}
\begin{bmatrix}
	1 & 1 \\ 1& -1
\end{bmatrix}
\quad \textrm{and} \quad
\mathrm{U}_{\phi ,\theta}  = 
\begin{bmatrix}
	1 & 0 \\ 0 & e^{i (\phi-\theta)}
\end{bmatrix}.
\label{umatrix}
\end{equation}

Under this adaptive quantum estimation scheme, an ensemble of $N$ qubits is injected sequentially into the circuit in randomly chosen quantum states, either in state $\ket{0}$ or state $\ket{1}$, and their outcome is measured to prepare the value of the controllable phase shifter for the next incoming qubit. The input state of the quantum circuit takes the form $\ket{0}_{1}\otimes \ket{1}_{2} \otimes \ket{1}_{3} \otimes .... \otimes \ket{0}_{N}$, which is manifestly separable. After the last $N^{th}$ qubit is injected and its outcome measured, the final phase value of $\theta$ is considered to be the estimated value of $\phi$. The initial states of the qubits can be represented in Dirac notation as
\begin{equation*}
\ket{0} = \begin{bmatrix} 1 \\0 \end{bmatrix}
\quad \textrm{and} \quad
\ket{1} = \begin{bmatrix} 0 \\1 \end{bmatrix}.
\end{equation*} 

The idea is to have at the end of the process an estimated controllable phase value $\theta$ as close as possible to the value of the unknown phase $\phi$. The state of each qubit after the second Hadamard gate is
\begin{equation*}
\ket{\psi_{\pm}} = \frac{1}{2} \left[ 1 \pm  e^{i (\phi - \theta)} \right] \ket{0} + \frac{1}{2} \left[ 1 \mp  e^{i (\phi - \theta)} \right] \ket{1},
\end{equation*}
where the upper sign corresponds to a qubit whose initial state was $\ket{0}$ and the lower sign to one whose initial state was $\ket{1}$. This yields measurement results $\zeta \in \{0,1\}$ with probabilities
\begin{equation}
\mathcal{P}_{+}(\zeta = 0|\phi , \theta ) =
\mathcal{P}_{-}(\zeta = 1|\phi , \theta ) =
\cos^{2}\frac{\phi - \theta }{2},
\label{eq:prob0}
\end{equation}
and
\begin{equation}
\mathcal{P}_{+}(\zeta = 1|\phi , \theta ) =
\mathcal{P}_{-}(\zeta = 0|\phi , \theta ) =
\sin^{2}\frac{\phi - \theta}{2}.
\label{eq:prob1}
\end{equation}

If the phase difference $\phi - \theta$ is zero, then $\theta$ approximates very well the unknown phase $\phi$ and a qubit prepared in state $\ket{0}$ will also exit in state $\ket{0}$. Similarly, the same logic can be applied for a qubit injected in the quantum state $\ket{1}$.

In order to estimate the phase $\phi$ at every step $m$ of the algorithm the controllable phase $\theta_{m}$ has to be updated based on the previous result $\zeta_{m-1}$ of the measurement. Here we adopt a standard updating rule
\begin{equation}
\theta_m = \theta_{m-1} + (-1)^{\zeta_m} x_m,
\label{eq:update}
\end{equation}
which has already been employed successfully in similar setups \cite{palittpongarnpim2016single, palittapongarnpim2016controlling, palittapongarnpim2017learning}, and where the initial value of $\theta$ can be fixed to $\theta_0=0$ without any loss of generality.

Note that the term $x_m$ in Eq. (\ref{eq:update}) represents the update policy which will be determined by the machine learning algorithms. This update rule can be viewed as a decision tree where for each qubit $m$, the controllable phase shifter must update its value $\theta_m$ either by adding or by subtracting $x_m$ depending on the measured outcome state of the previous qubit $\zeta_{m} \in \{0,1\}$. Consequently, for $N$ qubits the number of possible $\theta_m$ values increases exponentially as $2^{(N+1)}-1$. This is a Markovian feedback, since the new controllable phase $\theta_m$ depends only on the latest measured outcome state $\zeta_{m}$.

To evaluate the performance of a policy, we use the Holevo variance \cite{berry2009perform,berry2001optimal,berry2000optimal}, defined as
\begin{equation}
V_{\rm H} = S^{-2}-1 = |\langle e^{i(\phi-\theta)}\rangle|^{-2}-1,
\label{eq:holevo}
\end{equation}
where $\langle e^{i(\phi-\theta)} \rangle$ represents the average value of $e^{i(\phi-\theta)}$ for different phase values $\phi$ and their respective estimates $\theta$ considered in the learning process of the machine learning algorithms. Here we make the notation abbreviation $\theta = \theta_N$, since the Holevo variance of a policy can only be calculated after the last qubit $N$ is injected into the circuit and its outcome measured. The quantity $S = \langle e^{i(\phi-\theta)} \rangle \in [0,1]$ is called the sharpness of the phase distribution, where the value $S=1$ corresponds to a perfect estimation of $\phi$. For periodically bounded variables such as the phase, the Holevo variance is a direct measure of the standard deviation by $(\Delta\phi)^2=V_{\rm H}$. Therefore we have $V_{\rm H} \sim 1/N$ for the SQL. 
It is also important for the model of the designed adaptive quantum phase estimation scheme to include the possibility of errors and imperfections that occur in a real experimental situation. This provides an important test to the robustness of the algorithms to relatively general sources of noise which can be encountered on most experimental platforms.

The first source of noises to be considered are the noises in the application of the controlled unitary phase shift transformation $\mathrm{U}_{\phi - \theta}$, namely the Gaussian and the Random Telegraph noise. Note that the Random Telegraph noise is particularly relevant for experiments involving solid-state qubits.

In the scenario of Gaussian noise, the noise follows a normal distribution parametrized by the standard deviation $\sigma$. Letting $\theta^{{\scriptscriptstyle(\mathrm{GSN})}}$ represent the actual value of the controllable phase shifter subjected to the noise, the Gaussian noise distribution can be defined as:
\begin{equation}
p(\theta^{{\scriptscriptstyle\mathrm{GSN}}}_{m}) = \frac{1}{\sqrt{2\pi}\sigma} 
\exp\left[-\frac{1}{2\sigma^2}(\theta^{{\scriptscriptstyle\mathrm{GSN}}}_{m}-\theta_{m})^2\right].
\label{eq:gaussian}
\end{equation}

In the scenario of Random Telegraph noise, the noise follows a discrete distribution where at each time step the the controllable phase shifter value can be randomly offset by a fixed valued $\lambda$ with a probability $\eta$. Letting $\theta^{{\scriptscriptstyle (\mathrm{RTN})}}$ represent the value of the controllable phase shifter subject to this source of noise, the Random Telegraph noise distribution can be described as:
\begin{equation}
p(\theta^{{\scriptscriptstyle\mathrm{RTN}}}_{m}) = 
\begin{cases}
	1 - \eta ,         & \text{$\theta^{{\scriptscriptstyle\mathrm{RTN}}}_{m} = \theta$}_{m}\\
	\eta  ,  & \text{$\theta^{{\scriptscriptstyle\mathrm{RTN}}}_{m} = \theta_{m} + \lambda$}.
\end{cases}
\label{eq:rtn}
\end{equation}

The other important type of noise to be considered affects the qubit itself. This modifies the unitary evolution generated by a Hermitian Hamiltonian $\mathcal{H}$ to a non-unitary evolution given by the Lindblad master equation
\begin{equation}
\dot{\rho} = -\frac{i}{\hbar} [\mathcal{H}, \rho ] + \Gamma_{1} \mathcal{D}[\sigma_{-}](\rho ) +
\frac{\Gamma_{\varphi}}{2} \mathcal{D}[\sigma_{z}](\rho )
\label{eq:Lindblad}
\end{equation}
where $\Gamma_{1}$ is the relaxation rate, $\Gamma_{\varphi}$ is the pure dephasing rate and 
the dissipation superoperator $\mathcal{D}$ acting on the density matrix $\rho$ is defined by
$\mathcal{D}[A](\rho ) = A \rho A^{\dag} - \frac{1}{2} \{A^{\dag}A, \rho \}$. It is also useful to introduce the standard notations $T_{1}$ and $T_{2}$ for the relaxation and decoherence time, 
$T_1 = 1/\Gamma_{1}$ and $T_{2} = 1/\left ( \Gamma_{1}/2 + \Gamma_{\varphi}\right)$. 

To implement the phase gate $U_{\phi - \theta_{m}}$ from Eq. (\ref{umatrix}) the Hamiltonian must be of $\sigma_z$ type, with a component for the unknown phase $\phi$ and another one for the control $\theta_{m}$. Typically, for Ramsey experiments with a phase accumulation time $\tau$, we have $\mathcal{H}_{m}=\frac{\hbar}{2}(\phi/\tau - \theta_{m}/\tau)\sigma_{z}$ at the step $m$, with $U_{\phi - \theta_{m}} = \exp [- i \mathcal{H}_{m} \tau /\hbar]$, up to a global phase factor. The solution of Eq. (\ref{eq:Lindblad}) is a $2 \times 2$ matrix with elements $\rho_{00}(\tau ) = 1- \exp (-\tau /T_{1})\rho_{11}(0)$, $\rho_{01}(\tau ) = \exp (-i \phi + i \theta_{m} -\tau /T_{2})\rho_{01}(0)$, $\rho_{10}(\tau ) = \exp (i \phi - i \theta_m  -\tau /T_{2})\rho_{10}(0)$ and $\rho_{11}(\tau ) = \exp (- \tau /T_{1})\rho_{11}(0)$. If the state at $\tau=0$ is prepared by the action of the Hadamard gate from either $|0\rangle$ or $|1\rangle$, corresponding respectively to the $+$ and $-$ signs below, and at the final time $\tau$ we apply again a Hadamard gate, we obtain that at every step $m$ of the algorithm the probabilities are modified as
\begin{equation}
P_{\pm}(\zeta_{m}|\phi, \theta_{m}) = \frac{1}{2} \left[1 \pm  (-1)^{\zeta_{m}} \nu \cos (\phi - \theta_{m}) \right],
\label{eq:decoherence}
\end{equation}
where $\nu = \exp( - \tau /T_{2})$ is called interference visibility. One can check that for maximum visibility, $\nu=1$, we recover Eqs. (\ref{eq:prob0}) and (\ref{eq:prob1}). Further considerations can be found in Appendix \ref{apx:holevo}.

\section{Machine Learning Algorithms}

The problem of quantum phase estimation relies on a sequential and cumulative set of measurements to drive the estimation process, thus making it an ideal problem for reinforcement learning algorithms. In this work, we considered the Differential Evolution (DE) \cite{storn1997differential, price2006differential} and the Particle Swarm Optimization (PSO) \cite{kennedy2011particle, eberhart1995new, shi1998modified}, among other reinforcement learning algorithms, as they are the most commonly employed for similar tasks in literature \cite{Sanders2010, Sanders2011, Lovett2013, palittpongarnpim2016single, palittapongarnpim2016controlling, palittapongarnpim2017learning, palittapongarnpim2018robustness}.

These algorithms employ a direct search method to the exploration of the search space generated by all the possible policy configurations. Direct search methods use a greedy criterion to drive their exploration. Such methods guarantee fairly fast convergence times, even though such fast convergence times often come at the expense of becoming trapped in a local minimum. This comes as result of the greedy criterion promoting decisions that lead to shorter and more immediate term rewards usually in detriment of fully exploring the total search space. To avoid this scenario, it is important to perform a thorough study on the controllable parameters of each of the mentioned algorithms before applying them to the quantum phase estimation task. To evaluate the performance of an algorithm there are two fundamental criteria: its ability to converge within the imposed number of iterations to a solution and its ability to converge to a valid solution.

\subsection{Differential Evolution}

The implementation of the DE algorithm to the current problem starts with a set of $P$ populations, each representing a candidate solution for the adaptive scheme update policy. Each of these populations is a vector of size $N$, with each entry representing a new phase instruction to prepare the controllable phase shifter for each of the qubits being injected into the system. The DE algorithm at each iteration will employ its direct search method to $P$ $N$-dimensional vectors $x^{G}_{i,j}$, where $i \in \lbrace 1, 2, ..., P \rbrace$ and $j \in \lbrace 1, 2, ..., N \rbrace$ represent each entry of the candidate solutions vectors and $G$ represents the generation of the population vectors. 

Each of these vectors is initialized with random values for each entry in the interval $x^0_{i,j} \in [0, 2\pi]$. Afterwards, at each iteration, the DE algorithm generates possible new candidate solution vectors for the next generation by adding the weighted difference between four population vectors to a fifth vector. This process is referred to as mutation:
\begin{equation*}
\tilde{u}^{G+1}_{i,j} =  x^{G}_{r_1,j} + F \cdot (x^{G}_{r_2,j} + x^{G}_{r_3,j} - x^{G}_{r_4,j} - x^{G}_{r_5,j}).
\end{equation*}

Here $F$ represents a constant value in the interval $[0,1]$ which controls the amplification of the difference between the considered populations. Hence, $F$ will be referred to as the amplification parameter. Note as well that all indexes $\lbrace r_1, ..., r_5\rbrace$ are randomly chosen integer values always different between themselves. At this point, the entries of the newly mutated vectors $\tilde{u}^{G+1}_{i,j}$ are randomly mixed with the originally corresponding vectors to increase their diversity. This process is referred to as crossover:
\begin{equation*}
\tilde{x}^{G+1}_{i,j} = \left\{
\begin{array}{ll}
	x^{G}_{i,j} \quad \quad \text{, if $R_1 > C$ and $j \neq R_2$}\\
	\tilde{u}^{G+1}_{i,j} \quad \; \text{, if $R_1 \le C$ or $j = R_2$}.
\end{array}
\right.
\end{equation*}

This process is controlled by the crossover parameter $C$ which can take any value in the interval $[0,1]$. Hence, a crossover only occurs if the random value $R_1$, which is generated for each population member at each iteration, is below or equal to the chosen crossover parameter. The value of $R_2$ is an integer randomly chosen for each population at each iteration of the evaluation process to ensure that at least one entry from the newly mutated vectors $\tilde{u}^{G+1}_{i,j}$ is passed to the trial vector for the next generation $\tilde{x}^{G+1}_{i,j}$. 

Finally, the new trial vectors are compared against the population vectors of the previous generation to see which perform best against the cost function of the problem and, as a result, become a member of the next generation of populations. This process is referred to as selection:
\begin{equation*}
x^{G+1}_{i,j} = \left\{
\begin{array}{ll}
	x^{G}_{i,j} \quad \quad \text{, if $f\big(x^{G}_{i,j}\big) < f\big(\tilde{x}^{G+1}_{i,j}\big)$}\\
	\tilde{x}^{G+1}_{i,j} \quad \; \text{, if $f\big(x^{G}_{i,j}\big) \ge f\big(\tilde{x}^{G+1}_{i,j}\big)$}.
\end{array}
\right.
\end{equation*}

\begin{figure}[h!]
	\centering
	\includegraphics[width=300 pt]{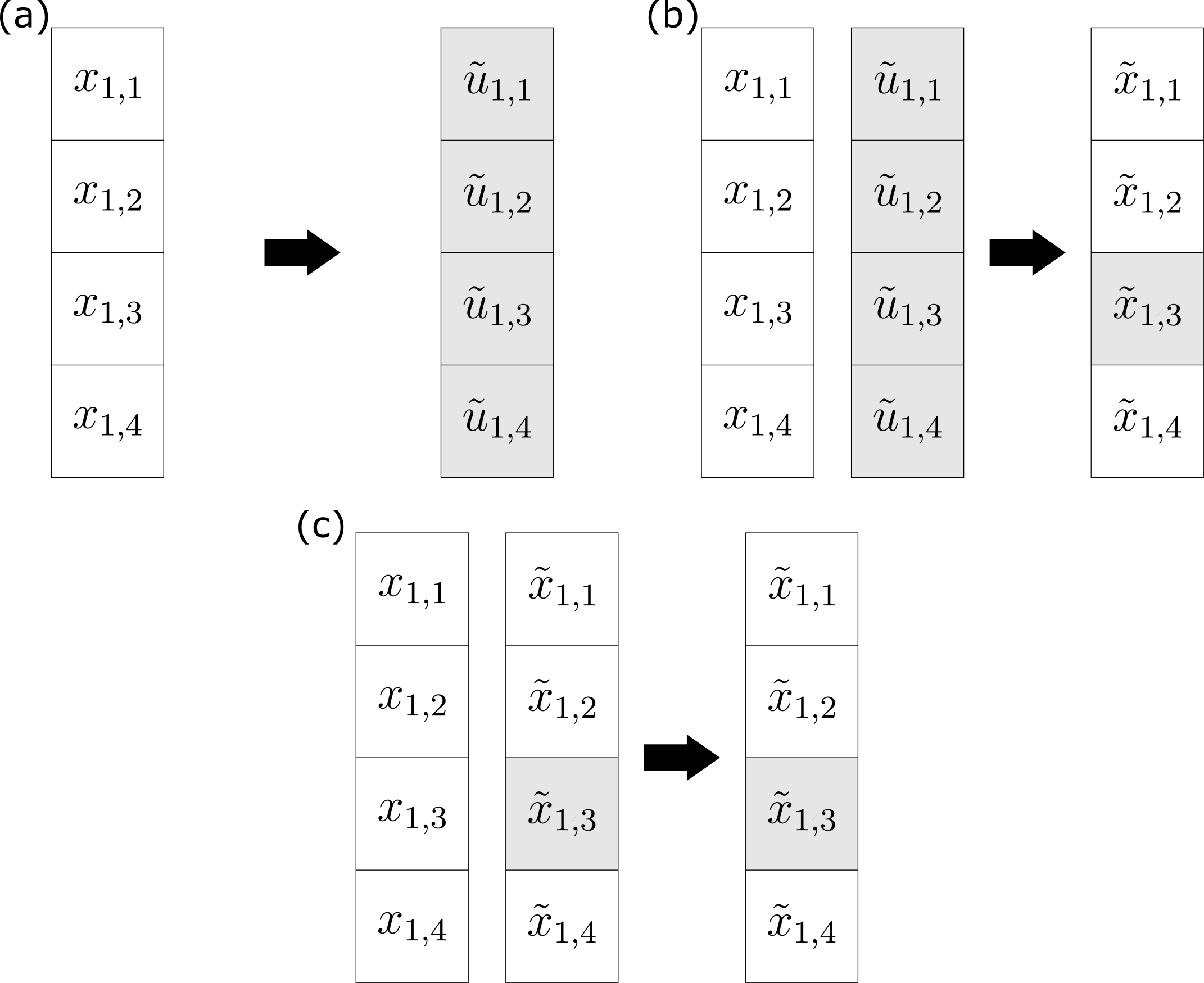}
	\caption{Overview of the three main stages of the DE algorithm. (a) Mutation: each entry of the candidate solution vector is mutated, generating a new mutated candidate solution vector; (b) Crossover: a new candidate solution vector is created with entries from the original and the newly created mutated vector; (c) Selection: the new and the original candidate solution vector are tested against the cost function and the one with the best results is propagated for the next generation.}
	\label{fig:DEstages}
\end{figure}

Here $f(\cdot)$ represents the cost function associated to the quantum phase estimation which is defined by the Holevo variance in Eq. (\ref{eq:holevo}). Therefore, if the new trial vectors $\tilde{x}^{G+1}_{i,j}$ minimize this cost function when compared to the previous generation vectors $x^{G}_{i,j}$, then they become part of the next generation of populations. Otherwise, the previous generation populations survive for the next generation. This entire process illustrates the adapted DE algorithm implemented in this work and is schematically represented in Fig. \ref{fig:DEstages}.

\subsection{Particle Swarm Optimization}

The implementation of the PSO algorithm starts with a set of $P$ particles, each representing an individual candidate solution to the update policy of the adaptive scheme. Each particle can move in the $N$-dimensional search space associated with the $N$ different phase instructions of the controllable phase shifter for each of the input qubits. Therefore, each particle will be represented by a position vector $x^{G}_{i,j}$ and velocity vector $v^{G}_{i,j}$, where $i \in \lbrace 1, 2, ..., P \rbrace$ and $j \in \lbrace 1, 2, ..., N \rbrace$ represent each entry and $G$ the generation of the vectors. Note that the position vector $x^{G}_{i,j}$ corresponds to a candidate solution vector, while the velocity vector $v^{G}_{i,j}$ represents the change in direction of the corresponding position vector in the search space.

\begin{figure}[h!]
	\centering
	\includegraphics[width=350pt]{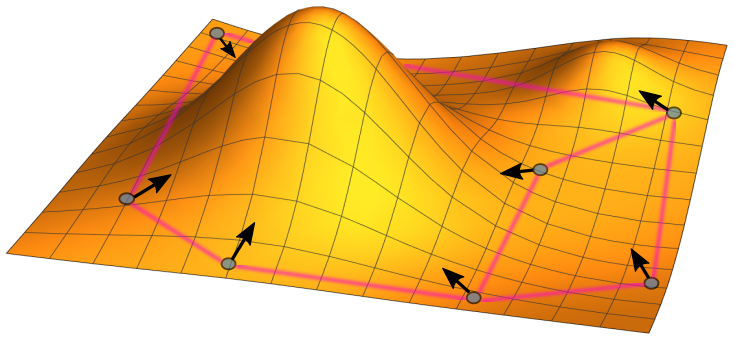}
	\caption{Visual representation of the PSO algorithm. Initially, all the particles representing the different candidate solution vectors are initialized at random positions with random velocities. At each iteration the particles explore the search space of the problem, eventually converging around the collectively found global optimum. The particles are able to share knowledge regarding the best-found position with other particles, which is generically represented by the red fuzzy lines. In our implementation the topology of inter-particle communication is such that all particles are able to share information with every other particle.}
	\label{fig:PSOillustration}
\end{figure}

All entries of these vectors are initialized with random values in the interval $[0, 2\pi]$. At each iteration, each particle evaluates its current position according to the cost function of the search problem, defined by the Holevo variance in Eq. (\ref{eq:holevo}), and compares its value with the positions previously visited by itself and with the positions previously visited by the entire collective of particles. If the current position is better than its own previously visited positions, the particle stores it in a variable $pbest_i$, where $i \in [1, ..., P]$ is the identifier of that particle. If, in addition, the current position is better than all of the other previously visited positions by the entire collective ensemble, the same position is stored in a variable $gbest$ shared among all other particles. This process is illustrated in Fig. \ref{fig:PSOillustration}.

Both of these variables will determine the entire exploration of the search space by the $P$ particle candidate solutions. After each iteration, each particle will use this collective knowledge to adjust its displacement for the next turn according to
\begin{equation*}
x^{G+1}_{i,j} = x^{G}_{i,j} + w \cdot v^{G+1}_{i,j}
\end{equation*}
and
\begin{equation*}
v^{G+1}_{i,j} = v^{G}_{i,j} + \alpha \cdot R_a \cdot (pbest_{i,j}-x_{i,j}) + \beta \cdot R_b \cdot (gbest_j-x_{i,j})\textrm{.}
\end{equation*}

Here the parameter $\alpha$  controls the desirability of each particle to move towards its best found position, while the parameter $\beta$ controls the desirability of each particle to move towards the best found solution by the entire collective. Both $R_a$ and $R_b$ are uniformly distributed random values in the interval $[0,1]$. In addition, the parameter $w$ works as a damping weight controlling the change of direction imposed by the new velocity vector at the current position of each particle. 

To steer the direction of each particle to a converging candidate policy solution and to avoid overstepping the found minima, an additional parameter $v_{max}$ is imposed to the algorithm which determines the maximum value that each entry in the velocity vector may take. As the algorithm advances in its search for the optimal policy solution, this parameter will decay with the number of iterations, consequently reducing the step size of each particle and forcing them to converge to a solution. This final adjustment completes the description of the adapted PSO algorithm implemented in this work.

\subsection{Parameters Analysis}

It is important to understand the behaviour and performance of these two algorithms under the different possible configurations of their controllable parameters. To verify the convergence of the solutions, it is important to remember that at each iteration of both algorithms there are $P$ different candidate solutions, each one represented by a given population of $N$ phase value policies. As the algorithm iterates, the candidate solutions should move closer to each other until converging to a final solution. Thus, one way of inferring this convergence value is by calculating the deviation from the average value of each population for each entry of their candidate solution and then averaging these values. To do so, the convergence value $L$ is defined as
\begin{equation}
L = \sum^N_j\frac{1}{N}\left(\sum^P_i\frac{\bar{x}_j-x_{i,j}}{P}\right)
\textrm{,}
\label{eq:dispersion}
\end{equation}
where $\bar{x}_j$ corresponds to the average value of entry $j$ over all the candidate solutions and $x_{i,j}$ corresponds to the entry $j$ of the candidate solution vector $i$. Therefore, lower values of $L$ occur when all the candidate solutions are relatively close to each other and the algorithm has converged to a solution. On the other hand, larger values of $L$ indicate that the algorithm was not able to converge to a solution at the given iteration.

The algorithms should also converge to a valid solution. It is not enough that the algorithms converge to a solution, if it is not the correct one. Letting $K$ represent the total number of different phase values of $\phi$ considered in the learning task imposed to the machine learning algorithms, the performance of a policy can be evaluated by the Holevo variance in Eq.(\ref{eq:holevo}). This equation, however, is computationally expensive in its current form. So instead, we can approximate it numerically \cite{palittapongarnpim2017learning,berry2000optimal,Sanders2011} to reduce computational time. A more efficient evaluation of the Holevo variance can be described as
\begin{equation}
V_{H} =
S^{-2} - 1 =
\bigg\vert \frac{1}{K} \sum^K_{k=1} e^{i\left[\phi^{(k)}-\theta_{N}^{(k)}\right]}\bigg\vert^{-2}-1
\textrm{,}
\label{eq:holevo_simplified}
\end{equation}
where values for $\phi^{(k)}-\theta_{N}^{(k)}$ close to zero signify lower values of imprecision (sharpness $S\approx 1$) and therefore better performance. 

Additionally, the performance $V_H$ of each candidate policy vector is evaluated $M=5$ separate times and the results averaged in order to smoothen small fluctuations. Repeating the simulation multiple times allows for a more accurate representation of the performance of each policy and thus more consistent results. The number of training instances $K$ can be any arbitrarily large enough number given that it does not hinder the computational time. A large number of training instances $K$ also ensures a faithful representation of $\phi$ in the interval $[0, 2\pi[$, since they are sampled uniformly and randomly from that same interval. A reasonable choice that satisfies these criteria is $K=10N^2$, a number sufficiently large to guarantee the convergence of the algorithms \cite{palittapongarnpim2017learning}.  Overall, the time complexity of the algorithms scales with $\mathcal{O}(P \cdot G \cdot N \cdot K \cdot M) \sim \mathcal{O}(N^3)$ which is polynomial in time. 

At this point it is possible to completely evaluate the performance of each algorithm according to the different possible configurations of each of their controllable parameters and choose those that achieve better results. While Eq. (\ref{eq:dispersion}) provides a measurement for the convergence of the algorithms, Eq. (\ref{eq:holevo_simplified}) shows the precision to which they are able to estimate the value of the unknown phase $\phi$.

A thorough study on the performance of both algorithms under the different possible controllable parameter configurations can be found in appendix \ref{apx:evolution} and appendix \ref{apx:optimization}. The optimal parameter configuration obtained for each algorithm is summarized in Table \ref{tab:optimal_parameters}.

\begin{table}[H]
\centering
\renewcommand{\arraystretch}{1.4}

\begin{tabularx}{0.75\textwidth}{c|cccccc}
	\hline\hline
	Algorithm & $F$ & $C$ & $\beta$ & $\alpha$ & $w$ & $v_{max}$ \\
	\hline
	Differential Evolution & $0.7$ & $0.8$ & - & - & - & - \\
	Particle Swarm Optimization & - & - & $0.8$ & $0.8$ & $0.8$ & $0.2$\\
	\hline\hline
\end{tabularx}

\caption{Optimal parameters for the DE and PSO algorithms obtained through the analysis described in Appendices \ref{apx:evolution} and \ref{apx:optimization}.}
\label{tab:optimal_parameters}
\end{table}

Along with a fixed configuration for each algorithm, it is also important to ensure that all the other variable parameters remain the same in order to draw comparable results. Thus, it is important that for the same number of input qubits $N$ being injected into the system, the population number of candidate solution vectors $P$ and the number of training instances $K$ remain the same under all different scenarios. As previously mentioned, the number of training instances was set to $K=10N^2$, while the number of populations was defined as $P=20+2\cdot {\rm int}(N/10)-1$, where ${\rm int}(\cdot)$ represents the integer part of the division. Note that for an increasing number of qubits being injected into the system, the population size of candidate solution vectors $P$ and the number of training instances $K$ must also increase to accommodate the increasing complexity of the problem search space. 

Ideally, both algorithms would be allowed to run until all the different candidate solution vectors would have converged to a successful policy vector. However, due to time constraints the number of iterations for which both algorithms were allowed to run for each number $N$ of input qubits was set to $G=100$, regardless of having reached convergence or not. Thus, both algorithms would stop either when they had converged to a solution, or when they reached iteration $G=100$, and the final policy vector would be the average of all the different candidate solution vectors at that point.

\section{Results}

In order to provide a benchmark for the two machine learning algorithms discussed above, we first introduce a non-adaptive protocol that can also be run in the presence of noise. This protocol has the advantage of simplicity and we find that for moderate noise values it yields results that are better or comparable with machine learning algorithms. On the other hand, for increased noise values the machine learning protocols show better results.

We start by discussing the ideal configuration where any source of noise or quantum decoherence is neglected, then we consider the Gaussian and Random Telegraph noise configurations and, finally, the visibility loss due to decoherence. These different sources of imperfection were applied to the quantum phase estimation configuration independently. The number of varying qubits used under all the different scenarios was set in the interval $N \in [5,25]$ and all the remaining parameters were left the same across all the different scenarios. For $N>20$ we found that the algorithms were already taking more than five days to arrive at a successful policy vector, which ultimately made any larger value of $N$ computationally impracticable under a reasonable amount of time. 

The break in performance in non-ideal systems is clearer for larger values of noise, since increasing the value of noise in the estimation process consequently leads to an increased complexity of the search space for both algorithms. Thus, for larger values of $N$ this added level of complexity becomes even more evident in the scattering of the obtained results. However, it is important to stress that when both algorithms had enough time to converge to a successful policy, they were able to perform with precision values close to the SQL, thus attesting their resilience to noise in the quantum phase estimation scheme.

\subsection{A Non-Adaptive Protocol and the Standard Quantum Limit Benchmark}

The SQL is defined strictly speaking in the limit of large $N$. Since we do not work in this asymptotic regime, it is important to devise a non-adaptive protocol that reproduces the SQL for $N \gg 1$, yet it yields results also at $N \approx 5-25$.

This protocol can be outlined as follows. We consider the random phase that should be estimated $\phi$ and a fixed control phase $\theta$. Based on these values, we calculate the probability $\mathcal{P}_{\pm}(0\vert \phi, \theta) $, see Eqs. (\ref{eq:prob0}), (\ref{eq:prob1}) or Eq. (\ref{eq:decoherence}) with $\nu =1$. Then, for each $N$ we generate $N$ uniformly random numbers in the interval $[0,1]$. If the random number is less than $\mathcal{P}_{\pm}(0\vert \phi, \theta)$ we add it into the $0$ bin, otherwise we add it to $1$. Next, we count the number of elements in the $0$ stack, $N_{\pm}(0)$. Finally, we find an estimation for the phase $\phi^{est}$, as $\phi^{est}=\theta + \arccos(\pm 2N_{\pm}(0)/N\mp 1)$, and we calculate the Holevo variance with the exponent in Eq. (\ref{eq:holevo_simplified}) as $\exp[i(\phi-\phi^{est})]$.

\begin{figure}[h!]
	\centering
	\includegraphics[width=380pt]{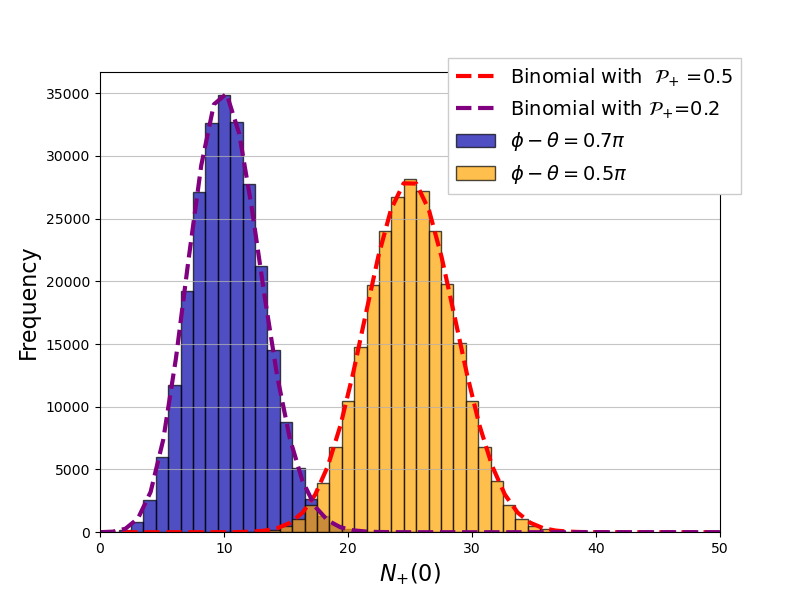}
	\caption{Distribution of $N_{+}(0)$ outcomes for 25000 experiments of fixed phase estimation for two values of $\mathcal{P}_{+}$ .}
	\label{fig:Binomial}
\end{figure}

The number of elements $N_{+}(0)$ and $N_{+}(1)=N-N_{+}(0)$ follows the binomial distribution as shown in Fig. \ref{fig:Binomial}. This distribution is obtained from two constant phase estimation by 50 measurements, repeated 250000 times.

This procedure is repeated $K=10 N^2$ times for each phase $\phi$ uniformly distributed in the interval $[0,2\pi]$. The resulting Holevo variance is represented by the blue squares in Fig. \ref{fig:res_clean}. We have verified numerically that the non-adaptive method reproduces asymptotically the SQL. For example, even at $N=100$ the difference
between  simulated  $V_H$ and $1/N$ is 10 \% and reaches  4 \% at $N=800$. Also, as it is seen from the inset at Fig. \ref{fig:res_clean}, the slope for this non-adaptive protocol is equal to $-1.0265$, while for the ideal SQL this slope should be $-1$.

The same procedure is used to simulate the variance in the presence of different noises, as it is shown below.

\subsection{Adaptive Protocols in the Ideal Noiseless Configuration}

\begin{figure}[h!]
	\centering
	\includegraphics[width=350pt]{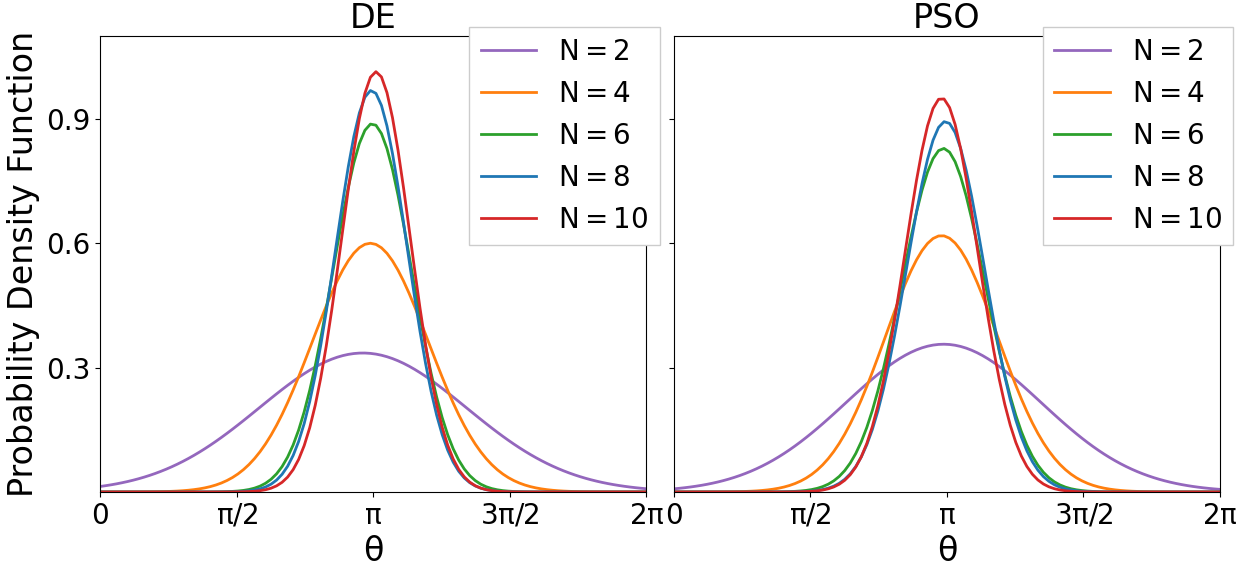}
	\caption{Probability density function of the DE and PSO algorithms as a function of the phase value $\theta$.}
	\label{fig:probabilities}
\end{figure}

Before analysing the evolution of the performance obtained by the DE and PSO algorithms under a increasing number of qubits $N$, it is important to show that this increase in $N$ does indeed lead to better estimation values. To do so, the two algorithms were first allowed to converge to a given value of the unknown phase $\phi$ for different values of $N$. This experiment was repeated $1000$ times for each value of $N$ and, at each time, the value of the estimated phase $\theta$ was recorded. Finally, these $1000$ different results of $\theta$ for each value of $N$ were fitted to a probability density function (PDF) and centered all around the same value $\theta = \pi$ to better compare the results. The results for each algorithm are displayed side by side in Fig. \ref{fig:probabilities}.

\begin{figure}[h!]
	\centering
	\includegraphics[width=380pt]{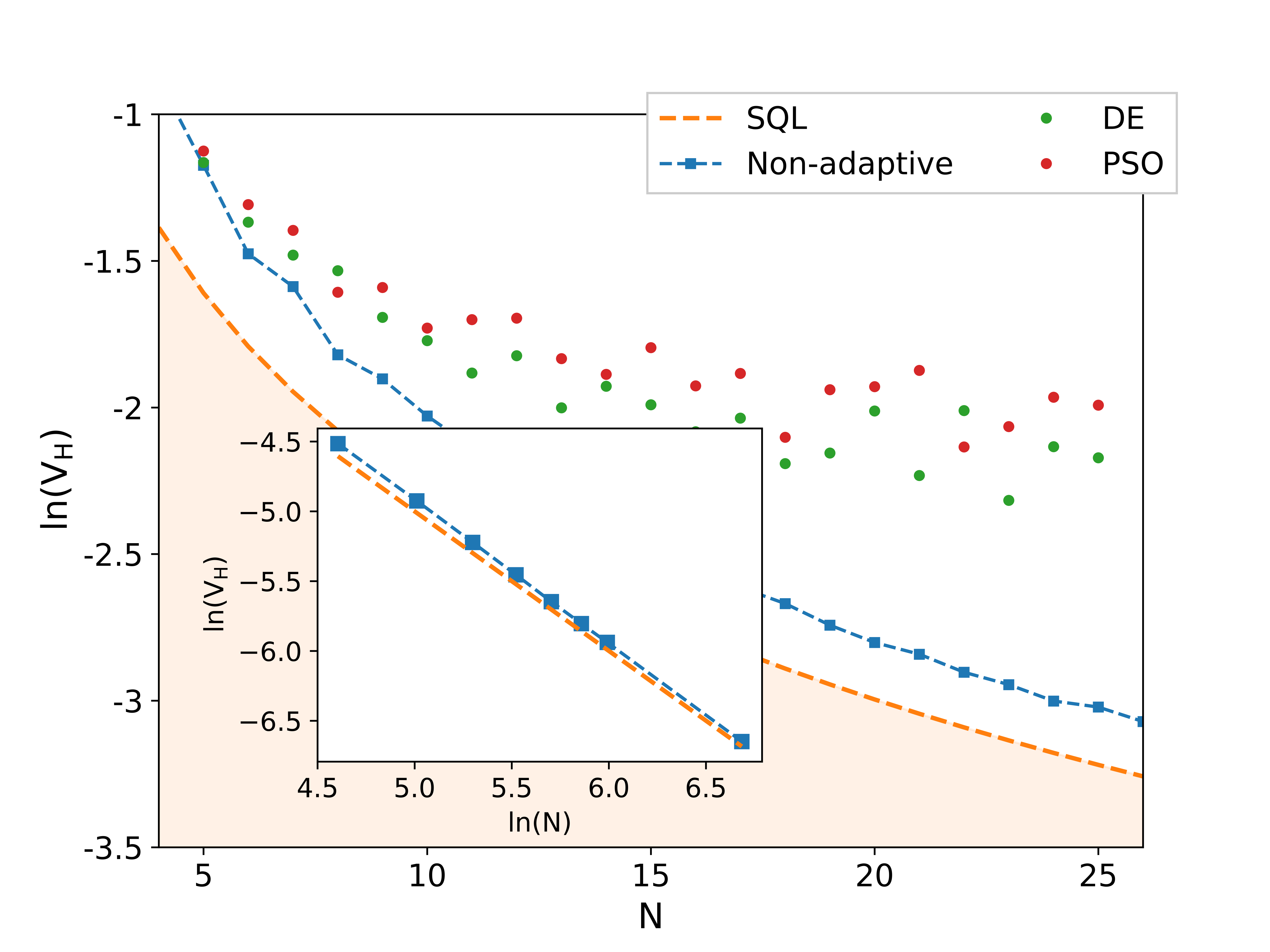}
	\caption{Estimation precision based on the logarithm of the Holevo variance $\ln(V_H)$ of both the DE and PSO algorithms under the ideal configuration for $N\in[5,25]$ qubits.}
	\label{fig:res_clean}
\end{figure}

Considering the results obtained in Fig. \ref{fig:probabilities} it is indeed possible to see that by increasing the number of qubits $N$ at the input level the estimation process leads to values of $\theta$ converging further towards the given value of the unknown phase $\phi$. It is also possible to see that as $N$ goes higher, this increase in precision starts to be less evident as the algorithms start running into convergence issues. Given that the complexity of the search space of each algorithm scales polynomial in $N$ and that due to time constraints the algorithms were not allowed to run for more than $G=100$, a decrease in performance under this time limitation for large values of $N$ is expected.

Having arrived at this conclusion, the final results obtained under the ideal scenario, where there is no source of noise or quantum decoherence, are shown in Fig. \ref{fig:res_clean}. Observing the results it is possible to see that both algorithms were able to follow the SQL for the ideal-case scenario. It is also possible to reinforce the previous found conclusion that an increase in $N$ leads to better results. In fact, for $N\in[5,15]$ the scaling of the Holevo variance is well approximated by a power law $V_{H}\sim N^{-\alpha}$, where $\alpha_{\rm DE}=0.75$ for the DE algorithm and $\alpha_{\rm PSO}=0.64$ for the PSO algorithm, while the the corresponding value for the SQL is $\alpha_{\rm SQL}=1$. This also shows that the DE performs slightly better than PSO at reaching low variances, which is consistent with results obtained in other contexts, such as estimating the bias in the quantum walk algorithm \cite{Lovett2013}. This scaling in precision close to the SQL for $N\in[5,15]$ is consistent with what has been numerically observed in other algorithms \cite{berry2009perform}.

The fact that the machine learning algorithms do not reach the SQL is also consistent with the known results that the adaptive measurements cannot saturate the Cram\'er-Rao inequality even for optimal measurements \cite{PhysRevA.94.022334,Garcia2020}. It is also noticeable that for larger values of $N$ the performance starts to break due to the limiting number of iterations $G=100$ imposed for the convergence of the algorithms. This is not a malfunction of the algorithms, but a direct consequence of the restricted time available. As the complexity of the search space of the algorithms increases, so does the number of generations required to converge to a successful policy.

\subsection{Configurations with Noise}

Considering the results obtained under the ideal configuration, it is important to study the resilience of the algorithms to different sorts of imperfections that can be found in an experimental quantum phase estimation scheme. First, the performance of the algorithms was evaluated in the presence of Gaussian noise, followed by Random Telegraph noise and finally in the presence of quantum decoherence.

\subsubsection{Gaussian Noise}

\begin{figure}[h!]
	\centering
	\includegraphics[width=380pt]{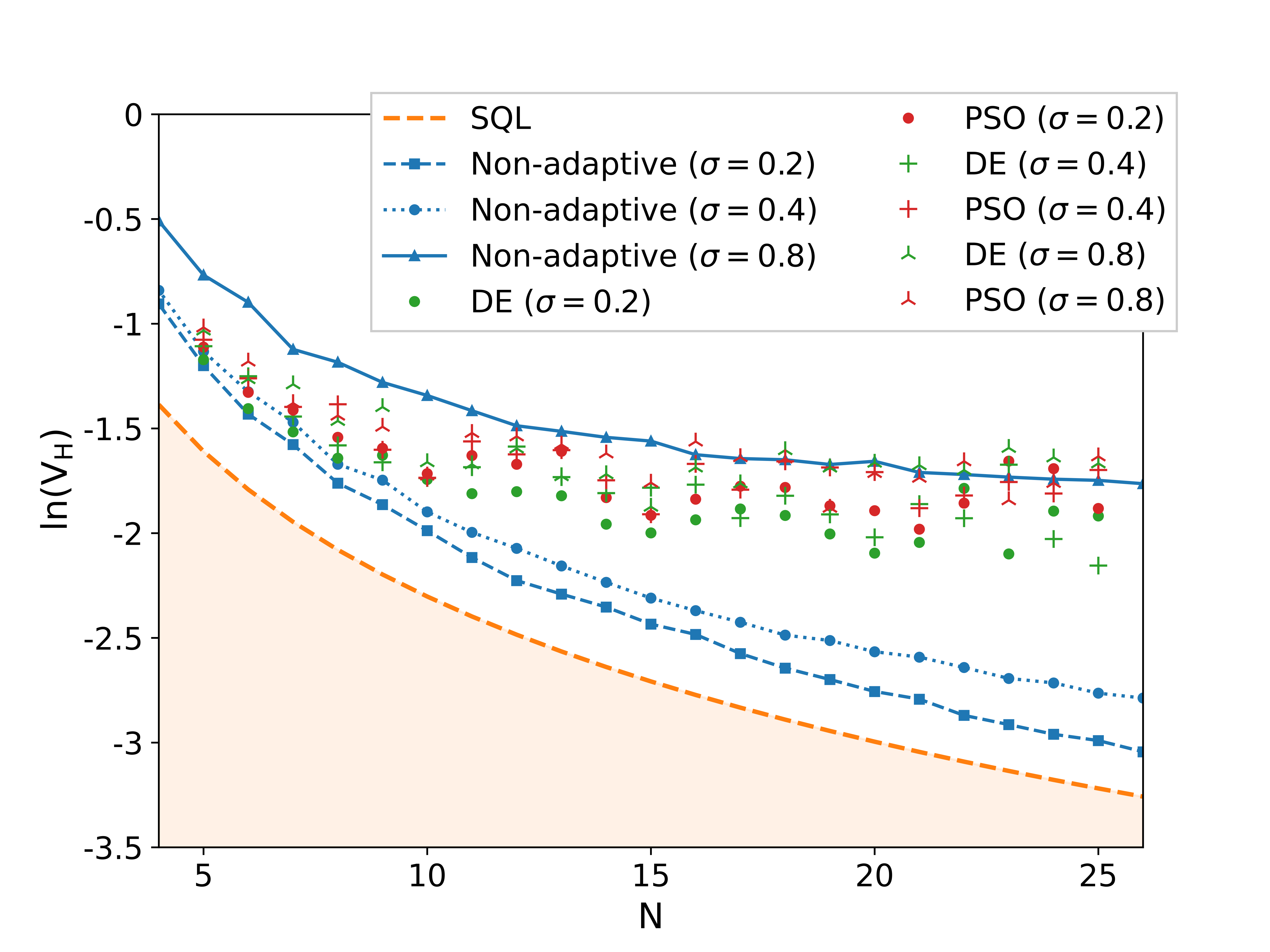}
	\caption{Estimation precision based on the logarithm of the Holevo variance $\ln(V_H)$ of both the DE and PSO algorithms under increasing values of Gaussian noise $\sigma$ for $N \in [5,25]$ qubits.}
	\label{fig:res_noise_gaussian}
\end{figure}

In this scenario the algorithms were tested for increasing amounts of Gaussian noise $\sigma=\{0.2,0.4,0.8\}$ when dealing with the controllable phase shifter $\theta$ according to Eq. (\ref{eq:gaussian}). The results obtained under these conditions are presented in Fig. \ref{fig:res_noise_gaussian}.

It is possible to see that as the noise fluctuations increase the precision of both adaptive and non-adaptive algorithms starts to diminish. This is, nevertheless, expected for any estimation process being conducted under increasing values of noise. The break in performance is also perceptible for larger values of $N$. However, we see from Fig. \ref{fig:res_noise_gaussian} that the adaptive algorithms are more sensitive to increasing values of Gaussian noise and that for $\sigma=0.8$ the policies obtained from machine learning become clearly superior to the non-adaptive protocol.

\subsubsection{Random Telegraph Noise}

\begin{figure}[h!]
	\centering
	\includegraphics[width=380pt]{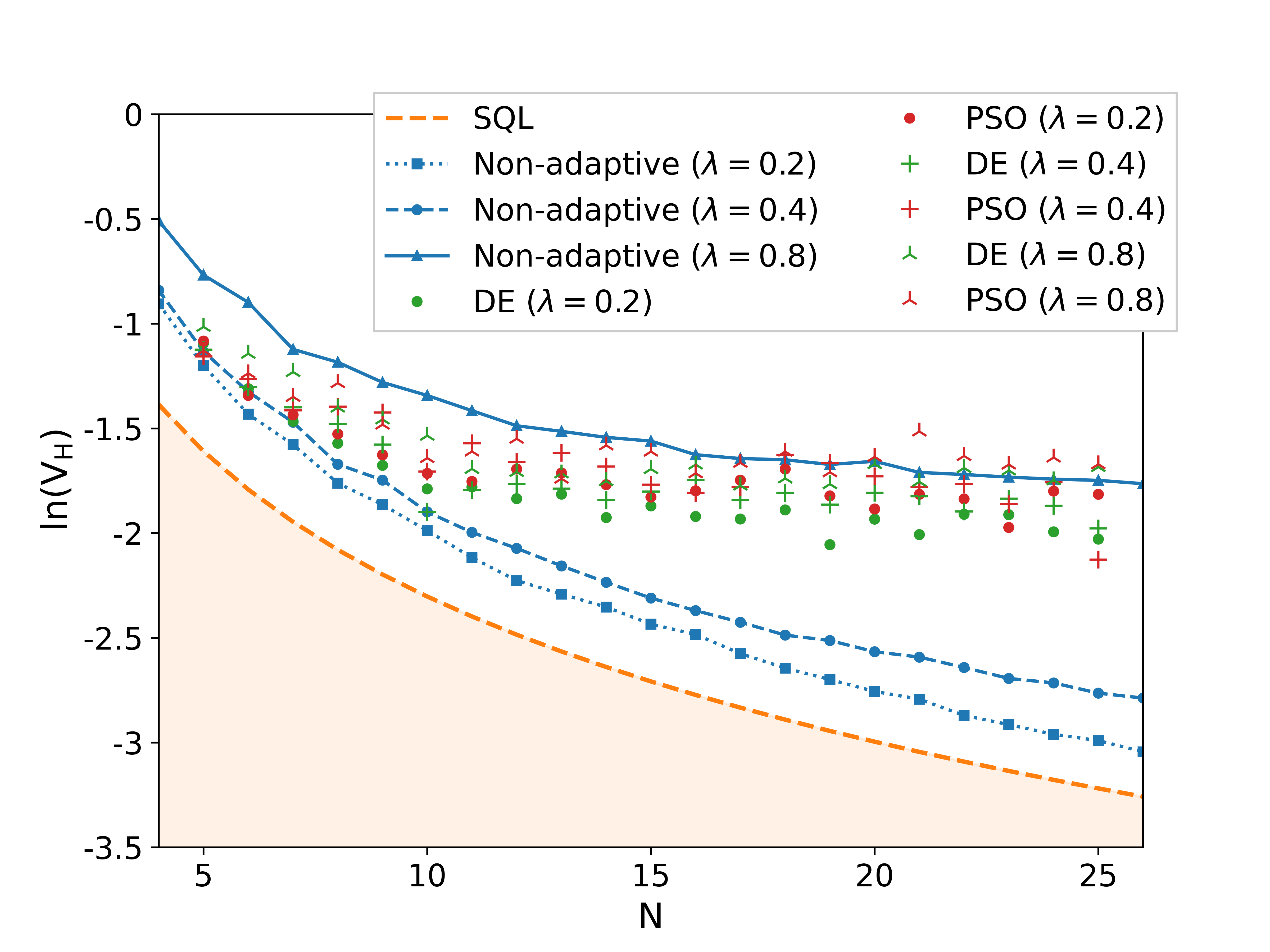}
	\caption{Estimation precision based on the logarithm of the Holevo variance $\ln(V_H)$ of both the DE and PSO algorithms under increasing values of Random Telegraph noise $\lambda$ with a fixed probability of $\eta=0.4$ for $N \in [5,25]$ qubits.}
	\label{fig:res_noise_rtn}
\end{figure}

Considering now the scenario where the estimation scheme was subject to a Random Telegraph noise following Eq. (\ref{eq:rtn}), the algorithms were tested against increasing values of $\lambda=\{0.2,0.4,0.8\}$ while keeping the probability of switching to the erroneous phase value fixed at $\eta=0.4$. The results obtained by the algorithms under these configurations are displayed in Fig. \ref{fig:res_noise_rtn}.

Similarly to the results obtained under the Gaussian noise, the results obtained in Fig. \ref{fig:res_noise_rtn} show that both adaptive and non-adaptive algorithms partly follow the SQL curve even in the presence of Random Telegraph noise. The poorer performance for larger values of noise, as well as the break in performance for larger values of $N$, is also evident under this scenario for the same reason as before. We can also see that for larger values of $\lambda$ the machine learning adaptive algorithms are more robust than the non-adaptive algorithm.

\subsubsection{Quantum Decoherence}

\begin{figure}[h!]
	\centering
	\includegraphics[width=380pt]{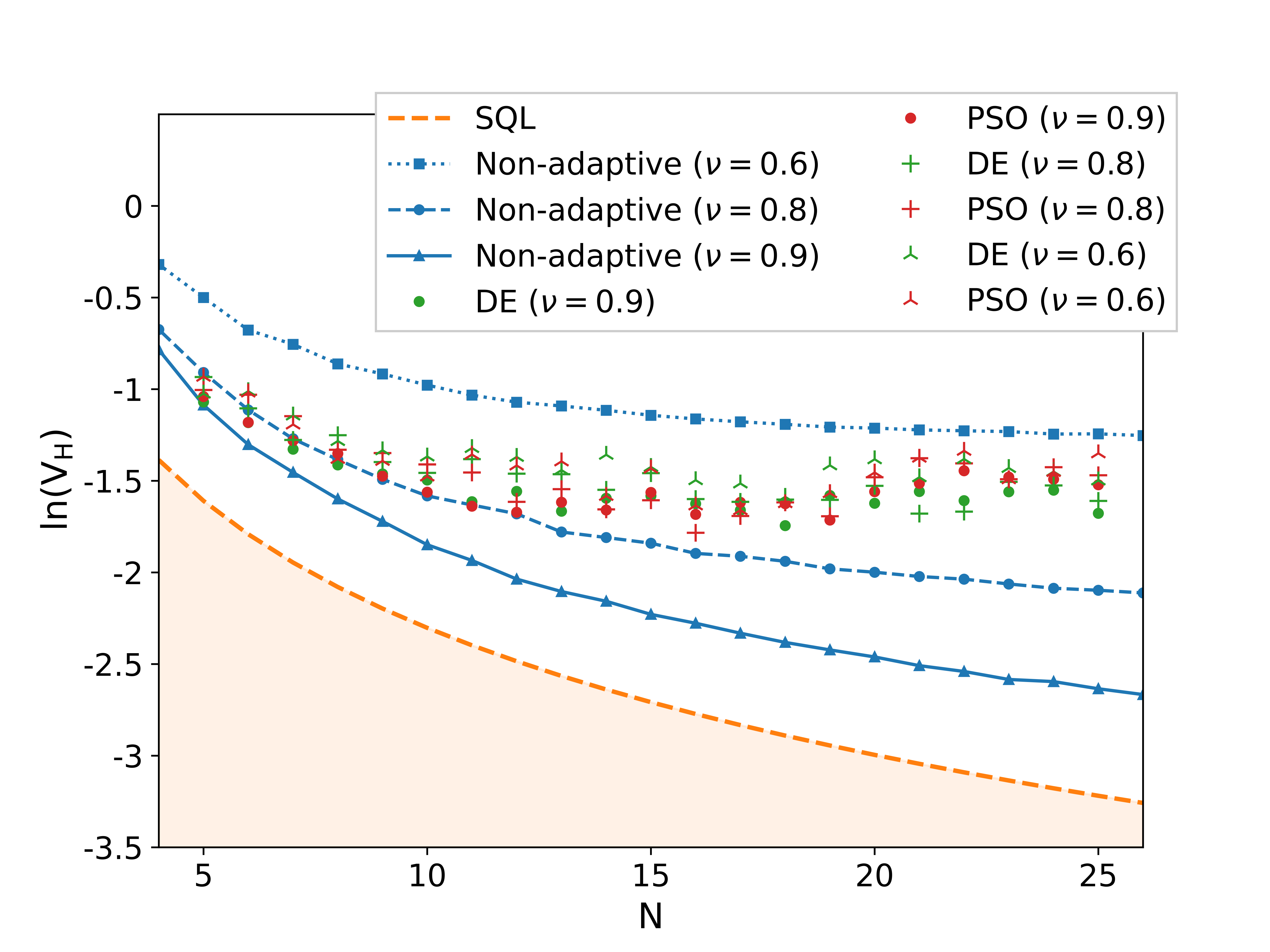}
	\caption{Estimation precision based on the logarithm of the Holevo variance $\ln(V_H)$ of both the DE and PSO algorithms under decreasing values of visibility $\nu=\{0.9, 0.8, 0.6\}$ for $N \in [5,25]$ qubits.}
	\label{fig:res_decoherence}
\end{figure}

In this last scenario, we study the impact of quantum decoherence in the performance of both algorithms. This is a key source of noise in all non-optical qubits (\textit{e.g.} trapped ions, superconducting qubits, NV centers). The algorithms were tested against decreasing values of visibility $\nu=\{0.9,0.8,0.6\}$ according to Eq. (\ref{eq:decoherence}). Note that unlike the ideal scenario, the reduced visibility also impacts the SQL, see Eq. (\ref{eq:SQL}) in Appendix A. Also note that the value $\nu = e^{-0.5} \approx 0.6$ appears in certain tasks of non-adaptive parameter estimation as the visibility corresponding to an optimal phase accumulation time $\tau$ of half the decoherence time \cite{degen_2017}. The results obtained under this configuration are shown in Fig. \ref{fig:res_decoherence}.

Considering the results obtained in Fig. \ref{fig:res_decoherence}, it is evident that reduced values of visibility have a significant impact on the performance of the algorithms. This is an expected behaviour, since it is known that quantum enhancement for systems operated at around the decoherence rate is asymptotically limited to only a constant factor improvement over the SQL \cite{sekatski2017quantum}. This can be confirmed by an analysis of the Fisher information:
\begin{eqnarray}
\mathcal{F}_{\phi ,\theta_{m}} = 
\overline{[\partial_{\phi}\ln P_{\pm} (\zeta_{m}|\phi, \theta_{m})]^{2}} =
\sum_{\zeta_{m} = 0,1} \frac{\left[\partial_{\phi} P_{\pm} (\zeta_{m}|\phi, \theta_{m})\right]^2}{P_{\pm} (\zeta_{m}|\phi, \theta_{m})} =
\frac{\nu^2 \sin^2 (\phi - \theta_{m})}{1- \nu^2 \cos^2 (\phi - \theta_{m})}
\label{eq:Fisher}
\end{eqnarray}

The Fisher information quantifies the knowledge gained at each measurements. It can also be extended from projective to positive operator-valued measurements, see \cite{Paraoanu_2011}, and it is used to define the SQL (see Appendix \ref{apx:holevo}). Indeed, from Eq. (\ref{eq:Fisher}) it can be observed that $\nu$ reduces the information extracted after each measurement from its maximum value $\mathcal{F}_{\phi ,\theta_{m}} =1$ at $\nu =1$. 

Similarly here we observe that relatively large values of decoherence (visibility around 0.6) result in the suppression of precision obtained by the non-adaptive protocol, while the adaptive DE and PSO are less sensitive.

\section{Experimental Implementations}

Summing up the theoretical results obtained so far, we can see that even for noisy systems it is possible to reduce the Holevo variance to values in a band of typically $[-1.2, -1.7]$ (standard deviations between 0.5 and 0.4) with a moderate numbers of $N \in [10,25]$ qubits. This protocol can be implemented on several experimental platforms, using either optical or microwave frequencies. We describe here the main experimental platforms and we show how the problem of optimizing the precision in the presence of noise can be addressed using the machine learning framework.

The most straightforward implementation is optical Mach-Zehnder interferometry. In this case, the operator $U_{\theta-\phi}$ can be realized by placing a phase shifter $\phi$ in one branch of the interferometer and a variable-phase shifter $\theta$ in the other branch. The latter can be realized physically as a half-wave plate placed on a rotation stage controlled by a computer \cite{Higgins2007}. The states $|0\rangle$ and $|1\rangle$ correspond to a single photon in one branch or the other of the interferometer (dual-rail encoding). Our results can be compared also with those obtained from running the more powerful Kitaev algorithm, which for the same number of resources (from 10 to 25) results in standard deviations ranging from 0.35 to 0.16 \cite{Higgins2007}, only marginally better than the data reported here. Also, our results are consistent with the theoretical limits reported in \cite{Pryde2018} for the case of non-entangled initial states ($V_{H} = 0.5609$, $\ln V_{H} =-0.25$). We obtain approximately a factor of 2 improvement in the standard deviation. For implementations employing optical interferometry, the visibility is typically very close to 1, and the main noise sources in the feedback loop concern changes in the refractive index of the beam-splitters and variable phase shifter, as well as variations in the optical paths typically caused by temperature. 

More recently, an experiment with optical photons has tested the PSO algorithm with sequential non-entangled photons \cite{Lumino2018}. The single photons were obtained non-deterministically, by generating entangled pairs through spontaneus parametric down-conversion in a  2 mm long beta-barium borate (BBO) crystal and heralding over detection events in another detector. The unknown and the control phases in the two arms of the Mach-Zehnder interferometer were realized with liquid crystal devices, where the relative phase between the horizontal and vertical polarization can be changed depending on the applied electric field. The results with the PSO algorithm obtained agree with our analysis: PSO is quite efficient at reaching values approaching the SQL, especially when the number of resources (number of photons $N$) is limited. Similarly to our findings, for $N$ exceeding approximately 15 photons the Holevo variance tends to saturate. The robustness with respect to Gaussian phase noise and dephasing noise was also demonstrated by artificially adding errors to the feedback phase.

In the case of qubits based on discrete levels (solid-state and ions/atoms), Mach-Zehnder interferometry corresponds to Ramsey interference \cite{paraoanu2006,danilin_2018}. To understand how the phase information is embedded in this system, consider the following generic qubit Hamiltonian driven by a microwave field \cite{Silveri2017}, 
\begin{equation}
\mathcal{H} = \frac{\hbar}{2}\omega_{01}\sigma_{z} + \hbar\Omega \cos \omega t \sigma_{x}.
\end{equation}
In a frame defined by the unitary $\exp (i \omega t \sigma_{z})$ (a clockwise rotation 
around the $z$-axis), using $\exp (i \omega t \sigma_{z})\sigma_{x}\exp (-i \omega t \sigma_{z}) = \sigma_{x}\cos \omega t - \sigma_{y} \sin \omega t$ and by
further applying the rotating-wave approximation we get the effective Hamiltonian
\begin{equation}
\mathcal{H} = \frac{\hbar}{2}(\omega_{01}-\omega) \sigma_{z}+ \frac{\hbar\Omega}{2}\sigma_{x}.
\end{equation}
A non-zero Rabi frequency $\Omega \neq 0$ can be thus used to create the two Hadamard gate, while the time in-between is spend for the actual sensing. Indeed, if $\Omega = 0$ for a sensing time $\tau$, the resulting unitary becomes
\begin{equation}
\mathrm{U}_{\phi ,\theta} =  \begin{bmatrix}
	1 & 0 \\ 0 & e^{i (\omega_{01} - \omega)\tau} \end{bmatrix},
\end{equation}
which is exactly Eq. (\ref{umatrix}) up to an overall irrelevant phase and the identification $\phi = \omega_{01}\tau$, $\theta = \omega \tau$.

Consider now a concrete problem, that of evaluating the magnetic field using a superconducting qubit, a trapped ion, 
or a nitrogen-vacancy (NV) center in diamond. In these cases, 
for typical experiments using Ramsey interferometry, the probability is given by Eq. (\ref{eq:decoherence}), where the dependence of the visibility $\nu$ on the sensing time $\tau$ can be sometimes different from a simple exponential decay $\nu(\tau) = \exp[-\tau/T_{2}]$,
{\hl valid for the case of Gaussian noise (see the Lindblad equation used in Section II)}.
For example, another dependence is
$\nu(\tau) = \exp[-(\tau/T_{2})^2]$ if the noise experienced by the qubit is 1/f, see e.g. \cite{Silveri2017,danilin_2018,Hanson2016}. Since $\tau$ is bounded by $T_{2}$, in these setups one might attempt to increase the precision by increasing the sensitivity of $\omega_{01}$ to the magnetic field. However, this means increasing the coupling to the magnetic field, which
at the same time this will increase the exposure of the qubit to noise. Thus, a tradeoff must be reached between these two competing effects. Our results demonstrate that the increase in noise can be mitigated successfully by the use of machine learning strategies.

In the case of a superconducting qubit in the symmetric transmon design as used in recent magnetometry experiments \cite{danilin_2018,shlyakhov_2018,Danilin2021}, the information about magnetic field is embedded in the energy level separation as
\begin{equation}
\omega_{01}(B) = \frac{1}{\hbar} \left( \sqrt{8 E_{\rm C} E_{\rm J\Sigma} \cos\left| \pi \frac{BS}{\Phi_{0}}\right|} - E_{\rm C} \right),
\end{equation}
where $B$ is the magnetic field piercing the SQUID area $S$ of the transmon and modulating the total Josephson energy $E_{\rm J\Sigma}$. In order to evaluate $B$, we can keep $\tau$ fixed and adjust the frequency $\omega$ (generated by an external signal generator) at every step. In the case of superconducting qubits, with relatively standard values $T_{2}$ = 10 $\mu$s and $\tau=1$ $\mu$s we obtain $\nu = 0.9$ (one of the values used in Sec. IV) if the noise is Gaussian and $\nu = 0.99$ if it is 1/f noise. 

In order to increase the precision of determining the magnetic field, we can use a higher excited state \cite{shlyakhov_2018,Perelshein2021} - for example, the second excited state $|2\rangle$, and considering the harmonic approximation for the transmon, the relevant accumulated phase will be $\approx 2 \omega_{01}\tau$; or we can bias the transmon to a magnetic field value where the slope $d\omega_{01}/dB$ is larger. Both situations result in an increase in the noise. In the first case, this is due to the fact that higher energy levels have higher couplings to the electromagnetic environment. This causes an increase in $T_{2}$, cause due to both an increase in the $T_{1}$ time and an increase in the pure dephasing time. For example, if the second excited state is used, the decay rate of the $1-2$ transition is twice that of the $0-1$ transition, which for $T_{1}$-limited qubits results in a significant reduction of the interferometric visibility. In the second situation, due to biasing far from the sweet point. In the latter case, let us restrict ourselves for simplicity to the interval $BS \in [-\Phi_{0}/2, \Phi_{0}/2 ]$. We then have 
\begin{equation*}
\frac{d\omega_{01}}{dB} = - \frac{\pi S \omega_{01}}{\Phi_{0}}\tan \left( \pi \frac{BS}{\Phi_{0}}\right).
\end{equation*}
This slope is infinite for $BS = \pm \Phi_{0}/2$, apparently allowing us to measure $B$ with an infinite precision. However, the displacement of the bias point from the sweet point is accompanied by a  significant increase in noise, since the qubit is no longer protected against linear-oder fluctuations. This results again in a visibility $\nu$ below unity. 

The noises in the control phase $\theta_{m}$ are typically not caused by electronics in the feedback look, but they result from uncontrollable frequency shifts that are poorly understood and controlled. Experimentally, these shifts  are of the order of a few MHz. If $\tau$ is of the order of microseconds, then the resulting values of $\lambda$ and $\sigma$ are well below those considered in this work. Therefore, this type of noise will not affect the performance of our protocols when run on a superconducting qubit.

In the case of trapped ions, sensitive magnetometry using the well-known $^{171}\mathrm{Yb}^{+}$ ion has been demonstrated \cite{PhysRevLett.116.240801}. This uses four hyperfine states,  $\ket{\mathrm{F} = 0, m_{\mathrm{F}} = 0}$, $\ket{\mathrm{F} = 1, m_{\mathrm{F}} = 1}$,  $\ket{\mathrm{F} = 1, m_{\mathrm{F}} = -1}$, and 
$\ket{\mathrm{F} = 1, m_{\mathrm{F}} = 0}$ belonging to the $^{2}S_{1/2}$ manifold. 
The latter three states are degenerate and they are separated by the hyperfine splitting $\omega_{\rm hf}/(2\pi) = 12.642$ GHz from the first state. The degeneracy of these three states can be lifted by the application of a magnetic field. In the first order in magnetic field, the state  $\ket{\mathrm{F} = 1, m_{\mathrm{F}} = 0}$ remains unmodified but 
$\ket{\mathrm{F} = 1, m_{\mathrm{F}} = \pm 1}$ acquires frequency shifts of $\pm (g_{e} \mu_{\rm B}/2\hbar) B$, where $g_{e} \approx 2$ is the $g$-factor of the electron and $\mu_{\textrm{B}}$ is the Bohr magnetron. Thus, for magnetic field detection one could in principle use the state $\ket{\mathrm{F} = 0, m_{\mathrm{F}} = 0}$ and either of the magnetic-sensitive states $\ket{\mathrm{F} = 1, m_{\mathrm{F}} = \pm 1}$ and drive resonant Ramsey $\pi/2$ microwave pulses at around 12 GHz with $\tau$ time separation.  Then the information about magnetic field is obtained from the phase $\phi =  (\omega_{\rm hf} \pm g_{e} \mu_{\rm B}B/2\hbar )\tau$. These states would be exposed not only to the magnetic field that we would like to sense, but also to magnetic field noise, making our results for the noisy case relevant. 
Further improvements may be achieved by the use of a continuous dynamical decoupling technique, where one could identify the  
$\ket{\mathrm{F} = 1, m_{\mathrm{F}} = 0} \equiv \ket{0}$ and the dark state $\frac{1}{\sqrt{2}}\left(\ket{\mathrm{F} = 1, m_{\mathrm{F}} = - 1} + \ket{\mathrm{F} = 1, m_{\mathrm{F}} = 1}\right) \equiv \ket{1}$ as a dressed-qubit states, with a $T_{2}$ time exceeding one second \cite{Wunderlich2011}, three orders of magnitude more than the bare atomic states which is a few miliseconds.

Similar ideas can be applied to NV centers in diamond. These defects have very long decay times, of the order of tens of miliseconds, and total decoherence times are of the order of microseconds and can be extended to hundreds of microseconds by the use of dynamical decoupling pulses \cite{Lukin2008}. 
Single NV centers have a triplet $S=1$ ground-state structure, 
with the states $m_{\rm S}=0$ and $m_{\rm S}=\pm 1$ separated by the so-called zero-field splitting
$D = 2.87$ GHz. By applying a magnetic field along the crystal axis, the levels $m_{\rm S}=\pm 1$ can be further split by the Zeeman effect \cite{Walsworth2020}; the resulting energy level difference between these levels is $2 g_{e} \mu_{\rm B} B/\hbar = 2 \gamma_{e}B $ where $\mu_{\textrm{B}}$ is the Bohr magnetron and $g_{e}$ is the electronic g-factor, $g_{e}\approx 2$. The gyromagnetic ratio $\gamma_{e}$ is defined as $\gamma_{e} = g_{e} \mu_{\rm B}/\hbar$. 
Because the triplet states can be selectively addressed with microwaves \cite{Walsworth2020} we can immediately identify
for example $|0\rangle = |m_{\rm S}=0 \rangle $ and $|1\rangle = |m_{\rm S}= 1 \rangle$
as the basis corresponding to our formalism, and we obtain $\omega_{01} = 2\pi D + \gamma_{e} B$. The encoding of the magnetic field in frequency and subsequently in the phase is linear, $\Phi = 
(2\pi D + \gamma_{e} B)\tau$.

To summarize, each of these experimental systems involved relatively well characterized imperfections, and an evaluation of our protocols for any specific setup can be realized by inspecting the general results obtained for various types of noise in Section IV. We find that the precision can be optimized by using machine learning protocols to mitigate the noise resulting from the increase in sensitivity. We also note that this tradeoff is expected to occur also if machine learning algorithms are included as subroutines in protocols that use highly entangled states to increase the precision, since highly entangled states are also very susceptible to noise.

\section{Conclusion}

The objective of this work was to study the performance of machine learning algorithms, namely the DE and PSO algorithm, in a quantum phase estimation scheme and determine their ability to come close to the SQL without resorting to multi-particle entanglement. To this end, both algorithms were tested against different configuration scenarios and were able to follow the SQL curve up to a given value of input qubits $N$. Under the constraint of $G=100$ it was  possible to notice that the algorithms start to loose performance for larger values of $N$. This becomes even more relevant in the scenario with quantum decoherence. However, it is important to reiterate that this is not a deficiency of the algorithms, but a direct consequence of the time and computational resources available.

These limitations can be overcome in future works by optimizing the code and making it more time efficient in order to allow both algorithms a larger number of iterations before converging to a valid solution. An immediate improvement would be to fully vectorize the code which would make it significantly faster. Another direct improvement would be to parallelize the code so that it could leverage the power of stronger computers with higher number of cores. This would allow for the different independent simulation threads of both algorithms to be conducted in parallel, making the estimation process even more time efficient. One can also  explore recurrent neural networks for the quantum phase estimation task, as they are particularly well suited for regression-based prediction models on time series events. It would be interesting to see how architectures such as the Long Short-Term Memory (LSTM), the Gated-Recurrent Unit (GRU) and the Temporal Convolutional Networks (TCN) would compare against our reinforcement learning algorithms.

Overall, we have shown that machine learning algorithms can provide noise-robust policies and we benchmarked their performances for various types of noise in close connection to experimental realizations. Above a certain critical value of noise, we found that machine learning based protocols show better results than non-adaptive methods. This opens a door to future developments that may push even further the precision of quantum phase estimation with classical machine learning algorithms without resorting to preparation and measurement of highly entangled states.

\appendix
\section*{Appendix}
\subsection{Theoretical background}
\label{apx:holevo}

\paragraph{Probabilities} Here we present analytical results related to the calculation of the Holevo variance. 
Given a qubit fed into the adaptive scheme with the adjustable phase set to $\theta_{m}$ and the unknown phase $\phi$ at step $m$, the probability of obtaining the result $\zeta_m$ is given by Eq. (\ref{eq:decoherence})
\begin{equation}
P_{\pm}(\zeta_{m}\vert \phi, \theta_{m} ) = \frac{1}{2}\pm \frac{\nu}{2}(-1)^{\zeta_{m}}\cos(\phi - \theta_{m}), \label{eq:probability}
\end{equation}
where the $+$ and $-$ signs correspond to the initial qubit in the state $\ket{0}$ respectively $\ket{1}$, and $\nu$ is the visibility, $\nu\in [0,1]$.

To simulate numerically $P_{\pm}(\zeta_{m}\vert \phi, \theta_{m} )$ we draw a random number from the interval $[0,1]$. If this number is smaller than $P^{(\nu )}_{\pm}(0\vert \phi, \theta_{m} )$, then the result $\zeta_{m}$ of the measurement is recorded as 0; if it is larger, it is recorded as 1. 
With these notations, we can understand easily the origin of the SQL. Indeed, the information about phase is contained in the probabilities Eq. (\ref{eq:probability}). However, the evaluation of this probabilities is affected by measurement projection noise.

\paragraph{Standard Quantum Limit} Suppose now that we try a simple non-adaptive strategy where we just repeat the Ramsey experiment for $N$ times with an initial states $|0\rangle$ and the same control phase $\theta$, selecting the results with $\zeta =1$. The precision $\delta \phi$ that we can achieve is given by $\delta P_{+}(1\vert \phi , \theta)=
(1/2) \nu \sin( \theta -\phi ) \delta \phi$. To achieve a small $\delta \phi$ it is advantageous to measure at phase parameters where the sine in this expression is maximal, that is, at around $P_{+}(1\vert \phi, \theta ) = 1/2 $.
At the same time, the measurement results have a binomial distribution, with $[\Delta P_{+}(1\vert \phi, \theta )]^2 = (1/N)P^{(\nu )}_{+}(1\vert \phi , \theta)[1-
P_{+}(1\vert \phi , \theta )]$, which results in $\Delta P_{+}(1\vert \phi, \theta ) = 1/(2\sqrt{N})$ around the region of maximal sensitivity $P_{+}(1\vert \phi, \theta )=1/2$.
In order to obtain a signal-to-noise ratio of at least 1, the uncertainty noise should be at most the same as the signal, $\delta P_{+}(1\vert \phi, \theta )\geq \Delta P_{+}(1\vert \phi, \theta )
$, which results in 
\begin{equation}
\delta \phi \geq (\Delta \phi)_{\rm SQL} = \frac{1}{\nu\sqrt{N}}.
\label{eq:SQL}
\end{equation}
Thus, the precision is limited by the SQL. As expected, a small visibility $\nu\neq 1$ results in an increased $(\Delta \phi)_{\rm SQL}$, corresponding to a loss in precision.

We can make this argument more rigorous by using the Cram\'er-Rao bound,
\begin{equation}
\delta \phi \geq \frac{1}{\sqrt{N\mathcal{F}_{\phi ,\theta}}} = \frac{1}{\sqrt{N}} \frac{\sqrt{1-\nu^2 \cos^2 (\phi - \theta )}}{\nu |\sin (\phi - \theta )|}.\label{eq:smu}
\end{equation}
The maximum value of $\mathcal{F}_{\phi ,\theta}$ is obtained for $\phi - \theta  = \pi/2, 3\pi /2$ and equals $\nu^2$. We then obtain the relation Eq. (\ref{eq:SQL}). 
An alternative derivation consists of calculating the fluctuations of the $\Sigma_{z} = \sum_{i=1}^{N}\sigma_{z}$ operator, 
namely $(\Delta \Sigma_{z})^{2} = \langle \Sigma_{z}^{2} \rangle - \langle \Sigma_{z} \rangle^{2} = N (\langle \sigma_{z}^2 \rangle -\langle\sigma_z\rangle^2)$, reflecting the fact that the fluctuations of the $\sigma_z$'s are uncorrelated. The average $\langle \sigma_z \rangle = P_{+}(0|\phi, \theta) - P_{+}(1|\phi,\theta)$, and using the error propagation rule
\begin{equation}
\Delta \phi = \frac{\Delta \sigma_{z}^N}{\vert \partial_{\phi} \langle\sigma_{z}^{N}\rangle\vert},
\end{equation}
we get the same result as in Eq. (\ref{eq:smu}).

\paragraph{Adaptive measurements}
In an adaptive measurement, the control phase is adjusted depending on the previous measurement values. Here, in order to simplify the notation, we will not write anymore the $\pm$ subscript for probabilities. Then, the probability of a sequence $\vec{\zeta}_{m}\equiv\{\zeta_{1}, \zeta_{2}, ..., \zeta_{m}\}$ of results given the phase adjustments $\vec{\theta}_{m}\equiv\{\theta_1, \theta_2, ....,  \theta_m\}$
is obtained as 
\begin{equation}
P(\vec{\zeta}_{m}|\phi , \theta)  = P(\zeta_{m}|\phi, \theta_{m})P(\zeta_{m-1}|\phi, \theta_{m-1}) ....P(\zeta_{1}|\phi, \theta_{1}).
\end{equation}
From the Bayes' theorem, we have at the $m$'th measurement
\begin{eqnarray}
P(\phi |\vec{\zeta}_{m}, \vec{\theta}_{m})=\frac{1}{P(\vec{\zeta}_{m})
}P(\vec{\zeta}_{m}|\phi ,\vec{\theta}_{m}) P(\phi). \label{eq:final_relation}
\end{eqnarray}
The prior $P(\phi)$ is taken uniformly distributed $P(\phi ) = 1/(2 \pi )$ and the marginalization $P(\vec{\zeta}_{m}) = \int_{0}^{2\pi} d\phi  P(\vec{\zeta}_{m}|\phi ,\vec{\theta}_{m}) P(\phi)$.

To further understand how the Holevo variance changes at every step of the algorithm, we present
a recursive calculation procedure based on the Fourier transform. We start by examining the sharpness appearing in the Holevo variance for a given depth $m$,
\begin{equation}
S  = 
\left| \int_{0}^{2\pi}d \phi e^{i \phi}P(\phi |\vec{\zeta}_{m},\vec{\theta}_{m}) \right|.
\label{eq:sharpness}
\end{equation}

From Eq. (\ref{eq:final_relation}) and Eq. (\ref{eq:probability}) we notice that the probability density is a harmonic function of $\phi, 2\phi, 3\phi ...$ therefore we can use the Fourier transform
\begin{equation}
P([k]|\vec{\zeta}_{m}, \vec{\theta}_{m}) = \frac{1}{2\pi} \int_{0}^{2\pi} d\phi e^{-i k\phi } P(\phi|\vec{\zeta}_{m}, \vec{\theta}_{m}),
\end{equation}
where for clarity we use $[k]$ as the index of the $k$'th Fourier component and  
with inverse given by
\begin{equation}
P(\phi|\vec{u}_{m}, \vec{\theta}_{m}) = \sum_{k=-\infty}^{\infty} e^{i k \phi}P([k]|\vec{u}_{m}, \vec{\theta}_{m}).
\end{equation}
From Eq. (\ref{eq:sharpness}) we get
\begin{equation}
S = |2 \pi P([k=-1]|\vec{\zeta}_{m}, \vec{\theta}_{m})|;
\end{equation}
in other words the Holevo variance can be seen as the $k=-1$ Fourier coefficient of the probability distribution. Moreover, the expectation value of $\phi$ is obtained from the average of $e^{i\phi}$
\begin{equation}
\langle\phi\rangle = \mathrm{arg} \langle e^{i\phi}\rangle = 2 \pi \mathrm{arg} P([k=-1]|\vec{\zeta}_{m}, \vec{\theta}_{m}).
\end{equation}

The Fourier coefficients can be calculated recursively. To see this, we apply the Bayes rule Eq. (\ref{eq:final_relation})
\begin{eqnarray}
P(\phi|\vec{\zeta}_{m}, \vec{\theta}_{m}) &=& \frac{1}{P(\vec{\zeta}_{m})
}P(\zeta_{m}|\phi ,\vec{\theta}_{m}) P(\vec{\zeta}_{m-1}|\phi ,\vec{\theta}_{m}) P(\phi) \nonumber \\
&=& P(\phi|\vec{\zeta}_{m-1}, \vec{\theta}_{m-1})P(\vec{\zeta}_{m}|\phi, \vec{\theta}_{m}) \frac{P(\vec{\zeta}_{m-1})}{P(\vec{\zeta}_{m})}, \nonumber
\end{eqnarray}
where the last fraction is a normalization factor.  Thus, if we consider non-normalized probabilities $p$, we have the Bayes relation  $p(\phi|\vec{\zeta}_{m}, \vec{\theta}_{m}) = p(\phi|\vec{\zeta}_{m-1}, \vec{\theta}_{m-1})p(\vec{\zeta}_{m}|\phi, \vec{\theta}_{m})$, from which we find the Fourier recurrence relation
\begin{eqnarray}
p([k]|\vec{\zeta}_{m}, \vec{\theta}_{m}) &=& \frac{1}{2} p([k]|\vec{\zeta}_{m-1}, \vec{\theta}_{m-1}) + \nonumber \\
& & \frac{1}{4}(-1)^{\zeta_{m}} \nu \left[e^{-i \theta_{m}}   p([k-1]|\vec{\zeta}_{m-1}, \vec{\theta}_{m-1})
+ \right. \nonumber \\
& & \left.  e^{i \theta_{m}}   p([k+1]|\vec{\zeta}_{m-1}, \vec{\theta}_{m-1}) 
\right].
\end{eqnarray}
To recover the correct normalization we simply impose $\int_{0}^{2\pi} d\phi P(\phi|\vec{\zeta}_{m}, \vec{\theta}_{m}) = 2 \pi P([k=0]|\vec{\zeta}_{m}, \vec{\theta}_{m})=1$, therefore we identify the normalization factor as 
$2\pi p([k=0]|\vec{\zeta}_{m})$, with $P([k]|\vec{\zeta}_{m}, \vec{\theta}_{m}) =   p([k]|\vec{\zeta}_{m}, \vec{\theta}_{m})/(2\pi p([k=0]|\vec{\zeta}_{m},  \vec{\theta}_{m}))$.

However, even with the use of these analytical results the evaluation of probabilities would require significant computational resources especially at large $N$. The reason is that, as it is clear from Eq. (\ref{eq:sharpness}), the evaluation needs to be done for all $2^{N}$ vectors $\zeta_{N}$, requiring exponentially more resources as $N$ increases.

\subsection{Differential Evolution}
\label{apx:evolution}

To study the convergence of the DE algorithm, two parameter control tests were done varying the controllable parameters of interest, $F$ and $C$, while keeping all the other values constant. The tests were conducted for $N=10$ qubits, $P=20$ populations, $G=50$ generations and $K=1000$ training instances. All sources of noise were neglected during the test, since they are not needed to study the overall performance of the different controllable parameters configurations and would only slow down the learning process.

The first parameter control test concerned the ability of the algorithm to converge to a solution, whether it was a valid one or not, with different possible parameter configurations. Referring to Eq. (\ref{eq:dispersion}) to evaluate the convergence of the algorithm the results obtained are displayed in Table \ref{tab:evo_dispersion1}. 

\begin{table}[H]
\renewcommand{\arraystretch}{1.2}

\begin{tabularx}{0.75\textwidth}{c|cccccc}
	\hline\hline
	\diagbox[height=2\line]{$F$}{$C$} & $0$ & $0.2$ & $0.4$ & $0.6$ & $0.8$ & $1$ \\
	\hline
	$0$	& $1.0166$ & $\textbf{0.0696}$ & $\textbf{0.0261}$ & $\textbf{0.0224}$ & $\textbf{0.0019}$ & $\textbf{0.0010}$ \\
	$0.2$ & $0.9334$ & $0.3726$ & $\textbf{0.0326}$ & $\textbf{0.0116}$ & $\textbf{0.0204}$ & $\textbf{0.0094}$ \\
	$0.4$ & $1.0897$ & $0.7491$ & $0.7231$ & $\textbf{0.0235}$ & $\textbf{0.0202}$ & $\textbf{0.0269}$ \\
	$0.6$ & $1.1647$ & $0.8364$ & $0.7764$ & $0.5226$ & $\textbf{0.0251}$ & $\textbf{0.0193}$ \\
	$0.8$ & $1.2183$ & $1.1356$ & $0.9333$ & $1.0901$ & $1.0133$ & $\textbf{0.0691}$ \\
	$1$	& $1.3101$ & $1.2460$ & $1.2433$ & $1.3081$ & $1.1359$ & $1.2798$\\
	\hline\hline
\end{tabularx}

\caption{Convergence values $L(F,C)$ for the DE algorithm. Values in bold represent configurations where convergence was achieved. Convergence was considered only for values of $L \leq 0.1256$, which corresponds to a maximum dispersion of approximately $2\%$ of the entire $2\pi$ search space for each entry of the different candidate solution vector.}
\label{tab:evo_dispersion1}
\end{table}

Inspecting the results in Table \ref{tab:evo_dispersion1} it is possible to find a pattern where only the configurations where the amplification constant $F$ is strictly smaller than the crossover constant $C$ lead to the convergence of the algorithm within the given number of generations. In other words: $F < C$ guarantees the convergence of the DE algorithm. This newfound rule can be understood from a physical interpretation of the process. Larger values for the crossover parameter $C$ encourage the different candidate solutions to experience more configurations, thus covering a larger area of the entire search space. This increased exploration of the search space, leads to a more exhaustive pursuit of the best policy solution. However, by increasing the amplification constant $F$, the step of this search process is also increased. This may lead the algorithm to overlook possibly desirable candidate solutions, which ultimately may not allow the algorithm to converge. Hence, smaller values of $F$ for larger values of $C$ ensure a more careful examination of the search space at each iteration of the algorithm.

The second parameter control test was concerned with the ability of the algorithm to converge to a valid a solution based on Eq. (\ref{eq:holevo}). Recalling that lower values of the Holevo variance correspond to candidate solutions with a higher degree of accuracy and keeping exactly the same configuration setup as before, the results obtained are shown in Table \ref{tab:evo_performance1}.

\begin{table}[H]
\centering
\renewcommand{\arraystretch}{1.2}

\begin{tabularx}{0.8\textwidth}{c|cccccc}
	\hline\hline
	\diagbox[height=2\line]{$F$}{$C$} & $0$ & $0.2$ & $0.4$ & $0.6$ & $0.8$ & $1$ \\ 
	\hline
	$0$	& $-0.3735$ & $\textbf{-1.2012}$ & $\textbf{-0.7845}$ & $\textbf{-0.7758}$ & $\textbf{-0.7083}$ & $\textbf{-0.4196}$ \\
	$0.2$ & $ 5.2111$ & $-1.6266$ & $\textbf{-1.3737}$ & $\textbf{-1.4401}$ & $\textbf{-1.0244}$ & $\textbf{-1.0948}$ \\
	$0.4$ & $-0.3897$ & $-0.2398$ & $ 0.6181$ & $\textbf{-1.5613}$ & $\textbf{-1.2976}$ & $\textbf{-1.5611}$ \\
	$0.6$ & $ 0.3131$ & $-1.1896$ & $-1.1589$ & $-1.3696$ & $\textbf{-1.8321}$ & $\textbf{-1.5453}$ \\
	$0.8$ & $ 0.6326$ & $-1.0395$ & $ 0.3694$ & $-0.6645$ & $0.8763$ & $\textbf{-1.6784}$ \\
	$1$	& $-0.1786$ & $ 1.2460$ & $ 2.5564$ & $ 0.6131$ & $ 4.2364$ & $ 0.0864$\\
	\hline\hline
\end{tabularx}

\caption{Performance results based on the logarithm of the Holevo variance $\ln(V_H(F,C))$ for the DE algorithm. Values in bold represent the configurations which have achieved convergence in Table \ref{tab:evo_dispersion1}. Lower variance values are obtained by solutions that lead to more precise estimations of the unknown parameter that is being measured.}
\label{tab:evo_performance1}
\end{table}

It is possible to see in Table \ref{tab:evo_performance1} that the parameter configurations that previously lead the algorithm to converge to a solution also lead the algorithm to converge to a valid solution if and only if $F>0$. Hence, the previously found rule can be made more specific: $F<C \; \forall \; F>0$ guarantees the convergence of the DE algorithm to a valid solution. Note, however, that the algorithm also obtained good performance results even with configurations that did not follow the previously found rule. These are misleading results and can be quickly disregarded, since the algorithm is told to stop regardless of having converged to a solution or not. Therefore, in these situations the resulting averaged solution among all the populations may haphazardly coincide with a valid one. These situations must, nevertheless, be regarded as fortuitously events and not be considered as possible valid solutions as the algorithm was not indeed able to converge to any solution.

\begin{figure}[h!]
	\centering
	\includegraphics[width=380pt]{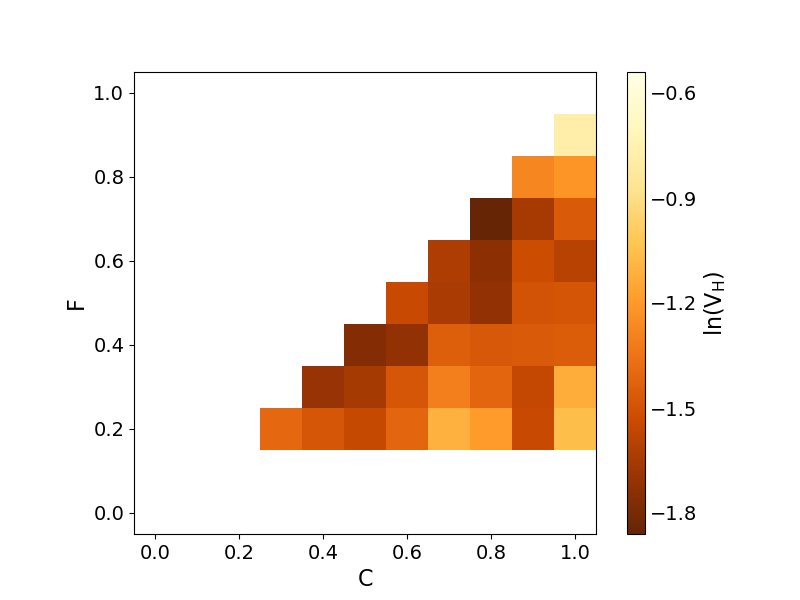}
	\caption{Visual representation of the performance values obtained for the DE algorithm based on the logarithm of the Holevo variance $\ln(V_H(F,C))$ with the parameter configurations following the rule $F<C \; \forall \; F>0$. Better configurations lead to lower values of $\ln(V_H(F,C))$.}
	\label{fig:evo_performance2}
\end{figure}

Having arrived at the conclusion that $F$ must be strictly smaller than $C$ while still being bigger than zero for the DE algorithm to work as desired, it is important to find the exact parameter configuration that guarantees the most promising results. A more thorough control test regarding the performance of the algorithm for each configuration following this rule was conducted three times and its results averaged and presented in Fig. \ref{fig:evo_performance2}. Note that this time, the step between the different parameter configurations was reduced to avoid overlooking possible optimal parameter configurations.

Looking at the results in Fig. \ref{fig:evo_performance2} it is possible to notice that the best results are obtained among the configurations where $C$ is only slightly bigger than $F$. It is also possible to notice that these results seem to be stronger for values of $F$ in the interval $[0.3,0.7]$. Ensuring that the step size at each iteration of the algorithm is not too big as to overstep a possible valid solution, it should not be too small either to avoid getting stuck in a local minimum. This explains why these values of $F$ lead to better results. Among these parameter configurations, the ones where where the value of $C$ is in the interval $[0.4,0.8]$ and is only slightly bigger than $F$ lead to the overall best results. Hence, the optimal parameter configuration considered for the DE algorithm is: $F = 0.7$ and $C = 0.8$.

\subsection{Particle Swarm Optimization}
\label{apx:optimization}

The PSO algorithm has four controllable parameters of interest. Two of these parameters, $\alpha$ and $\beta$, are mostly concerned with the ability of the algorithm to converge to a valid solution, while the other two, $w$ and $v_{max}$, regulate the speed at which the parameter will converge to a solution.

As before, the first step is to study the overall ability of the algorithm to converge to a solution with the different possible parameter configurations. Keeping in mind that only the parameters $\alpha$ and $\beta$ have a direct impact on the convergence process, all the remaining parameters were set to a constant value and the results obtained are presented in Table \ref{tab:opt_dispersion1}.

\begin{table}[H]
\centering
\renewcommand{\arraystretch}{1.2}

\begin{tabularx}{0.8\textwidth}{c|cccccc}
	\hline\hline
	\diagbox[height=2\line]{$\alpha$}{$\beta$} & $0$ & $0.2$ & $0.4$ & $0.6$ & $0.8$ & $1$ \\
	\hline
	$0$	& $1.3231$ & $1.3292$ & $1.3139$ & $1.3341$ & $1.4045$ & $1.3105$ \\
	$0.2$ & $\textbf{0.1239}$ & $\textbf{0.1072}$ & $0.2193$ & $0.2379$ & $0.4736$ & $0.4580$ \\
	$0.4$ & $\textbf{0.1014}$ & $\textbf{0.0998}$ & $\textbf{0.0974}$ & $0.1742$ & $0.2328$ & $0.4280$ \\
	$0.6$ & $\textbf{0.0771}$ & $\textbf{0.0788}$ & $\textbf{0.0808}$ & $\textbf{0.0902}$ & $0.1296$ & $0.2410$ \\
	$0.8$ & $\textbf{0.0821}$ & $\textbf{0.0821}$ & $\textbf{0.1072}$ & $\textbf{0.0734}$ & $\textbf{0.0806}$ & $0.1589$ \\
	$1$	& $\textbf{0.0835}$ & $\textbf{0.0976}$ & $\textbf{0.1008}$ & $\textbf{0.0887}$ & $\textbf{0.0708}$ & $\textbf{0.0933}$\\
	\hline\hline
\end{tabularx}
\caption{Convergence values $L(\alpha,\beta)$ for the PSO algorithm. Convergence was considered only for values of $L \leq 0.1256$, which corresponds to a maximum dispersion of approximately $2\%$ of the entire $2\pi$ search space for each entry of the different candidate solution vector.}
\label{tab:opt_dispersion1}
\end{table}

Considering the results obtained, it is possible to see that as long as $\beta>0$ the algorithm is able to converge to a solution, even if it may take more than $G=50$ iterations. This can be intuitively understood by remembering that the $\beta$ parameter defines the desirability of each individual to follow the best found solution of the entire group until that point. Thus, as long as this value is bigger than zero, each individual will feel the urge to move towards the best globally found solution. In other words: $\beta > 0$ guarantees the convergence of the PSO algorithm.

Additionally, it is also possible to notice that for values of $\beta$ equal or larger than $\alpha$ the algorithm is always able to converge to a solution within the stipulated number of iterations. This leads to a second conclusion: $\beta \geq \alpha \; \forall \; \beta > 0$ guarantees the convergence of the PSO algorithm within the given number of iterations. However, for $\beta$ values much larger than $\alpha$ the algorithm converges within very few iterations as it gets stuck in local minima. These configurations privilege moving towards the already found best global solution in detriment of exploring other new possible solutions.

Keeping the same configurations as before, it is necessary to study which of these parameter configurations lead to valid solutions. Referring to Eq. (\ref{eq:holevo}) to evaluate their performance, the results obtained are shown in Table \ref{tab:opt_performance1}.

\begin{table}[H]
\centering
\renewcommand{\arraystretch}{1.2}

\begin{tabularx}{0.8\textwidth}{c|cccccc}
	\hline\hline
	\diagbox[height=2\line]{$\alpha$}{$\beta$} & $0$ & $0.2$ & $0.4$ & $0.6$ & $0.8$ & $1$ \\
	\hline
	$0$	& $ 1.3082$ & $ 2.3717$ & $ 0.7076$ & $ 1.5727$ & $ 1.7540$ & $ 6.0563$ \\
	$0.2$ & $\textbf{-1.5475}$ & $\textbf{-1.3651}$ & $-1.6623$ & $-1.4530$ & $-0.8383$ & $-1.0281$ \\
	$0.4$ & $\textbf{-1.4016}$ & $\textbf{-1.5598}$ & $\textbf{-1.5295}$ & $-1.6831$ & $-1.6005$ & $ 1.2126$ \\
	$0.6$ & $\textbf{-1.6229}$ & $\textbf{-1.5110}$ & $\textbf{-1.9324}$ & $\textbf{-1.3650}$ & $-1.6605$ & $-1.5408$ \\
	$0.8$ & $\textbf{-1.7227}$ & $\textbf{-1.4539}$ & $\textbf{-1.5615}$ & $\textbf{-1.5828}$ & $\textbf{-1.8361}$ & $-1.2132$ \\
	$1$	& $\textbf{-1.6869}$ & $\textbf{-1.3290}$ & $\textbf{-1.3956}$ & $\textbf{-1.6726}$ & $\textbf{-1.5909}$ & $\textbf{-1.4826}$\\
	\hline\hline
\end{tabularx}

\caption{Performance results based on the logarithm of the Holevo variance $\ln(V_H(\alpha,\beta))$ for the PSO algorithm. Values in bold represent the configurations which have achieved convergence in Table \ref{tab:opt_dispersion1}. Lower variance values are obtained by solutions that lead to more precise estimations of the unknown parameter that is being measured.}
\label{tab:opt_performance1}
\end{table}

Considering the previously found rule and the results obtained in Table \ref{tab:opt_performance1} is is possible to see that for $\beta \geq \alpha$ while keeping $\beta > 0$ not only guarantees the convergence of the algorithm, but also guarantees the convergence of the algorithm to a valid solution. Therefore, it is possible to come to the conclusion that: $\beta \geq \alpha \; \forall \; \beta > 0$ guarantees the convergence of the PSO algorithm within the given number of iterations to a valid solution.

\begin{figure}[h!]
	\centering
	\includegraphics[width=380pt]{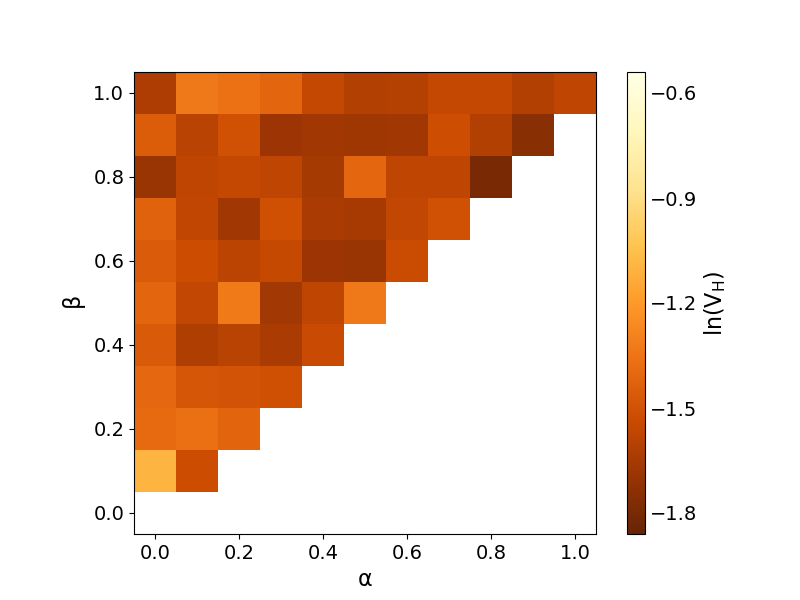}
	\caption{Visual representation of the performance values obtained for the PSO algorithm based on the logarithm of the Holevo variance $\ln(V_H(\alpha,\beta))$ with the parameter configurations following the rule $\beta \geq \alpha \; \forall \; \beta > 0$. Better configurations lead to lower values of $\ln(V_H(\alpha,\beta))$.}
	\label{fig:pso_performance2}
\end{figure}

Having arrived to this conclusion it is still important to determine which parameter configuration leads to the best results. As before, a more thorough control test on the performance of the algorithm under these conditions was conducted three times and its results averaged and displayed in Fig. \ref{fig:pso_performance2}. Notice that, once again, there is a smaller increment between each parameter step in order to avoid overlooking any possibly better parameter configuration.

Unlike with the DE algorithm, the results obtained for the PSO algorithm are less contrasting. Even though configurations where the value of $\alpha$ and $\beta$ are similar to each other deliver slightly better results, it appears that having $\beta \geq \alpha \; \forall \; \beta > 0$ is a condition strong enough to arrive at optimal candidate solutions. This robustness of the algorithm against different possible parameter configurations would have allowed for a multiple number of pair of configurations to be chosen. Therefore, the first two optimal controllable parameters chosen for the PSO algorithm are $\alpha=0.8$ and $\beta=0.8$ as they appear to stand out even if only slightly.

To study how the remaining two controllable parameters, the update weight $w$ and the maximum velocity $v_{max}$, might impact the performance of the algorithm a similar study was also made for them. Keeping all of the other parameters constant and already using the optimal configuration $\alpha=0.8$ and $\beta=0.8$, the performance of the algorithm was studied for the different possible configurations of $w$ and $v_{max}$. However, this time all the different configurations of these two parameters have lead the algorithm to converge to a solution and, most importantly, to a valid one. This comes as no surprise since these parameters, from a physical viewpoint of their behaviour on the search space, were not expected to have any influence on the ability of the algorithm to converge to a solution. Their impact was mostly expected on converging times.

Having arrived to an optimal configuration for the parameters $\alpha$ and $\beta$, the algorithm is quite robust to any configuration of the parameters $w$ and $v_{max}$ and is able to converge to a solution in the available number of iterations. It is nevertheless important to bare in mind that these parameter analysis was made for $N=10$. For larger values of $N$, the convergence speed of the algorithms becomes more relevant as the search space scales polynomially with $N$. Unfortunately, such a parameter evaluation would also have been too computationally expensive and time consuming. Keeping this in mind, the optimal configuration considered for the PSO algorithm is: $\alpha = 0.8$, $\beta = 0.8$, $w = 0.8$ and $v_{max} = 0.2$.


\begin{backmatter}

\section*{Acknowledgements}
The authors would like to thank João Seixas, José Leitão, Pantita Palittapongarnpim, Marco Antonio Rodríguez García and Barry C. Sanders  for useful discussions on machine learning methods.

\section*{Funding}
This project has received funding from the European Union's Horizon
2020 research and innovation programme under grant agreement No. 862644 (FET-Open  QUARTET).
In addition, YO would like to thank for support from Funda\c{c}\~{a}o para a Ci\^{e}ncia e a Tecnologia (Portugal), namely through project UIDB/50008/2020, as well as from projects TheBlinQC and QuantHEP supported by the EU H2020 QuantERA ERA-NET Cofund in Quantum Technologies and by FCT (QuantERA/0001/2017 and QuantERA/0001/2019, respectively), and from the EU H2020 Quantum Flagship project QMiCS (820505).  AS and GSP acknowledge additional support from the Foundational Questions Institute Fund (FQXi) via  grant no. FQXi-IAF19-06, as well as from the Academy of Finland through RADDESS grant no. 328193 and through the "Finnish
Center of Excellence in Quantum Technology QTF" grants nos. 312296 and  336810.

\section*{Abbreviations}
SQL, Standard Quantum Limit; HL, Heisenberg Limit; DE, Differential Evolution; PSO, Particle Swarm Optimisation; NISQ, Noisy Intermediate-Scale Quantum; RTN, Random Telegraph Noise; LSTM, Long Short-Term Memory, GRU Gated-Recurrent Unit; TCN, Temporal Convolutional Networks.

.

\section*{Availability of data and materials}
The code used in this paper is available at \underline{\href{https://github.com/nelsonfilipecosta/Adaptive-Quantum-Phase-Estimation}{GitHub}}, or from the authors upon reasonable requests, for anyone interested in simulating these scenarios either for experimental purposes or simply for gaining theoretical insights.


\section*{Competing interests}
The authors declare that they have no competing interests.


\section*{Authors' contributions}
Conceptualization: GSP and YO. Code development for machine learning: NC. Code development for non-adaptive protocols: A. S. Analytical results and connection to experiments: GSP. Writing of the paper: NC and GSP, with additional contributions from YO and AS.
All authors have read and approved the final manuscript.

	

	\bibliographystyle{bmc-mathphys} 
	\bibliography{EPJ_biblio}      

\providecommand{\noopsort}[1]{}\providecommand{\singleletter}[1]{#1}%

\begin{thebibliography}{71}
\ifx \bisbn   \undefined \def \bisbn  #1{ISBN #1}\fi
\ifx \binits  \undefined \def \binits#1{#1}\fi
\ifx \bauthor  \undefined \def \bauthor#1{#1}\fi
\ifx \batitle  \undefined \def \batitle#1{#1}\fi
\ifx \bjtitle  \undefined \def \bjtitle#1{#1}\fi
\ifx \bvolume  \undefined \def \bvolume#1{\textbf{#1}}\fi
\ifx \byear  \undefined \def \byear#1{#1}\fi
\ifx \bissue  \undefined \def \bissue#1{#1}\fi
\ifx \bfpage  \undefined \def \bfpage#1{#1}\fi
\ifx \blpage  \undefined \def \blpage #1{#1}\fi
\ifx \burl  \undefined \def \burl#1{\textsf{#1}}\fi
\ifx \doiurl  \undefined \def \doiurl#1{\textsf{#1}}\fi
\ifx \betal  \undefined \def \betal{\textit{et al.}}\fi
\ifx \binstitute  \undefined \def \binstitute#1{#1}\fi
\ifx \binstitutionaled  \undefined \def \binstitutionaled#1{#1}\fi
\ifx \bctitle  \undefined \def \bctitle#1{#1}\fi
\ifx \beditor  \undefined \def \beditor#1{#1}\fi
\ifx \bpublisher  \undefined \def \bpublisher#1{#1}\fi
\ifx \bbtitle  \undefined \def \bbtitle#1{#1}\fi
\ifx \bedition  \undefined \def \bedition#1{#1}\fi
\ifx \bseriesno  \undefined \def \bseriesno#1{#1}\fi
\ifx \blocation  \undefined \def \blocation#1{#1}\fi
\ifx \bsertitle  \undefined \def \bsertitle#1{#1}\fi
\ifx \bsnm \undefined \def \bsnm#1{#1}\fi
\ifx \bsuffix \undefined \def \bsuffix#1{#1}\fi
\ifx \bparticle \undefined \def \bparticle#1{#1}\fi
\ifx \barticle \undefined \def \barticle#1{#1}\fi
\ifx \bconfdate \undefined \def \bconfdate #1{#1}\fi
\ifx \botherref \undefined \def \botherref #1{#1}\fi
\ifx \url \undefined \def \url#1{\textsf{#1}}\fi
\ifx \bchapter \undefined \def \bchapter#1{#1}\fi
\ifx \bbook \undefined \def \bbook#1{#1}\fi
\ifx \bcomment \undefined \def \bcomment#1{#1}\fi
\ifx \oauthor \undefined \def \oauthor#1{#1}\fi
\ifx \citeauthoryear \undefined \def \citeauthoryear#1{#1}\fi
\ifx \endbibitem  \undefined \def \endbibitem {}\fi
\ifx \bconflocation  \undefined \def \bconflocation#1{#1}\fi
\ifx \arxivurl  \undefined \def \arxivurl#1{\textsf{#1}}\fi
\csname PreBibitemsHook\endcsname

\bibitem{giovannetti2004quantum}
\begin{barticle}
\bauthor{\bsnm{Giovannetti}, \binits{V.}},
\bauthor{\bsnm{Lloyd}, \binits{S.}},
\bauthor{\bsnm{Maccone}, \binits{L.}}:
\batitle{Quantum-enhanced measurements: beating the standard quantum limit}.
\bjtitle{Science}
\bvolume{306}(\bissue{5700}),
\bfpage{1330}--\blpage{1336}
(\byear{2004})
\end{barticle}
\endbibitem

\bibitem{giovannetti2006quantum}
\begin{barticle}
\bauthor{\bsnm{Giovannetti}, \binits{V.}},
\bauthor{\bsnm{Lloyd}, \binits{S.}},
\bauthor{\bsnm{Maccone}, \binits{L.}}:
\batitle{Quantum metrology}.
\bjtitle{Physical review letters}
\bvolume{96}(\bissue{1}),
\bfpage{010401}
(\byear{2006})
\end{barticle}
\endbibitem

\bibitem{giovannetti2011advances}
\begin{barticle}
\bauthor{\bsnm{Giovannetti}, \binits{V.}},
\bauthor{\bsnm{Lloyd}, \binits{S.}},
\bauthor{\bsnm{Maccone}, \binits{L.}}:
\batitle{Advances in quantum metrology}.
\bjtitle{Nature photonics}
\bvolume{5}(\bissue{4}),
\bfpage{222}
(\byear{2011})
\end{barticle}
\endbibitem

\bibitem{toth2014quantum}
\begin{barticle}
\bauthor{\bsnm{T{\'o}th}, \binits{G.}},
\bauthor{\bsnm{Apellaniz}, \binits{I.}}:
\batitle{Quantum metrology from a quantum information science perspective}.
\bjtitle{Journal of Physics A: Mathematical and Theoretical}
\bvolume{47}(\bissue{42}),
\bfpage{424006}
(\byear{2014})
\end{barticle}
\endbibitem

\bibitem{helstrom1969quantum}
\begin{barticle}
\bauthor{\bsnm{Helstrom}, \binits{C.W.}}:
\batitle{Quantum detection and estimation theory}.
\bjtitle{Journal of Statistical Physics}
\bvolume{1}(\bissue{2}),
\bfpage{231}--\blpage{252}
(\byear{1969})
\end{barticle}
\endbibitem

\bibitem{holevo2011probabilistic}
\begin{bbook}
\bauthor{\bsnm{Holevo}, \binits{A.S.}}:
\bbtitle{Probabilistic and Statistical Aspects of Quantum Theory}
vol. \bseriesno{1}.
\bpublisher{Springer}, \blocation{???}
(\byear{2011})
\end{bbook}
\endbibitem

\bibitem{cramer2016mathematical}
\begin{bbook}
\bauthor{\bsnm{Cram{\'e}r}, \binits{H.}}:
\bbtitle{Mathematical Methods of Statistics (PMS-9)}
vol. \bseriesno{9}.
\bpublisher{Princeton university press}, \blocation{???}
(\byear{2016})
\end{bbook}
\endbibitem

\bibitem{braginski1975}
\begin{barticle}
\bauthor{\bsnm{Braginski{\u{\i}}}, \binits{V.B.}},
\bauthor{\bsnm{Vorontsov}, \binits{Y.I.}}:
\batitle{Quantum-mechanical limitations in macroscopic experiments and modern
  experimental technique}.
\bjtitle{Soviet Physics Uspekhi}
\bvolume{17}(\bissue{5}),
\bfpage{644}--\blpage{650}
(\byear{1975}).
doi:\doiurl{10.1070/pu1975v017n05abeh004362}
\end{barticle}
\endbibitem

\bibitem{braginsky1980quantum}
\begin{barticle}
\bauthor{\bsnm{Braginsky}, \binits{V.B.}},
\bauthor{\bsnm{Vorontsov}, \binits{Y.I.}},
\bauthor{\bsnm{Thorne}, \binits{K.S.}}:
\batitle{Quantum nondemolition measurements}.
\bjtitle{Science}
\bvolume{209}(\bissue{4456}),
\bfpage{547}--\blpage{557}
(\byear{1980})
\end{barticle}
\endbibitem

\bibitem{braginsky1995quantum}
\begin{bbook}
\bauthor{\bsnm{Braginsky}, \binits{V.B.}},
\bauthor{\bsnm{Braginsky}, \binits{V.B.}},
\bauthor{\bsnm{Khalili}, \binits{F.Y.}}:
\bbtitle{Quantum Measurement}.
\bpublisher{Cambridge University Press}, \blocation{???}
(\byear{1995})
\end{bbook}
\endbibitem

\bibitem{ozawa1989realization}
\begin{bchapter}
\bauthor{\bsnm{Ozawa}, \binits{M.}}:
\bctitle{Realization of measurement and the standard quantum limit}.
In: \bbtitle{Squeezed and Nonclassical Light},
pp. \bfpage{263}--\blpage{286}.
\bpublisher{Springer}, \blocation{???}
(\byear{1989})
\end{bchapter}
\endbibitem

\bibitem{Buluta_2011}
\begin{barticle}
\bauthor{\bsnm{Buluta}, \binits{I.}},
\bauthor{\bsnm{Ashhab}, \binits{S.}},
\bauthor{\bsnm{Nori}, \binits{F.}}:
\batitle{Natural and artificial atoms for quantum computation}.
\bjtitle{Rep. Prog. Phys.}
\bvolume{74}(\bissue{10}),
\bfpage{104401}
(\byear{2011}).
doi:\doiurl{10.1088/0034-4885/74/10/104401}
\end{barticle}
\endbibitem

\bibitem{Pryde_2019}
\begin{barticle}
\bauthor{\bsnm{Slussarenko}, \binits{S.}},
\bauthor{\bsnm{Pryde}, \binits{G.J.}}:
\batitle{Photonic quantum information processing: A concise review}.
\bjtitle{Applied Physics Reviews}
\bvolume{6}(\bissue{4}),
\bfpage{041303}
(\byear{2019}).
doi:\doiurl{10.1063/1.5115814}.
\arxivurl{https://doi.org/10.1063/1.5115814}
\end{barticle}
\endbibitem

\bibitem{Flamini_2018}
\begin{barticle}
\bauthor{\bsnm{Flamini}, \binits{F.}},
\bauthor{\bsnm{Spagnolo}, \binits{N.}},
\bauthor{\bsnm{Sciarrino}, \binits{F.}}:
\batitle{Photonic quantum information processing: a review}.
\bjtitle{Reports on Progress in Physics}
\bvolume{82}(\bissue{1}),
\bfpage{016001}
(\byear{2018}).
doi:\doiurl{10.1088/1361-6633/aad5b2}
\end{barticle}
\endbibitem

\bibitem{Wrachtrup_2006}
\begin{barticle}
\bauthor{\bsnm{Wrachtrup}, \binits{J.}},
\bauthor{\bsnm{Jelezko}, \binits{F.}}:
\batitle{Processing quantum information in diamond}.
\bjtitle{Journal of Physics: Condensed Matter}
\bvolume{18}(\bissue{21}),
\bfpage{807}--\blpage{824}
(\byear{2006}).
doi:\doiurl{10.1088/0953-8984/18/21/s08}
\end{barticle}
\endbibitem

\bibitem{Prawer_2018}
\begin{bbook}
\bauthor{\bsnm{Prawer}, \binits{S.}},
\bauthor{\bsnm{Aharonovich}, \binits{I.}}:
\bbtitle{Quantum Information Processing with Diamond: Principles and
  Applications},
\bedition{1st} edn.
\bpublisher{Woodhead Publishing, Limited}, \blocation{???}
(\byear{2018}).
doi:\doiurl{10.5555/3312180}
\end{bbook}
\endbibitem

\bibitem{Lange2012}
\begin{bbook}
\bauthor{\bsnm{Lange}, \binits{W.}}:
In: \beditor{\bsnm{Meyers}, \binits{R.A.}} (ed.)
\bbtitle{Quantum Computing with Trapped Ions},
pp. \bfpage{2406}--\blpage{2436}.
\bpublisher{Springer},
\blocation{New York, NY}
(\byear{2012}).
doi:\doiurl{10.1007/978-1-4614-1800-9\_149}.
\burl{https://doi.org/10.1007/978-1-4614-1800-9\_149}
\end{bbook}
\endbibitem

\bibitem{Sage_2019}
\begin{barticle}
\bauthor{\bsnm{Bruzewicz}, \binits{C.D.}},
\bauthor{\bsnm{Chiaverini}, \binits{J.}},
\bauthor{\bsnm{McConnell}, \binits{R.}},
\bauthor{\bsnm{Sage}, \binits{J.M.}}:
\batitle{Trapped-ion quantum computing: Progress and challenges}.
\bjtitle{Applied Physics Reviews}
\bvolume{6}(\bissue{2}),
\bfpage{021314}
(\byear{2019}).
doi:\doiurl{10.1063/1.5088164}.
\arxivurl{https://doi.org/10.1063/1.5088164}
\end{barticle}
\endbibitem

\bibitem{Paraoanu2014}
\begin{barticle}
\bauthor{\bsnm{Paraoanu}, \binits{G.S.}}:
\batitle{Recent progress in quantum simulation using superconducting circuits}.
\bjtitle{J. Low Temp. Phys.}
\bvolume{175}(\bissue{5}),
\bfpage{633}--\blpage{654}
(\byear{2014}).
doi:\doiurl{10.1007/s10909-014-1175-8}
\end{barticle}
\endbibitem

\bibitem{Nori_review_2011}
\begin{barticle}
\bauthor{\bsnm{You}, \binits{J.Q.}},
\bauthor{\bsnm{Nori}, \binits{F.}}:
\batitle{Atomic physics and quantum optics using superconducting circuits}.
\bjtitle{Nature}
\bvolume{474},
\bfpage{589}
(\byear{2011}).
doi:\doiurl{10.1038/nature101222}
\end{barticle}
\endbibitem

\bibitem{degen_2017}
\begin{botherref}
\oauthor{\bsnm{Degen}, \binits{C.â.}},
\oauthor{\bsnm{Reinhard}, \binits{F.}},
\oauthor{\bsnm{Cappellaro}, \binits{P.}}:
Quantum sensing.
Reviews of Modern Physics
\textbf{89}(3)
(2017).
doi:\doiurl{10.1103/revmodphys.89.035002}
\end{botherref}
\endbibitem

\bibitem{paris2009quantum}
\begin{barticle}
\bauthor{\bsnm{Paris}, \binits{M.G.}}:
\batitle{Quantum estimation for quantum technology}.
\bjtitle{International Journal of Quantum Information}
\bvolume{7}(\bissue{supp01}),
\bfpage{125}--\blpage{137}
(\byear{2009})
\end{barticle}
\endbibitem

\bibitem{biercuk_2010}
\begin{barticle}
\bauthor{\bsnm{Biercuk}, \binits{M.J.}},
\bauthor{\bsnm{Uys}, \binits{H.}},
\bauthor{\bsnm{Britton}, \binits{J.W.}},
\bauthor{\bsnm{VanDevender}, \binits{A.P.}},
\bauthor{\bsnm{Bollinger}, \binits{J.J.}}:
\batitle{Ultrasensitive detection of force and displacement using trapped
  ions}.
\bjtitle{Nature Nanotechnology}
\bvolume{5}(\bissue{9}),
\bfpage{646}--\blpage{650}
(\byear{2010}).
doi:\doiurl{10.1038/nnano.2010.165}
\end{barticle}
\endbibitem

\bibitem{Scerri_2020}
\begin{barticle}
\bauthor{\bsnm{Scerri}, \binits{E.}},
\bauthor{\bsnm{Gauger}, \binits{E.M.}},
\bauthor{\bsnm{Bonato}, \binits{C.}}:
\batitle{Extending qubit coherence by adaptive quantum environment learning}.
\bjtitle{New Journal of Physics}
\bvolume{22}(\bissue{3}),
\bfpage{035002}
(\byear{2020}).
doi:\doiurl{10.1088/1367-2630/ab7bf3}
\end{barticle}
\endbibitem

\bibitem{brownnutt_2015}
\begin{barticle}
\bauthor{\bsnm{Brownnutt}, \binits{M.}},
\bauthor{\bsnm{Kumph}, \binits{M.}},
\bauthor{\bsnm{Rabl}, \binits{P.}},
\bauthor{\bsnm{Blatt}, \binits{R.}}:
\batitle{Ion-trap measurements of electric-field noise near surfaces}.
\bjtitle{Reviews of Modern Physics}
\bvolume{87}(\bissue{4}),
\bfpage{1419}--\blpage{1482}
(\byear{2015}).
doi:\doiurl{10.1103/revmodphys.87.1419}
\end{barticle}
\endbibitem

\bibitem{review_Naderi}
\begin{botherref}
\oauthor{\bsnm{Motazedifard}, \binits{A.}},
\oauthor{\bsnm{Dalafi}, \binits{A.}},
\oauthor{\bsnm{Naderi}, \binits{M.H.}}:
Ultra-precision quantum sensing and measurement based on nonlinear hybrid
  optomechanical systems containing ultracold atoms or atomic-be.
arXiv:2011.01336
(2020)
\end{botherref}
\endbibitem

\bibitem{abadie2011gravitational}
\begin{barticle}
\bauthor{\bsnm{Abadie}, \binits{J.}},
\bauthor{\bsnm{Abbott}, \binits{B.}},
\bauthor{\bsnm{Abbott}, \binits{R.}},
\bauthor{\bsnm{Abbott}, \binits{T.}},
\bauthor{\bsnm{Abernathy}, \binits{M.}},
\bauthor{\bsnm{Adams}, \binits{C.}},
\bauthor{\bsnm{Adhikari}, \binits{R.}},
\bauthor{\bsnm{Affeldt}, \binits{C.}},
\bauthor{\bsnm{Allen}, \binits{B.}},
\bauthor{\bsnm{Allen}, \binits{G.}}, \betal:
\batitle{A gravitational wave observatory operating beyond the quantum
  shot-noise limit}.
\bjtitle{Nature Physics}
\bvolume{7}(\bissue{12}),
\bfpage{962}
(\byear{2011})
\end{barticle}
\endbibitem

\bibitem{Smerzi_RMP}
\begin{barticle}
\bauthor{\bsnm{Pezz\`e}, \binits{L.}},
\bauthor{\bsnm{Smerzi}, \binits{A.}},
\bauthor{\bsnm{Oberthaler}, \binits{M.K.}},
\bauthor{\bsnm{Schmied}, \binits{R.}},
\bauthor{\bsnm{Treutlein}, \binits{P.}}:
\batitle{Quantum metrology with nonclassical states of atomic ensembles}.
\bjtitle{Rev. Mod. Phys.}
\bvolume{90},
\bfpage{035005}
(\byear{2018}).
doi:\doiurl{10.1103/RevModPhys.90.035005}
\end{barticle}
\endbibitem

\bibitem{berry2000optimal}
\begin{barticle}
\bauthor{\bsnm{Berry}, \binits{D.}},
\bauthor{\bsnm{Wiseman}, \binits{H.}}:
\batitle{Optimal states and almost optimal adaptive measurements for quantum
  interferometry}.
\bjtitle{Physical Review Letters}
\bvolume{85}(\bissue{24}),
\bfpage{5098}
(\byear{2000})
\end{barticle}
\endbibitem

\bibitem{berry2001optimal}
\begin{barticle}
\bauthor{\bsnm{Berry}, \binits{D.W.}},
\bauthor{\bsnm{Wiseman}, \binits{H.}},
\bauthor{\bsnm{Breslin}, \binits{J.}}:
\batitle{Optimal input states and feedback for interferometric phase
  estimation}.
\bjtitle{Physical Review A}
\bvolume{63}(\bissue{5}),
\bfpage{053804}
(\byear{2001})
\end{barticle}
\endbibitem

\bibitem{berry2009perform}
\begin{barticle}
\bauthor{\bsnm{Berry}, \binits{D.W.}},
\bauthor{\bsnm{Higgins}, \binits{B.L.}},
\bauthor{\bsnm{Bartlett}, \binits{S.D.}},
\bauthor{\bsnm{Mitchell}, \binits{M.W.}},
\bauthor{\bsnm{Pryde}, \binits{G.J.}},
\bauthor{\bsnm{Wiseman}, \binits{H.M.}}:
\batitle{How to perform the most accurate possible phase measurements}.
\bjtitle{Physical Review A}
\bvolume{80}(\bissue{5}),
\bfpage{052114}
(\byear{2009})
\end{barticle}
\endbibitem

\bibitem{Wiseman1995}
\begin{barticle}
\bauthor{\bsnm{Wiseman}, \binits{H.M.}}:
\batitle{Adaptive phase measurements of optical modes: Going beyond the
  marginal $q$ distribution}.
\bjtitle{Phys. Rev. Lett.}
\bvolume{75},
\bfpage{4587}--\blpage{4590}
(\byear{1995}).
doi:\doiurl{10.1103/PhysRevLett.75.4587}
\end{barticle}
\endbibitem

\bibitem{Mabuchi2002}
\begin{barticle}
\bauthor{\bsnm{Armen}, \binits{M.A.}},
\bauthor{\bsnm{Au}, \binits{J.K.}},
\bauthor{\bsnm{Stockton}, \binits{J.K.}},
\bauthor{\bsnm{Doherty}, \binits{A.C.}},
\bauthor{\bsnm{Mabuchi}, \binits{H.}}:
\batitle{Adaptive homodyne measurement of optical phase}.
\bjtitle{Phys. Rev. Lett.}
\bvolume{89},
\bfpage{133602}
(\byear{2002}).
doi:\doiurl{10.1103/PhysRevLett.89.133602}
\end{barticle}
\endbibitem

\bibitem{Fujiwara_2006}
\begin{barticle}
\bauthor{\bsnm{Fujiwara}, \binits{A.}}:
\batitle{Strong consistency and asymptotic efficiency for adaptive quantum
  estimation problems}.
\bjtitle{Journal of Physics A: Mathematical and General}
\bvolume{39}(\bissue{40}),
\bfpage{12489}--\blpage{12504}
(\byear{2006}).
doi:\doiurl{10.1088/0305-4470/39/40/014}
\end{barticle}
\endbibitem

\bibitem{Takeuchi2012}
\begin{barticle}
\bauthor{\bsnm{Okamoto}, \binits{R.}},
\bauthor{\bsnm{Iefuji}, \binits{M.}},
\bauthor{\bsnm{Oyama}, \binits{S.}},
\bauthor{\bsnm{Yamagata}, \binits{K.}},
\bauthor{\bsnm{Imai}, \binits{H.}},
\bauthor{\bsnm{Fujiwara}, \binits{A.}},
\bauthor{\bsnm{Takeuchi}, \binits{S.}}:
\batitle{Experimental demonstration of adaptive quantum state estimation}.
\bjtitle{Phys. Rev. Lett.}
\bvolume{109},
\bfpage{130404}
(\byear{2012}).
doi:\doiurl{10.1103/PhysRevLett.109.130404}
\end{barticle}
\endbibitem

\bibitem{Paris2010}
\begin{barticle}
\bauthor{\bsnm{Brivio}, \binits{D.}},
\bauthor{\bsnm{Cialdi}, \binits{S.}},
\bauthor{\bsnm{Vezzoli}, \binits{S.}},
\bauthor{\bsnm{Gebrehiwot}, \binits{B.T.}},
\bauthor{\bsnm{Genoni}, \binits{M.G.}},
\bauthor{\bsnm{Olivares}, \binits{S.}},
\bauthor{\bsnm{Paris}, \binits{M.G.A.}}:
\batitle{Experimental estimation of one-parameter qubit gates in the presence
  of phase diffusion}.
\bjtitle{Phys. Rev. A}
\bvolume{81},
\bfpage{012305}
(\byear{2010}).
doi:\doiurl{10.1103/PhysRevA.81.012305}
\end{barticle}
\endbibitem

\bibitem{PhysRevLett.76.3228}
\begin{barticle}
\bauthor{\bsnm{Griffiths}, \binits{R.B.}},
\bauthor{\bsnm{Niu}, \binits{C.-S.}}:
\batitle{Semiclassical fourier transform for quantum computation}.
\bjtitle{Phys. Rev. Lett.}
\bvolume{76},
\bfpage{3228}--\blpage{3231}
(\byear{1996}).
doi:\doiurl{10.1103/PhysRevLett.76.3228}
\end{barticle}
\endbibitem

\bibitem{higgins_2007}
\begin{barticle}
\bauthor{\bsnm{Higgins}, \binits{B.L.}},
\bauthor{\bsnm{Berry}, \binits{D.W.}},
\bauthor{\bsnm{Bartlett}, \binits{S.D.}},
\bauthor{\bsnm{Wiseman}, \binits{H.M.}},
\bauthor{\bsnm{Pryde}, \binits{G.J.}}:
\batitle{Entanglement-free heisenberg-limited phase estimation}.
\bjtitle{Nature}
\bvolume{450}(\bissue{7168}),
\bfpage{393}--\blpage{396}
(\byear{2007}).
doi:\doiurl{10.1038/nature06257}
\end{barticle}
\endbibitem

\bibitem{danilin2018quantum}
\begin{barticle}
\bauthor{\bsnm{Danilin}, \binits{S.}},
\bauthor{\bsnm{Lebedev}, \binits{A.}},
\bauthor{\bsnm{Veps{\"a}l{\"a}inen}, \binits{A.}},
\bauthor{\bsnm{Lesovik}, \binits{G.}},
\bauthor{\bsnm{Blatter}, \binits{G.}},
\bauthor{\bsnm{Paraoanu}, \binits{G.}}:
\batitle{Quantum-enhanced magnetometry by phase estimation algorithms with a
  single artificial atom}.
\bjtitle{npj Quantum Information}
\bvolume{4}(\bissue{1}),
\bfpage{29}
(\byear{2018})
\end{barticle}
\endbibitem

\bibitem{Hanson2016}
\begin{barticle}
\bauthor{\bsnm{Bonato}, \binits{C.}},
\bauthor{\bsnm{Blok}, \binits{M.S.}},
\bauthor{\bsnm{Dinani}, \binits{H.T.}},
\bauthor{\bsnm{Berry}, \binits{D.W.}},
\bauthor{\bsnm{Markham}, \binits{M.L.}},
\bauthor{\bsnm{Twitchen}, \binits{D.J.}},
\bauthor{\bsnm{R.}, \binits{H.}}:
\batitle{Optimized quantum sensing with a single electronspin using real-time
  adaptive measurements}.
\bjtitle{Nature Nanotechnology}
\bvolume{11},
\bfpage{247}--\blpage{252}
(\byear{2016}).
doi:\doiurl{10.1038/nnano.2015.26}
\end{barticle}
\endbibitem

\bibitem{Sanders2010}
\begin{barticle}
\bauthor{\bsnm{Hentschel}, \binits{A.}},
\bauthor{\bsnm{Sanders}, \binits{B.C.}}:
\batitle{Machine learning for precise quantum measurement}.
\bjtitle{Phys. Rev. Lett.}
\bvolume{104},
\bfpage{063603}
(\byear{2010}).
doi:\doiurl{10.1103/PhysRevLett.104.063603}
\end{barticle}
\endbibitem

\bibitem{Sanders2011}
\begin{barticle}
\bauthor{\bsnm{Hentschel}, \binits{A.}},
\bauthor{\bsnm{Sanders}, \binits{B.C.}}:
\batitle{Efficient algorithm for optimizing adaptive quantum metrology
  processes}.
\bjtitle{Phys. Rev. Lett.}
\bvolume{107},
\bfpage{233601}
(\byear{2011}).
doi:\doiurl{10.1103/PhysRevLett.107.233601}
\end{barticle}
\endbibitem

\bibitem{Lovett2013}
\begin{barticle}
\bauthor{\bsnm{Lovett}, \binits{N.B.}},
\bauthor{\bsnm{Crosnier}, \binits{C.}},
\bauthor{\bsnm{Perarnau-Llobet}, \binits{M.}},
\bauthor{\bsnm{Sanders}, \binits{B.C.}}:
\batitle{Differential evolution for many-particle adaptive quantum metrology}.
\bjtitle{Physical Review Letters}
\bvolume{110}(\bissue{1}),
\bfpage{220501}
(\byear{2013})
\end{barticle}
\endbibitem

\bibitem{palittpongarnpim2016single}
\begin{bchapter}
\bauthor{\bsnm{Palittpongarnpim}, \binits{P.}},
\bauthor{\bsnm{Wittek}, \binits{P.}},
\bauthor{\bsnm{Sanders}, \binits{B.C.}}:
\bctitle{Single-shot adaptive measurement for quantum-enhanced metrology}.
In: \bbtitle{Quantum Communications and Quantum Imaging XIV},
vol. \bseriesno{9980},
p. \bfpage{99800}
(\byear{2016}).
\bcomment{International Society for Optics and Photonics}
\end{bchapter}
\endbibitem

\bibitem{palittapongarnpim2016controlling}
\begin{bchapter}
\bauthor{\bsnm{Palittapongarnpim}, \binits{P.}},
\bauthor{\bsnm{Wittek}, \binits{P.}},
\bauthor{\bsnm{Sanders}, \binits{B.C.}}:
\bctitle{Controlling adaptive quantum phase estimation with scalable
  reinforcement learning}.
In: \bbtitle{24th European Symposium on Artificial Neural Networks, Bruges,
  April 27--29, 2016},
pp. \bfpage{327}--\blpage{332}
(\byear{2016})
\end{bchapter}
\endbibitem

\bibitem{palittapongarnpim2017learning}
\begin{barticle}
\bauthor{\bsnm{Palittapongarnpim}, \binits{P.}},
\bauthor{\bsnm{Wittek}, \binits{P.}},
\bauthor{\bsnm{Zahedinejad}, \binits{E.}},
\bauthor{\bsnm{Vedaie}, \binits{S.}},
\bauthor{\bsnm{Sanders}, \binits{B.C.}}:
\batitle{Learning in quantum control: High-dimensional global optimization for
  noisy quantum dynamics}.
\bjtitle{Neurocomputing}
\bvolume{268},
\bfpage{116}--\blpage{126}
(\byear{2017})
\end{barticle}
\endbibitem

\bibitem{palittapongarnpim2018robustness}
\begin{botherref}
\oauthor{\bsnm{Palittapongarnpim}, \binits{P.}},
\oauthor{\bsnm{Sanders}, \binits{B.C.}}:
Robustness of adaptive quantum-enhanced phase estimation.
arXiv preprint arXiv:1809.05525
(2018)
\end{botherref}
\endbibitem

\bibitem{Lumino2018}
\begin{barticle}
\bauthor{\bsnm{Lumino}, \binits{A.}},
\bauthor{\bsnm{Polino}, \binits{E.}},
\bauthor{\bsnm{Rab}, \binits{A.S.}},
\bauthor{\bsnm{Milani}, \binits{G.}},
\bauthor{\bsnm{Spagnolo}, \binits{N.}},
\bauthor{\bsnm{Wiebe}, \binits{N.}},
\bauthor{\bsnm{Sciarrino}, \binits{F.}}:
\batitle{Experimental phase estimation enhanced by machine learning}.
\bjtitle{Physical Review Applied}
\bvolume{10}(\bissue{1}),
\bfpage{044033}
(\byear{2018})
\end{barticle}
\endbibitem

\bibitem{Wossnig2020}
\begin{barticle}
\bauthor{\bsnm{Ciliberto}, \binits{C.}},
\bauthor{\bsnm{Rocchetto}, \binits{A.}},
\bauthor{\bsnm{Rudi}, \binits{A.}},
\bauthor{\bsnm{Wossnig}, \binits{L.}}:
\batitle{Statistical limits of supervised quantum learning}.
\bjtitle{Phys. Rev. A}
\bvolume{102},
\bfpage{042414}
(\byear{2020}).
doi:\doiurl{10.1103/PhysRevA.102.042414}
\end{barticle}
\endbibitem

\bibitem{NielsenChuang}
\begin{bbook}
\bauthor{\bsnm{Nielsen}, \binits{M.A.}},
\bauthor{\bsnm{Chuang}, \binits{I.}}:
\bbtitle{Quantum Computation and Quantum Information}.
\bpublisher{Cambridge University Press},
\blocation{Cambridge, U.K.}
(\byear{2000}).
doi:\doiurl{10.1007/978-1-4614-1800-9\_149}.
\burl{https://doi.org/10.1007/978-1-4614-1800-9\_149}
\end{bbook}
\endbibitem

\bibitem{storn1997differential}
\begin{barticle}
\bauthor{\bsnm{Storn}, \binits{R.}},
\bauthor{\bsnm{Price}, \binits{K.}}:
\batitle{Differential evolution--a simple and efficient heuristic for global
  optimization over continuous spaces}.
\bjtitle{Journal of global optimization}
\bvolume{11}(\bissue{4}),
\bfpage{341}--\blpage{359}
(\byear{1997})
\end{barticle}
\endbibitem

\bibitem{price2006differential}
\begin{bbook}
\bauthor{\bsnm{Price}, \binits{K.}},
\bauthor{\bsnm{Storn}, \binits{R.M.}},
\bauthor{\bsnm{Lampinen}, \binits{J.A.}}:
\bbtitle{Differential Evolution: a Practical Approach to Global Optimization}.
\bpublisher{Springer}, \blocation{???}
(\byear{2006})
\end{bbook}
\endbibitem

\bibitem{kennedy2011particle}
\begin{bchapter}
\bauthor{\bsnm{Kennedy}, \binits{J.}}:
\bctitle{Particle swarm optimization}.
In: \bbtitle{Encyclopedia of Machine Learning},
pp. \bfpage{760}--\blpage{766}.
\bpublisher{Springer}, \blocation{???}
(\byear{2011})
\end{bchapter}
\endbibitem

\bibitem{eberhart1995new}
\begin{bchapter}
\bauthor{\bsnm{Eberhart}, \binits{R.}},
\bauthor{\bsnm{Kennedy}, \binits{J.}}:
\bctitle{A new optimizer using particle swarm theory}.
In: \bbtitle{Micro Machine and Human Science, 1995. MHS'95., Proceedings of the
  Sixth International Symposium On},
pp. \bfpage{39}--\blpage{43}
(\byear{1995}).
\bcomment{IEEE}
\end{bchapter}
\endbibitem

\bibitem{shi1998modified}
\begin{bchapter}
\bauthor{\bsnm{Shi}, \binits{Y.}},
\bauthor{\bsnm{Eberhart}, \binits{R.}}:
\bctitle{A modified particle swarm optimizer}.
In: \bbtitle{Evolutionary Computation Proceedings, 1998. IEEE World Congress on
  Computational Intelligence., The 1998 IEEE International Conference On},
pp. \bfpage{69}--\blpage{73}
(\byear{1998}).
\bcomment{IEEE}
\end{bchapter}
\endbibitem

\bibitem{PhysRevA.94.022334}
\begin{barticle}
\bauthor{\bsnm{Chapeau-Blondeau}, \binits{F.}}:
\batitle{Optimizing qubit phase estimation}.
\bjtitle{Phys. Rev. A}
\bvolume{94},
\bfpage{022334}
(\byear{2016}).
doi:\doiurl{10.1103/PhysRevA.94.022334}
\end{barticle}
\endbibitem

\bibitem{Garcia2020}
\begin{botherref}
\oauthor{\bsnm{RodrÃ­guez-GarcÃ­a}, \binits{M.A.}},
\oauthor{\bsnm{1}, \binits{I.} \bsuffix{PÃ©rez~Castillo}},
\oauthor{\bsnm{Barberis-Blostein}, \binits{P.}}:
Efficient qubit phase estimation using adaptive measurements.
arXiv preprint arXiv:2012.11088
(2020)
\end{botherref}
\endbibitem

\bibitem{sekatski2017quantum}
\begin{barticle}
\bauthor{\bsnm{Sekatski}, \binits{P.}},
\bauthor{\bsnm{Skotiniotis}, \binits{M.}},
\bauthor{\bsnm{Ko{\l}ody{\'n}ski}, \binits{J.}},
\bauthor{\bsnm{D{\"u}r}, \binits{W.}}:
\batitle{Quantum metrology with full and fast quantum control}.
\bjtitle{Quantum}
\bvolume{1},
\bfpage{27}
(\byear{2017})
\end{barticle}
\endbibitem

\bibitem{Paraoanu_2011}
\begin{barticle}
\bauthor{\bsnm{Paraoanu}, \binits{G.S.}}:
\batitle{Generalized partial measurements}.
\bjtitle{{EPL} (Europhysics Letters)}
\bvolume{93}(\bissue{6}),
\bfpage{64002}
(\byear{2011}).
doi:\doiurl{10.1209/0295-5075/93/64002}
\end{barticle}
\endbibitem

\bibitem{Higgins2007}
\begin{barticle}
\bauthor{\bsnm{Higgins}, \binits{B.L.}},
\bauthor{\bsnm{Berry}, \binits{D.W.}},
\bauthor{\bsnm{Bartlett}, \binits{S.D.}},
\bauthor{\bsnm{Wiseman}, \binits{H.M.}},
\bauthor{\bsnm{Pryde}, \binits{G.J.}}:
\batitle{Adaptive single-shot phase measurements: The full quantum theory}.
\bjtitle{Nature}
\bvolume{450},
\bfpage{393}--\blpage{396}
(\byear{2007})
\end{barticle}
\endbibitem

\bibitem{Pryde2018}
\begin{barticle}
\bauthor{\bsnm{Daryanoosh}, \binits{S.}},
\bauthor{\bsnm{Slussarenko}, \binits{S.}},
\bauthor{\bsnm{Berry}, \binits{D.W.}},
\bauthor{\bsnm{Wiseman}, \binits{H.M.}},
\bauthor{\bsnm{Pryde}, \binits{G.J.}}:
\batitle{Experimental optical phase measurement approaching the exact
  heisenberg limit}.
\bjtitle{Nature Communications}
\bvolume{9},
\bfpage{4606}
(\byear{2018})
\end{barticle}
\endbibitem

\bibitem{paraoanu2006}
\begin{barticle}
\bauthor{\bsnm{Paraoanu}, \binits{G.S.}}:
\batitle{Interaction-free measurements with superconducting qubits}.
\bjtitle{Phys. Rev. Lett.}
\bvolume{97},
\bfpage{180406}
(\byear{2006}).
doi:\doiurl{10.1103/PhysRevLett.97.180406}
\end{barticle}
\endbibitem

\bibitem{danilin_2018}
\begin{botherref}
\oauthor{\bsnm{Danilin}, \binits{S.}},
\oauthor{\bsnm{Lebedev}, \binits{A.V.}},
\oauthor{\bsnm{VepsÃ¤lÃ¤inen}, \binits{A.}},
\oauthor{\bsnm{Lesovik}, \binits{G.B.}},
\oauthor{\bsnm{Blatter}, \binits{G.}},
\oauthor{\bsnm{Paraoanu}, \binits{G.S.}}:
Quantum-enhanced magnetometry by phase estimation algorithms with a single
  artificial atom.
npj Quantum Information
\textbf{4}(1)
(2018).
doi:\doiurl{10.1038/s41534-018-0078-y}
\end{botherref}
\endbibitem

\bibitem{Silveri2017}
\begin{barticle}
\bauthor{\bsnm{Silveri}, \binits{M.P.}},
\bauthor{\bsnm{Tuorila}, \binits{J.A.}},
\bauthor{\bsnm{Thuneberg}, \binits{E.V.}},
\bauthor{\bsnm{Paraoanu}, \binits{G.S.}}:
\batitle{Quantum systems under frequency modulation}.
\bjtitle{Rep. Prog. Phys.}
\bvolume{80},
\bfpage{056002}
(\byear{2017}).
doi:\doiurl{10.1088/1361-6633/aa5170}
\end{barticle}
\endbibitem

\bibitem{shlyakhov_2018}
\begin{botherref}
\oauthor{\bsnm{Shlyakhov}, \binits{A.R.}},
\oauthor{\bsnm{Zemlyanov}, \binits{V.V.}},
\oauthor{\bsnm{Suslov}, \binits{M.V.}},
\oauthor{\bsnm{Lebedev}, \binits{A.V.}},
\oauthor{\bsnm{Paraoanu}, \binits{G.S.}},
\oauthor{\bsnm{Lesovik}, \binits{G.B.}},
\oauthor{\bsnm{Blatter}, \binits{G.}}:
Quantum metrology with a transmon qutrit.
Physical Review A
\textbf{97}(2)
(2018).
doi:\doiurl{10.1103/physreva.97.022115}
\end{botherref}
\endbibitem

\bibitem{Danilin2021}
\begin{botherref}
\oauthor{\bsnm{Danilin~S.}, \binits{W.M.}}:
Quantum sensing with superconducting circuits.
arXiv:2103.11022
(2021)
\end{botherref}
\endbibitem

\bibitem{Perelshein2021}
\begin{barticle}
\bauthor{\bsnm{Perelshtein}, \binits{M.R.}},
\bauthor{\bsnm{Kirsanov}, \binits{N.S.}},
\bauthor{\bsnm{Zemlyanov}, \binits{V.V.}},
\bauthor{\bsnm{Lebedev}, \binits{A.V.}},
\bauthor{\bsnm{Blatter}, \binits{G.}},
\bauthor{\bsnm{Vinokur}, \binits{V.M.}},
\bauthor{\bsnm{Lesovik}, \binits{G.B.}}:
\batitle{Linear ascending metrological algorithm}.
\bjtitle{Phys. Rev. Research}
\bvolume{3},
\bfpage{013257}
(\byear{2021}).
doi:\doiurl{10.1103/PhysRevResearch.3.013257}
\end{barticle}
\endbibitem

\bibitem{PhysRevLett.116.240801}
\begin{barticle}
\bauthor{\bsnm{Baumgart}, \binits{I.}},
\bauthor{\bsnm{Cai}, \binits{J.-M.}},
\bauthor{\bsnm{Retzker}, \binits{A.}},
\bauthor{\bsnm{Plenio}, \binits{M.B.}},
\bauthor{\bsnm{Wunderlich}, \binits{C.}}:
\batitle{Ultrasensitive magnetometer using a single atom}.
\bjtitle{Phys. Rev. Lett.}
\bvolume{116},
\bfpage{240801}
(\byear{2016}).
doi:\doiurl{10.1103/PhysRevLett.116.240801}
\end{barticle}
\endbibitem

\bibitem{Wunderlich2011}
\begin{barticle}
\bauthor{\bsnm{Timoney}, \binits{N.}},
\bauthor{\bsnm{Baumgart}, \binits{I.}},
\bauthor{\bsnm{Johanning}, \binits{M.}},
\bauthor{\bsnm{Var{\`o}n}, \binits{A.F.}},
\bauthor{\bsnm{Plenio}, \binits{M.B.}},
\bauthor{\bsnm{A.}, \binits{R.}},
\bauthor{\bsnm{Ch.}, \binits{W.}}:
\batitle{Quantum gates and memory using microwave-dressed states}.
\bjtitle{Nature}
\bvolume{476},
\bfpage{185}--\blpage{188}
(\byear{2011}).
doi:\doiurl{10.1038/nature10319}
\end{barticle}
\endbibitem

\bibitem{Lukin2008}
\begin{barticle}
\bauthor{\bsnm{Taylor}, \binits{J.M.}},
\bauthor{\bsnm{Cappellaro}, \binits{P.}},
\bauthor{\bsnm{Childress}, \binits{L.}},
\bauthor{\bsnm{Jiang}, \binits{L.}},
\bauthor{\bsnm{Budker}, \binits{D.}},
\bauthor{\bsnm{Hemmer}, \binits{P.R.}},
\bauthor{\bsnm{Yacoby}, \binits{A.}},
\bauthor{\bsnm{Walsworth}, \binits{R.}},
\bauthor{\bsnm{Lukin}, \binits{M.D.}}:
\batitle{High-sensitivity diamond magnetometer with nanoscale resolution}.
\bjtitle{Nature Physics}
\bvolume{4},
\bfpage{810}--\blpage{816}
(\byear{2008})
\end{barticle}
\endbibitem

\bibitem{Walsworth2020}
\begin{barticle}
\bauthor{\bsnm{Barry}, \binits{J.F.}},
\bauthor{\bsnm{Schloss}, \binits{J.M.}},
\bauthor{\bsnm{Bauch}, \binits{E.}},
\bauthor{\bsnm{Turner}, \binits{M.J.}},
\bauthor{\bsnm{Hart}, \binits{C.A.}},
\bauthor{\bsnm{Pham}, \binits{L.M.}},
\bauthor{\bsnm{Walsworth}, \binits{R.L.}}:
\batitle{Sensitivity optimization for nv-diamond magnetometry}.
\bjtitle{Rev. Mod. Phys.}
\bvolume{92},
\bfpage{015004}
(\byear{2020}).
doi:\doiurl{10.1103/RevModPhys.92.015004}
\end{barticle}
\endbibitem

\end{thebibliography}

\newcommand{\BMCxmlcomment}[1]{}

\BMCxmlcomment{

<refgrp>

<bibl id="B1">
  <title><p>Quantum-enhanced measurements: beating the standard quantum
  limit</p></title>
  <aug>
    <au><snm>Giovannetti</snm><fnm>V</fnm></au>
    <au><snm>Lloyd</snm><fnm>S</fnm></au>
    <au><snm>Maccone</snm><fnm>L</fnm></au>
  </aug>
  <source>Science</source>
  <publisher>American Association for the Advancement of Science</publisher>
  <pubdate>2004</pubdate>
  <volume>306</volume>
  <issue>5700</issue>
  <fpage>1330</fpage>
  <lpage>-1336</lpage>
</bibl>

<bibl id="B2">
  <title><p>Quantum metrology</p></title>
  <aug>
    <au><snm>Giovannetti</snm><fnm>V</fnm></au>
    <au><snm>Lloyd</snm><fnm>S</fnm></au>
    <au><snm>Maccone</snm><fnm>L</fnm></au>
  </aug>
  <source>Physical review letters</source>
  <publisher>APS</publisher>
  <pubdate>2006</pubdate>
  <volume>96</volume>
  <issue>1</issue>
  <fpage>010401</fpage>
</bibl>

<bibl id="B3">
  <title><p>Advances in quantum metrology</p></title>
  <aug>
    <au><snm>Giovannetti</snm><fnm>V</fnm></au>
    <au><snm>Lloyd</snm><fnm>S</fnm></au>
    <au><snm>Maccone</snm><fnm>L</fnm></au>
  </aug>
  <source>Nature photonics</source>
  <publisher>Nature Publishing Group</publisher>
  <pubdate>2011</pubdate>
  <volume>5</volume>
  <issue>4</issue>
  <fpage>222</fpage>
</bibl>

<bibl id="B4">
  <title><p>Quantum metrology from a quantum information science
  perspective</p></title>
  <aug>
    <au><snm>T{\'o}th</snm><fnm>G</fnm></au>
    <au><snm>Apellaniz</snm><fnm>I</fnm></au>
  </aug>
  <source>Journal of Physics A: Mathematical and Theoretical</source>
  <publisher>IOP Publishing</publisher>
  <pubdate>2014</pubdate>
  <volume>47</volume>
  <issue>42</issue>
  <fpage>424006</fpage>
</bibl>

<bibl id="B5">
  <title><p>Quantum detection and estimation theory</p></title>
  <aug>
    <au><snm>Helstrom</snm><fnm>CW</fnm></au>
  </aug>
  <source>Journal of Statistical Physics</source>
  <publisher>Springer</publisher>
  <pubdate>1969</pubdate>
  <volume>1</volume>
  <issue>2</issue>
  <fpage>231</fpage>
  <lpage>-252</lpage>
</bibl>

<bibl id="B6">
  <title><p>Probabilistic and statistical aspects of quantum theory</p></title>
  <aug>
    <au><snm>Holevo</snm><fnm>AS</fnm></au>
  </aug>
  <publisher>Springer Science \& Business Media</publisher>
  <pubdate>2011</pubdate>
  <volume>1</volume>
</bibl>

<bibl id="B7">
  <title><p>Mathematical methods of statistics (PMS-9)</p></title>
  <aug>
    <au><snm>Cram{\'e}r</snm><fnm>H</fnm></au>
  </aug>
  <publisher>Princeton university press</publisher>
  <pubdate>2016</pubdate>
  <volume>9</volume>
</bibl>

<bibl id="B8">
  <title><p>Quantum-mechanical limitations in macroscopic experiments and
  modern experimental technique</p></title>
  <aug>
    <au><snm>Braginski{\u{\i}}</snm><fnm>VB</fnm></au>
    <au><snm>Vorontsov</snm><fnm>YI</fnm></au>
  </aug>
  <source>Soviet Physics Uspekhi</source>
  <publisher>{IOP} Publishing</publisher>
  <pubdate>1975</pubdate>
  <volume>17</volume>
  <issue>5</issue>
  <fpage>644</fpage>
  <lpage>-650</lpage>
  <url>https://doi.org/10.1070
</bibl>

<bibl id="B9">
  <title><p>Quantum nondemolition measurements</p></title>
  <aug>
    <au><snm>Braginsky</snm><fnm>VB</fnm></au>
    <au><snm>Vorontsov</snm><fnm>YI</fnm></au>
    <au><snm>Thorne</snm><fnm>KS</fnm></au>
  </aug>
  <source>Science</source>
  <publisher>American Association for the Advancement of Science</publisher>
  <pubdate>1980</pubdate>
  <volume>209</volume>
  <issue>4456</issue>
  <fpage>547</fpage>
  <lpage>-557</lpage>
</bibl>

<bibl id="B10">
  <title><p>Quantum measurement</p></title>
  <aug>
    <au><snm>Braginsky</snm><fnm>VB</fnm></au>
    <au><snm>Braginsky</snm><fnm>VB</fnm></au>
    <au><snm>Khalili</snm><fnm>FY</fnm></au>
  </aug>
  <publisher>Cambridge University Press</publisher>
  <pubdate>1995</pubdate>
</bibl>

<bibl id="B11">
  <title><p>Realization of measurement and the standard quantum
  limit</p></title>
  <aug>
    <au><snm>Ozawa</snm><fnm>M</fnm></au>
  </aug>
  <source>Squeezed and Nonclassical Light</source>
  <publisher>Springer</publisher>
  <pubdate>1989</pubdate>
  <fpage>263</fpage>
  <lpage>-286</lpage>
</bibl>

<bibl id="B12">
  <title><p>Natural and artificial atoms for quantum computation</p></title>
  <aug>
    <au><snm>Buluta</snm><fnm>I</fnm></au>
    <au><snm>Ashhab</snm><fnm>S</fnm></au>
    <au><snm>Nori</snm><fnm>F</fnm></au>
  </aug>
  <source>Rep. Prog. Phys.</source>
  <publisher>{IOP} Publishing</publisher>
  <pubdate>2011</pubdate>
  <volume>74</volume>
  <issue>10</issue>
  <fpage>104401</fpage>
  <url>https://doi.org/10.1088
</bibl>

<bibl id="B13">
  <title><p>Photonic quantum information processing: A concise
  review</p></title>
  <aug>
    <au><snm>Slussarenko</snm><fnm>S</fnm></au>
    <au><snm>Pryde</snm><fnm>GJ</fnm></au>
  </aug>
  <source>Applied Physics Reviews</source>
  <pubdate>2019</pubdate>
  <volume>6</volume>
  <issue>4</issue>
  <fpage>041303</fpage>
  <url>https://doi.org/10.1063/1.5115814</url>
</bibl>

<bibl id="B14">
  <title><p>Photonic quantum information processing: a review</p></title>
  <aug>
    <au><snm>Flamini</snm><fnm>F</fnm></au>
    <au><snm>Spagnolo</snm><fnm>N</fnm></au>
    <au><snm>Sciarrino</snm><fnm>F</fnm></au>
  </aug>
  <source>Reports on Progress in Physics</source>
  <publisher>{IOP} Publishing</publisher>
  <pubdate>2018</pubdate>
  <volume>82</volume>
  <issue>1</issue>
  <fpage>016001</fpage>
  <url>https://doi.org/10.1088/1361-6633/aad5b2</url>
</bibl>

<bibl id="B15">
  <title><p>Processing quantum information in diamond</p></title>
  <aug>
    <au><snm>Wrachtrup</snm><fnm>J</fnm></au>
    <au><snm>Jelezko</snm><fnm>F</fnm></au>
  </aug>
  <source>Journal of Physics: Condensed Matter</source>
  <publisher>{IOP} Publishing</publisher>
  <pubdate>2006</pubdate>
  <volume>18</volume>
  <issue>21</issue>
  <fpage>S807</fpage>
  <lpage>-S824</lpage>
  <url>https://doi.org/10.1088
</bibl>

<bibl id="B16">
  <title><p>Quantum Information Processing with Diamond: Principles and
  Applications</p></title>
  <aug>
    <au><snm>Prawer</snm><fnm>S</fnm></au>
    <au><snm>Aharonovich</snm><fnm>I</fnm></au>
  </aug>
  <publisher>Woodhead Publishing, Limited</publisher>
  <edition>1</edition>
  <pubdate>2018</pubdate>
</bibl>

<bibl id="B17">
  <title><p>Quantum Computing with Trapped Ions</p></title>
  <aug>
    <au><snm>Lange</snm><fnm>W</fnm></au>
  </aug>
  <source>Computational Complexity: Theory, Techniques, and
  Applications</source>
  <publisher>New York, NY: Springer New York</publisher>
  <editor>Meyers, Robert A.</editor>
  <pubdate>2012</pubdate>
  <fpage>2406</fpage>
  <lpage>-2436</lpage>
  <url>https://doi.org/10.1007/978-1-4614-1800-9\_149</url>
</bibl>

<bibl id="B18">
  <title><p>Trapped-ion quantum computing: Progress and challenges</p></title>
  <aug>
    <au><snm>Bruzewicz</snm><fnm>CD</fnm></au>
    <au><snm>Chiaverini</snm><fnm>J</fnm></au>
    <au><snm>McConnell</snm><fnm>R</fnm></au>
    <au><snm>Sage</snm><fnm>JM</fnm></au>
  </aug>
  <source>Applied Physics Reviews</source>
  <pubdate>2019</pubdate>
  <volume>6</volume>
  <issue>2</issue>
  <fpage>021314</fpage>
  <url>https://doi.org/10.1063/1.5088164</url>
</bibl>

<bibl id="B19">
  <title><p>Recent Progress in Quantum Simulation Using Superconducting
  Circuits</p></title>
  <aug>
    <au><snm>Paraoanu</snm><fnm>G. S.</fnm></au>
  </aug>
  <source>J. Low Temp. Phys.</source>
  <pubdate>2014</pubdate>
  <volume>175</volume>
  <issue>5</issue>
  <fpage>633</fpage>
  <lpage>-654</lpage>
  <url>https://doi.org/10.1007/s10909-014-1175-8</url>
</bibl>

<bibl id="B20">
  <title><p>Atomic physics and quantum optics using superconducting
  circuits</p></title>
  <aug>
    <au><snm>You</snm><fnm>J. Q.</fnm></au>
    <au><snm>Nori</snm><fnm>F</fnm></au>
  </aug>
  <source>Nature</source>
  <publisher>Macmillan Publishers Limited</publisher>
  <pubdate>2011</pubdate>
  <volume>474</volume>
  <fpage>589</fpage>
  <url>https://www.nature.com/articles/nature10122</url>
</bibl>

<bibl id="B21">
  <title><p>Quantum sensing</p></title>
  <aug>
    <au><snm>Degen</snm><fnm>C.âL.</fnm></au>
    <au><snm>Reinhard</snm><fnm>F.</fnm></au>
    <au><snm>Cappellaro</snm><fnm>P.</fnm></au>
  </aug>
  <source>Reviews of Modern Physics</source>
  <publisher>American Physical Society (APS)</publisher>
  <pubdate>2017</pubdate>
  <volume>89</volume>
  <issue>3</issue>
  <url>http://dx.doi.org/10.1103/RevModPhys.89.035002</url>
</bibl>

<bibl id="B22">
  <title><p>Quantum estimation for quantum technology</p></title>
  <aug>
    <au><snm>Paris</snm><fnm>MG</fnm></au>
  </aug>
  <source>International Journal of Quantum Information</source>
  <publisher>World Scientific</publisher>
  <pubdate>2009</pubdate>
  <volume>7</volume>
  <issue>supp01</issue>
  <fpage>125</fpage>
  <lpage>-137</lpage>
</bibl>

<bibl id="B23">
  <title><p>Ultrasensitive detection of force and displacement using trapped
  ions</p></title>
  <aug>
    <au><snm>Biercuk</snm><fnm>MJ</fnm></au>
    <au><snm>Uys</snm><fnm>H</fnm></au>
    <au><snm>Britton</snm><fnm>JW</fnm></au>
    <au><snm>VanDevender</snm><fnm>AP</fnm></au>
    <au><snm>Bollinger</snm><fnm>JJ</fnm></au>
  </aug>
  <source>Nature Nanotechnology</source>
  <publisher>Springer Science and Business Media LLC</publisher>
  <pubdate>2010</pubdate>
  <volume>5</volume>
  <issue>9</issue>
  <fpage>646â650</fpage>
  <url>http://dx.doi.org/10.1038/NNANO.2010.165</url>
</bibl>

<bibl id="B24">
  <title><p>Extending qubit coherence by adaptive quantum environment
  learning</p></title>
  <aug>
    <au><snm>Scerri</snm><fnm>E</fnm></au>
    <au><snm>Gauger</snm><fnm>EM</fnm></au>
    <au><snm>Bonato</snm><fnm>C</fnm></au>
  </aug>
  <source>New Journal of Physics</source>
  <publisher>{IOP} Publishing</publisher>
  <pubdate>2020</pubdate>
  <volume>22</volume>
  <issue>3</issue>
  <fpage>035002</fpage>
  <url>https://doi.org/10.1088
</bibl>

<bibl id="B25">
  <title><p>Ion-trap measurements of electric-field noise near
  surfaces</p></title>
  <aug>
    <au><snm>Brownnutt</snm><fnm>M.</fnm></au>
    <au><snm>Kumph</snm><fnm>M.</fnm></au>
    <au><snm>Rabl</snm><fnm>P.</fnm></au>
    <au><snm>Blatt</snm><fnm>R.</fnm></au>
  </aug>
  <source>Reviews of Modern Physics</source>
  <publisher>American Physical Society (APS)</publisher>
  <pubdate>2015</pubdate>
  <volume>87</volume>
  <issue>4</issue>
  <fpage>1419â1482</fpage>
  <url>http://dx.doi.org/10.1103/RevModPhys.87.1419</url>
</bibl>

<bibl id="B26">
  <title><p>Ultra-precision quantum sensing and measurement based on nonlinear
  hybrid optomechanical systems containing ultracold atoms or
  atomic-BE</p></title>
  <aug>
    <au><snm>Motazedifard</snm><fnm>A</fnm></au>
    <au><snm>Dalafi</snm><fnm>A.</fnm></au>
    <au><snm>Naderi</snm><fnm>M. H.</fnm></au>
  </aug>
  <source>arXiv:2011.01336</source>
  <pubdate>2020</pubdate>
</bibl>

<bibl id="B27">
  <title><p>A gravitational wave observatory operating beyond the quantum
  shot-noise limit</p></title>
  <aug>
    <au><snm>Abadie</snm><fnm>J</fnm></au>
    <au><snm>Abbott</snm><fnm>BP</fnm></au>
    <au><snm>Abbott</snm><fnm>R</fnm></au>
    <au><snm>Abbott</snm><fnm>TD</fnm></au>
    <au><snm>Abernathy</snm><fnm>M</fnm></au>
    <au><snm>Adams</snm><fnm>C</fnm></au>
    <au><snm>Adhikari</snm><fnm>R</fnm></au>
    <au><snm>Affeldt</snm><fnm>C</fnm></au>
    <au><snm>Allen</snm><fnm>B</fnm></au>
    <au><snm>Allen</snm><fnm>GS</fnm></au>
    <au><cnm>others</cnm></au>
  </aug>
  <source>Nature Physics</source>
  <publisher>Nature Publishing Group</publisher>
  <pubdate>2011</pubdate>
  <volume>7</volume>
  <issue>12</issue>
  <fpage>962</fpage>
</bibl>

<bibl id="B28">
  <title><p>Quantum metrology with nonclassical states of atomic
  ensembles</p></title>
  <aug>
    <au><snm>Pezz\`e</snm><fnm>L</fnm></au>
    <au><snm>Smerzi</snm><fnm>A</fnm></au>
    <au><snm>Oberthaler</snm><fnm>MK</fnm></au>
    <au><snm>Schmied</snm><fnm>R</fnm></au>
    <au><snm>Treutlein</snm><fnm>P</fnm></au>
  </aug>
  <source>Rev. Mod. Phys.</source>
  <publisher>American Physical Society</publisher>
  <pubdate>2018</pubdate>
  <volume>90</volume>
  <fpage>035005</fpage>
  <url>https://link.aps.org/doi/10.1103/RevModPhys.90.035005</url>
</bibl>

<bibl id="B29">
  <title><p>Optimal states and almost optimal adaptive measurements for quantum
  interferometry</p></title>
  <aug>
    <au><snm>Berry</snm><fnm>DW</fnm></au>
    <au><snm>Wiseman</snm><fnm>HM</fnm></au>
  </aug>
  <source>Physical Review Letters</source>
  <publisher>APS</publisher>
  <pubdate>2000</pubdate>
  <volume>85</volume>
  <issue>24</issue>
  <fpage>5098</fpage>
</bibl>

<bibl id="B30">
  <title><p>Optimal input states and feedback for interferometric phase
  estimation</p></title>
  <aug>
    <au><snm>Berry</snm><fnm>DW</fnm></au>
    <au><snm>Wiseman</snm><fnm>HM</fnm></au>
    <au><snm>Breslin</snm><fnm>JK</fnm></au>
  </aug>
  <source>Physical Review A</source>
  <publisher>APS</publisher>
  <pubdate>2001</pubdate>
  <volume>63</volume>
  <issue>5</issue>
  <fpage>053804</fpage>
</bibl>

<bibl id="B31">
  <title><p>How to perform the most accurate possible phase
  measurements</p></title>
  <aug>
    <au><snm>Berry</snm><fnm>DW</fnm></au>
    <au><snm>Higgins</snm><fnm>BL</fnm></au>
    <au><snm>Bartlett</snm><fnm>SD</fnm></au>
    <au><snm>Mitchell</snm><fnm>MW</fnm></au>
    <au><snm>Pryde</snm><fnm>GJ</fnm></au>
    <au><snm>Wiseman</snm><fnm>HM</fnm></au>
  </aug>
  <source>Physical Review A</source>
  <publisher>APS</publisher>
  <pubdate>2009</pubdate>
  <volume>80</volume>
  <issue>5</issue>
  <fpage>052114</fpage>
</bibl>

<bibl id="B32">
  <title><p>Adaptive Phase Measurements of Optical Modes: Going Beyond the
  Marginal $Q$ Distribution</p></title>
  <aug>
    <au><snm>Wiseman</snm><fnm>H. M.</fnm></au>
  </aug>
  <source>Phys. Rev. Lett.</source>
  <publisher>American Physical Society</publisher>
  <pubdate>1995</pubdate>
  <volume>75</volume>
  <fpage>4587</fpage>
  <lpage>-4590</lpage>
  <url>https://link.aps.org/doi/10.1103/PhysRevLett.75.4587</url>
</bibl>

<bibl id="B33">
  <title><p>Adaptive Homodyne Measurement of Optical Phase</p></title>
  <aug>
    <au><snm>Armen</snm><fnm>MA</fnm></au>
    <au><snm>Au</snm><fnm>JK</fnm></au>
    <au><snm>Stockton</snm><fnm>JK</fnm></au>
    <au><snm>Doherty</snm><fnm>AC</fnm></au>
    <au><snm>Mabuchi</snm><fnm>H</fnm></au>
  </aug>
  <source>Phys. Rev. Lett.</source>
  <publisher>American Physical Society</publisher>
  <pubdate>2002</pubdate>
  <volume>89</volume>
  <fpage>133602</fpage>
  <url>https://link.aps.org/doi/10.1103/PhysRevLett.89.133602</url>
</bibl>

<bibl id="B34">
  <title><p>Strong consistency and asymptotic efficiency for adaptive quantum
  estimation problems</p></title>
  <aug>
    <au><snm>Fujiwara</snm><fnm>A</fnm></au>
  </aug>
  <source>Journal of Physics A: Mathematical and General</source>
  <publisher>{IOP} Publishing</publisher>
  <pubdate>2006</pubdate>
  <volume>39</volume>
  <issue>40</issue>
  <fpage>12489</fpage>
  <lpage>-12504</lpage>
  <url>https://doi.org/10.1088/0305-4470/39/40/014</url>
</bibl>

<bibl id="B35">
  <title><p>Experimental Demonstration of Adaptive Quantum State
  Estimation</p></title>
  <aug>
    <au><snm>Okamoto</snm><fnm>R</fnm></au>
    <au><snm>Iefuji</snm><fnm>M</fnm></au>
    <au><snm>Oyama</snm><fnm>S</fnm></au>
    <au><snm>Yamagata</snm><fnm>K</fnm></au>
    <au><snm>Imai</snm><fnm>H</fnm></au>
    <au><snm>Fujiwara</snm><fnm>A</fnm></au>
    <au><snm>Takeuchi</snm><fnm>S</fnm></au>
  </aug>
  <source>Phys. Rev. Lett.</source>
  <publisher>American Physical Society</publisher>
  <pubdate>2012</pubdate>
  <volume>109</volume>
  <fpage>130404</fpage>
  <url>https://link.aps.org/doi/10.1103/PhysRevLett.109.130404</url>
</bibl>

<bibl id="B36">
  <title><p>Experimental estimation of one-parameter qubit gates in the
  presence of phase diffusion</p></title>
  <aug>
    <au><snm>Brivio</snm><fnm>D</fnm></au>
    <au><snm>Cialdi</snm><fnm>S</fnm></au>
    <au><snm>Vezzoli</snm><fnm>S</fnm></au>
    <au><snm>Gebrehiwot</snm><fnm>BT</fnm></au>
    <au><snm>Genoni</snm><fnm>MG</fnm></au>
    <au><snm>Olivares</snm><fnm>S</fnm></au>
    <au><snm>Paris</snm><fnm>MGA</fnm></au>
  </aug>
  <source>Phys. Rev. A</source>
  <publisher>American Physical Society</publisher>
  <pubdate>2010</pubdate>
  <volume>81</volume>
  <fpage>012305</fpage>
  <url>https://link.aps.org/doi/10.1103/PhysRevA.81.012305</url>
</bibl>

<bibl id="B37">
  <title><p>Semiclassical Fourier Transform for Quantum Computation</p></title>
  <aug>
    <au><snm>Griffiths</snm><fnm>RB</fnm></au>
    <au><snm>Niu</snm><fnm>CS</fnm></au>
  </aug>
  <source>Phys. Rev. Lett.</source>
  <publisher>American Physical Society</publisher>
  <pubdate>1996</pubdate>
  <volume>76</volume>
  <fpage>3228</fpage>
  <lpage>-3231</lpage>
  <url>https://link.aps.org/doi/10.1103/PhysRevLett.76.3228</url>
</bibl>

<bibl id="B38">
  <title><p>Entanglement-free Heisenberg-limited phase estimation</p></title>
  <aug>
    <au><snm>Higgins</snm><fnm>B. L.</fnm></au>
    <au><snm>Berry</snm><fnm>D. W.</fnm></au>
    <au><snm>Bartlett</snm><fnm>S. D.</fnm></au>
    <au><snm>Wiseman</snm><fnm>H. M.</fnm></au>
    <au><snm>Pryde</snm><fnm>G. J.</fnm></au>
  </aug>
  <source>Nature</source>
  <publisher>Springer Science and Business Media LLC</publisher>
  <pubdate>2007</pubdate>
  <volume>450</volume>
  <issue>7168</issue>
  <fpage>393â396</fpage>
  <url>http://dx.doi.org/10.1038/nature06257</url>
</bibl>

<bibl id="B39">
  <title><p>Quantum-enhanced magnetometry by phase estimation algorithms with a
  single artificial atom</p></title>
  <aug>
    <au><snm>Danilin</snm><fnm>S</fnm></au>
    <au><snm>Lebedev</snm><fnm>AV</fnm></au>
    <au><snm>Veps{\"a}l{\"a}inen</snm><fnm>A</fnm></au>
    <au><snm>Lesovik</snm><fnm>GB</fnm></au>
    <au><snm>Blatter</snm><fnm>G</fnm></au>
    <au><snm>Paraoanu</snm><fnm>GS</fnm></au>
  </aug>
  <source>npj Quantum Information</source>
  <publisher>Nature Publishing Group</publisher>
  <pubdate>2018</pubdate>
  <volume>4</volume>
  <issue>1</issue>
  <fpage>29</fpage>
</bibl>

<bibl id="B40">
  <title><p>Optimized quantum sensing with a single electronspin using
  real-time adaptive measurements</p></title>
  <aug>
    <au><snm>Bonato</snm><fnm>C.</fnm></au>
    <au><snm>Blok</snm><fnm>M.S.</fnm></au>
    <au><snm>Dinani</snm><fnm>H. T.</fnm></au>
    <au><snm>Berry</snm><fnm>D. W.</fnm></au>
    <au><snm>Markham</snm><fnm>M. L.</fnm></au>
    <au><snm>Twitchen</snm><fnm>D. J.</fnm></au>
    <au><snm>R.</snm><fnm>H</fnm></au>
  </aug>
  <source>Nature Nanotechnology</source>
  <publisher>Nature Publishing Group</publisher>
  <pubdate>2016</pubdate>
  <volume>11</volume>
  <fpage>247</fpage>
  <lpage>252</lpage>
  <url>https://www.nature.com/articles/nnano.2015.261</url>
</bibl>

<bibl id="B41">
  <title><p>Machine Learning for Precise Quantum Measurement</p></title>
  <aug>
    <au><snm>Hentschel</snm><fnm>A</fnm></au>
    <au><snm>Sanders</snm><fnm>BC</fnm></au>
  </aug>
  <source>Phys. Rev. Lett.</source>
  <publisher>American Physical Society</publisher>
  <pubdate>2010</pubdate>
  <volume>104</volume>
  <fpage>063603</fpage>
  <url>https://link.aps.org/doi/10.1103/PhysRevLett.104.063603</url>
</bibl>

<bibl id="B42">
  <title><p>Efficient Algorithm for Optimizing Adaptive Quantum Metrology
  Processes</p></title>
  <aug>
    <au><snm>Hentschel</snm><fnm>A</fnm></au>
    <au><snm>Sanders</snm><fnm>BC</fnm></au>
  </aug>
  <source>Phys. Rev. Lett.</source>
  <publisher>American Physical Society</publisher>
  <pubdate>2011</pubdate>
  <volume>107</volume>
  <fpage>233601</fpage>
  <url>https://link.aps.org/doi/10.1103/PhysRevLett.107.233601</url>
</bibl>

<bibl id="B43">
  <title><p>Differential Evolution for Many-Particle Adaptive Quantum
  Metrology</p></title>
  <aug>
    <au><snm>Lovett</snm><fnm>NB</fnm></au>
    <au><snm>Crosnier</snm><fnm>C</fnm></au>
    <au><snm>Perarnau Llobet</snm><fnm>M</fnm></au>
    <au><snm>Sanders</snm><fnm>BC</fnm></au>
  </aug>
  <source>Physical Review Letters</source>
  <publisher>APS</publisher>
  <pubdate>2013</pubdate>
  <volume>110</volume>
  <issue>1</issue>
  <fpage>220501</fpage>
</bibl>

<bibl id="B44">
  <title><p>Single-shot adaptive measurement for quantum-enhanced
  metrology</p></title>
  <aug>
    <au><snm>Palittpongarnpim</snm><fnm>P</fnm></au>
    <au><snm>Wittek</snm><fnm>P</fnm></au>
    <au><snm>Sanders</snm><fnm>BC</fnm></au>
  </aug>
  <source>Quantum Communications and Quantum Imaging XIV</source>
  <pubdate>2016</pubdate>
  <volume>9980</volume>
  <fpage>99800H</fpage>
</bibl>

<bibl id="B45">
  <title><p>Controlling adaptive quantum phase estimation with scalable
  reinforcement learning</p></title>
  <aug>
    <au><snm>Palittapongarnpim</snm><fnm>P</fnm></au>
    <au><snm>Wittek</snm><fnm>P</fnm></au>
    <au><snm>Sanders</snm><fnm>BC</fnm></au>
  </aug>
  <source>24th European Symposium on Artificial Neural Networks, Bruges, April
  27--29, 2016</source>
  <pubdate>2016</pubdate>
  <fpage>327</fpage>
  <lpage>-332</lpage>
</bibl>

<bibl id="B46">
  <title><p>Learning in quantum control: High-dimensional global optimization
  for noisy quantum dynamics</p></title>
  <aug>
    <au><snm>Palittapongarnpim</snm><fnm>P</fnm></au>
    <au><snm>Wittek</snm><fnm>P</fnm></au>
    <au><snm>Zahedinejad</snm><fnm>E</fnm></au>
    <au><snm>Vedaie</snm><fnm>S</fnm></au>
    <au><snm>Sanders</snm><fnm>BC</fnm></au>
  </aug>
  <source>Neurocomputing</source>
  <publisher>Elsevier</publisher>
  <pubdate>2017</pubdate>
  <volume>268</volume>
  <fpage>116</fpage>
  <lpage>-126</lpage>
</bibl>

<bibl id="B47">
  <title><p>Robustness of Adaptive Quantum-Enhanced Phase
  Estimation</p></title>
  <aug>
    <au><snm>Palittapongarnpim</snm><fnm>P</fnm></au>
    <au><snm>Sanders</snm><fnm>BC</fnm></au>
  </aug>
  <source>arXiv preprint arXiv:1809.05525</source>
  <pubdate>2018</pubdate>
</bibl>

<bibl id="B48">
  <title><p>Experimental Phase Estimation Enhanced by Machine
  Learning</p></title>
  <aug>
    <au><snm>Lumino</snm><fnm>A</fnm></au>
    <au><snm>Polino</snm><fnm>E</fnm></au>
    <au><snm>Rab</snm><fnm>AS</fnm></au>
    <au><snm>Milani</snm><fnm>G</fnm></au>
    <au><snm>Spagnolo</snm><fnm>N</fnm></au>
    <au><snm>Wiebe</snm><fnm>N</fnm></au>
    <au><snm>Sciarrino</snm><fnm>F</fnm></au>
  </aug>
  <source>Physical Review Applied</source>
  <publisher>APS</publisher>
  <pubdate>2018</pubdate>
  <volume>10</volume>
  <issue>1</issue>
  <fpage>044033</fpage>
</bibl>

<bibl id="B49">
  <title><p>Statistical limits of supervised quantum learning</p></title>
  <aug>
    <au><snm>Ciliberto</snm><fnm>C</fnm></au>
    <au><snm>Rocchetto</snm><fnm>A</fnm></au>
    <au><snm>Rudi</snm><fnm>A</fnm></au>
    <au><snm>Wossnig</snm><fnm>L</fnm></au>
  </aug>
  <source>Phys. Rev. A</source>
  <publisher>American Physical Society</publisher>
  <pubdate>2020</pubdate>
  <volume>102</volume>
  <fpage>042414</fpage>
  <url>https://link.aps.org/doi/10.1103/PhysRevA.102.042414</url>
</bibl>

<bibl id="B50">
  <title><p>Quantum Computation and Quantum Information</p></title>
  <aug>
    <au><snm>Nielsen</snm><fnm>MA</fnm></au>
    <au><snm>Chuang</snm><fnm>I</fnm></au>
  </aug>
  <source>Quantum Computation and Quantum Information</source>
  <publisher>Cambridge, U.K.: Cambridge University Press</publisher>
  <pubdate>2000</pubdate>
  <url>https://doi.org/10.1007/978-1-4614-1800-9\_149</url>
</bibl>

<bibl id="B51">
  <title><p>Differential evolution--a simple and efficient heuristic for global
  optimization over continuous spaces</p></title>
  <aug>
    <au><snm>Storn</snm><fnm>R</fnm></au>
    <au><snm>Price</snm><fnm>K</fnm></au>
  </aug>
  <source>Journal of global optimization</source>
  <publisher>Springer</publisher>
  <pubdate>1997</pubdate>
  <volume>11</volume>
  <issue>4</issue>
  <fpage>341</fpage>
  <lpage>-359</lpage>
</bibl>

<bibl id="B52">
  <title><p>Differential evolution: a practical approach to global
  optimization</p></title>
  <aug>
    <au><snm>Price</snm><fnm>K</fnm></au>
    <au><snm>Storn</snm><fnm>RM</fnm></au>
    <au><snm>Lampinen</snm><fnm>JA</fnm></au>
  </aug>
  <publisher>Springer Science \& Business Media</publisher>
  <pubdate>2006</pubdate>
</bibl>

<bibl id="B53">
  <title><p>Particle swarm optimization</p></title>
  <aug>
    <au><snm>Kennedy</snm><fnm>J</fnm></au>
  </aug>
  <source>Encyclopedia of machine learning</source>
  <publisher>Springer</publisher>
  <pubdate>2011</pubdate>
  <fpage>760</fpage>
  <lpage>-766</lpage>
</bibl>

<bibl id="B54">
  <title><p>A new optimizer using particle swarm theory</p></title>
  <aug>
    <au><snm>Eberhart</snm><fnm>R</fnm></au>
    <au><snm>Kennedy</snm><fnm>J</fnm></au>
  </aug>
  <source>Micro Machine and Human Science, 1995. MHS'95., Proceedings of the
  Sixth International Symposium on</source>
  <pubdate>1995</pubdate>
  <fpage>39</fpage>
  <lpage>-43</lpage>
</bibl>

<bibl id="B55">
  <title><p>A modified particle swarm optimizer</p></title>
  <aug>
    <au><snm>Shi</snm><fnm>Y</fnm></au>
    <au><snm>Eberhart</snm><fnm>R</fnm></au>
  </aug>
  <source>Evolutionary Computation Proceedings, 1998. IEEE World Congress on
  Computational Intelligence., The 1998 IEEE International Conference
  on</source>
  <pubdate>1998</pubdate>
  <fpage>69</fpage>
  <lpage>-73</lpage>
</bibl>

<bibl id="B56">
  <title><p>Optimizing qubit phase estimation</p></title>
  <aug>
    <au><snm>Chapeau Blondeau</snm><fnm>F.</fnm></au>
  </aug>
  <source>Phys. Rev. A</source>
  <publisher>American Physical Society</publisher>
  <pubdate>2016</pubdate>
  <volume>94</volume>
  <fpage>022334</fpage>
  <url>https://link.aps.org/doi/10.1103/PhysRevA.94.022334</url>
</bibl>

<bibl id="B57">
  <title><p>Efficient qubit phase estimation using adaptive
  measurements</p></title>
  <aug>
    <au><snm>RodrÃ­guez GarcÃ­a</snm><fnm>MA</fnm></au>
    <au><snm>1</snm><fnm>I</fnm></au>
    <au><snm>Barberis Blostein</snm><fnm>P.</fnm></au>
  </aug>
  <source>arXiv preprint arXiv:2012.11088</source>
  <pubdate>2020</pubdate>
</bibl>

<bibl id="B58">
  <title><p>Quantum metrology with full and fast quantum control</p></title>
  <aug>
    <au><snm>Sekatski</snm><fnm>P</fnm></au>
    <au><snm>Skotiniotis</snm><fnm>M</fnm></au>
    <au><snm>Ko{\l}ody{\'n}ski</snm><fnm>J</fnm></au>
    <au><snm>D{\"u}r</snm><fnm>W</fnm></au>
  </aug>
  <source>Quantum</source>
  <publisher>Verein zur F{\"o}rderung des Open Access Publizierens in den
  Quantenwissenschaften</publisher>
  <pubdate>2017</pubdate>
  <volume>1</volume>
  <fpage>27</fpage>
</bibl>

<bibl id="B59">
  <title><p>Generalized partial measurements</p></title>
  <aug>
    <au><snm>Paraoanu</snm><fnm>G. S.</fnm></au>
  </aug>
  <source>{EPL} (Europhysics Letters)</source>
  <publisher>{IOP} Publishing</publisher>
  <pubdate>2011</pubdate>
  <volume>93</volume>
  <issue>6</issue>
  <fpage>64002</fpage>
  <url>https://doi.org/10.1209/0295-5075/93/64002</url>
</bibl>

<bibl id="B60">
  <title><p>Adaptive single-shot phase measurements: The full quantum
  theory</p></title>
  <aug>
    <au><snm>Higgins</snm><fnm>B. L.</fnm></au>
    <au><snm>Berry</snm><fnm>D. W.</fnm></au>
    <au><snm>Bartlett</snm><fnm>S. D.</fnm></au>
    <au><snm>Wiseman</snm><fnm>H. M.</fnm></au>
    <au><snm>Pryde</snm><fnm>G. J.</fnm></au>
  </aug>
  <source>Nature</source>
  <publisher>NPG</publisher>
  <pubdate>2007</pubdate>
  <volume>450</volume>
  <fpage>393</fpage>
  <lpage>396</lpage>
</bibl>

<bibl id="B61">
  <title><p>Experimental optical phase measurement approaching the exact
  Heisenberg limit</p></title>
  <aug>
    <au><snm>Daryanoosh</snm><fnm>S.</fnm></au>
    <au><snm>Slussarenko</snm><fnm>S.</fnm></au>
    <au><snm>Berry</snm><fnm>D. W.</fnm></au>
    <au><snm>Wiseman</snm><fnm>H. M.</fnm></au>
    <au><snm>Pryde</snm><fnm>G. J.</fnm></au>
  </aug>
  <source>Nature Communications</source>
  <publisher>NPG</publisher>
  <pubdate>2018</pubdate>
  <volume>9</volume>
  <fpage>4606</fpage>
</bibl>

<bibl id="B62">
  <title><p>Interaction-Free Measurements with Superconducting
  Qubits</p></title>
  <aug>
    <au><snm>Paraoanu</snm><fnm>G. S.</fnm></au>
  </aug>
  <source>Phys. Rev. Lett.</source>
  <publisher>American Physical Society</publisher>
  <pubdate>2006</pubdate>
  <volume>97</volume>
  <fpage>180406</fpage>
  <url>https://link.aps.org/doi/10.1103/PhysRevLett.97.180406</url>
</bibl>

<bibl id="B63">
  <title><p>Quantum-enhanced magnetometry by phase estimation algorithms with a
  single artificial atom</p></title>
  <aug>
    <au><snm>Danilin</snm><fnm>S.</fnm></au>
    <au><snm>Lebedev</snm><fnm>A. V.</fnm></au>
    <au><snm>VepsÃ¤lÃ¤inen</snm><fnm>A.</fnm></au>
    <au><snm>Lesovik</snm><fnm>G. B.</fnm></au>
    <au><snm>Blatter</snm><fnm>G.</fnm></au>
    <au><snm>Paraoanu</snm><fnm>G. S.</fnm></au>
  </aug>
  <source>npj Quantum Information</source>
  <publisher>Springer Science and Business Media LLC</publisher>
  <pubdate>2018</pubdate>
  <volume>4</volume>
  <issue>1</issue>
  <url>http://dx.doi.org/10.1038/s41534-018-0078-y</url>
</bibl>

<bibl id="B64">
  <title><p>Quantum systems under frequency modulation</p></title>
  <aug>
    <au><snm>Silveri</snm><fnm>M. P.</fnm></au>
    <au><snm>Tuorila</snm><fnm>J. A.</fnm></au>
    <au><snm>Thuneberg</snm><fnm>E. V.</fnm></au>
    <au><snm>Paraoanu</snm><fnm>G. S.</fnm></au>
  </aug>
  <source>Rep. Prog. Phys.</source>
  <publisher>IOP Publishing</publisher>
  <pubdate>2017</pubdate>
  <volume>80</volume>
  <fpage>056002</fpage>
  <url>https://iopscience.iop.org/article/10.1088/1361-6633/aa5170/meta</url>
</bibl>

<bibl id="B65">
  <title><p>Quantum metrology with a transmon qutrit</p></title>
  <aug>
    <au><snm>Shlyakhov</snm><fnm>A. R.</fnm></au>
    <au><snm>Zemlyanov</snm><fnm>V. V.</fnm></au>
    <au><snm>Suslov</snm><fnm>M. V.</fnm></au>
    <au><snm>Lebedev</snm><fnm>A. V.</fnm></au>
    <au><snm>Paraoanu</snm><fnm>G. S.</fnm></au>
    <au><snm>Lesovik</snm><fnm>G. B.</fnm></au>
    <au><snm>Blatter</snm><fnm>G.</fnm></au>
  </aug>
  <source>Physical Review A</source>
  <publisher>American Physical Society (APS)</publisher>
  <pubdate>2018</pubdate>
  <volume>97</volume>
  <issue>2</issue>
  <url>http://dx.doi.org/10.1103/PhysRevA.97.022115</url>
</bibl>

<bibl id="B66">
  <title><p>Quantum sensing with superconducting circuits</p></title>
  <aug>
    <au><snm>Danilin S.</snm><fnm>WM</fnm></au>
  </aug>
  <source>arXiv:2103.11022</source>
  <pubdate>2021</pubdate>
</bibl>

<bibl id="B67">
  <title><p>Linear Ascending Metrological Algorithm</p></title>
  <aug>
    <au><snm>Perelshtein</snm><fnm>M. R.</fnm></au>
    <au><snm>Kirsanov</snm><fnm>N. S.</fnm></au>
    <au><snm>Zemlyanov</snm><fnm>V. V.</fnm></au>
    <au><snm>Lebedev</snm><fnm>A. V.</fnm></au>
    <au><snm>Blatter</snm><fnm>G.</fnm></au>
    <au><snm>Vinokur</snm><fnm>V. M.</fnm></au>
    <au><snm>Lesovik</snm><fnm>G. B.</fnm></au>
  </aug>
  <source>Phys. Rev. Research</source>
  <publisher>American Physical Society</publisher>
  <pubdate>2021</pubdate>
  <volume>3</volume>
  <fpage>013257</fpage>
  <url>https://link.aps.org/doi/10.1103/PhysRevResearch.3.013257</url>
</bibl>

<bibl id="B68">
  <title><p>Ultrasensitive Magnetometer using a Single Atom</p></title>
  <aug>
    <au><snm>Baumgart</snm><fnm>I.</fnm></au>
    <au><snm>Cai</snm><fnm>J. M.</fnm></au>
    <au><snm>Retzker</snm><fnm>A.</fnm></au>
    <au><snm>Plenio</snm><fnm>M. B.</fnm></au>
    <au><snm>Wunderlich</snm><fnm>C</fnm></au>
  </aug>
  <source>Phys. Rev. Lett.</source>
  <publisher>American Physical Society</publisher>
  <pubdate>2016</pubdate>
  <volume>116</volume>
  <fpage>240801</fpage>
  <url>https://link.aps.org/doi/10.1103/PhysRevLett.116.240801</url>
</bibl>

<bibl id="B69">
  <title><p>Quantum gates and memory using microwave-dressed states</p></title>
  <aug>
    <au><snm>Timoney</snm><fnm>N.</fnm></au>
    <au><snm>Baumgart</snm><fnm>I.</fnm></au>
    <au><snm>Johanning</snm><fnm>M.</fnm></au>
    <au><snm>Var{\`o}n</snm><fnm>A. F.</fnm></au>
    <au><snm>Plenio</snm><fnm>M.B.</fnm></au>
    <au><snm>A.</snm><fnm>R</fnm></au>
    <au><snm>Ch.</snm><fnm>W</fnm></au>
  </aug>
  <source>Nature</source>
  <publisher>Nature Publishing Group</publisher>
  <pubdate>2011</pubdate>
  <volume>476</volume>
  <fpage>185</fpage>
  <lpage>188</lpage>
  <url>https://www.nature.com/articles/nature10319</url>
</bibl>

<bibl id="B70">
  <title><p>High-sensitivity diamond magnetometer with nanoscale
  resolution</p></title>
  <aug>
    <au><snm>Taylor</snm><fnm>J. M.</fnm></au>
    <au><snm>Cappellaro</snm><fnm>P.</fnm></au>
    <au><snm>Childress</snm><fnm>L.</fnm></au>
    <au><snm>Jiang</snm><fnm>L.</fnm></au>
    <au><snm>Budker</snm><fnm>D.</fnm></au>
    <au><snm>Hemmer</snm><fnm>P. R.</fnm></au>
    <au><snm>Yacoby</snm><fnm>A.</fnm></au>
    <au><snm>Walsworth</snm><fnm>R.</fnm></au>
    <au><snm>Lukin</snm><fnm>M. D.</fnm></au>
  </aug>
  <source>Nature Physics</source>
  <publisher>NPG</publisher>
  <pubdate>2008</pubdate>
  <volume>4</volume>
  <fpage>810</fpage>
  <lpage>816</lpage>
</bibl>

<bibl id="B71">
  <title><p>Sensitivity optimization for NV-diamond magnetometry</p></title>
  <aug>
    <au><snm>Barry</snm><fnm>JF</fnm></au>
    <au><snm>Schloss</snm><fnm>JM</fnm></au>
    <au><snm>Bauch</snm><fnm>E</fnm></au>
    <au><snm>Turner</snm><fnm>MJ</fnm></au>
    <au><snm>Hart</snm><fnm>CA</fnm></au>
    <au><snm>Pham</snm><fnm>LM</fnm></au>
    <au><snm>Walsworth</snm><fnm>RL</fnm></au>
  </aug>
  <source>Rev. Mod. Phys.</source>
  <publisher>American Physical Society</publisher>
  <pubdate>2020</pubdate>
  <volume>92</volume>
  <fpage>015004</fpage>
  <url>https://link.aps.org/doi/10.1103/RevModPhys.92.015004</url>
</bibl>

</refgrp>
} 

\end{backmatter}
\end{document}